\documentclass[twocolumn]{aastex62}
\usepackage{bm}
\usepackage{hyperref}
\usepackage{breakcites} 
\usepackage{lineno}

\newcommand{\eqb}{\begin{eqnarray}}
\newcommand{\eqe}{\end{eqnarray}}
\newcommand{\bi}{\begin{itemize}}
\newcommand{\ei}{\end{itemize}}

\def\comp{\,c/\omega_{p}}

\newcommand{\sigeh}{\sigma_{e, h}}
\newcommand{\sigec}{\sigma_{e, c}}
\newcommand{\sigih}{\sigma_{i,h}} 
\newcommand{\sigic}{\sigma_{i,c}} 
\newcommand{\rLe}{\rho_{Le}}
\newcommand{\rLi}{\rho_{Li}}

\newcommand{\alf}{Alfv\'en}
\newcommand{\sect}[1]{Sect.~\ref{sec:#1}}
 
\newcommand{\tab}[1]{Table~\ref{tab:#1}}
\newcommand{\eq}[1]{Eq.~(\ref{eq:#1})}
\newcommand{\fign}[1]{Fig.~\ref{fig:#1}}

\newcommand{\nee}{n_{e^\pm}}

\graphicspath{{./}{figures/}}

\received{}
\revised{}
\accepted{}
\submitjournal{ApJ}

\shorttitle{Pair-proton reconnection}
\shortauthors{Petropoulou et al.}

\begin{document}

\title{Relativistic Magnetic Reconnection in Electron-Positron-Proton Plasmas:\\
Implications for Jets of Active Galactic Nuclei}

\correspondingauthor{Maria Petropoulou}
\email{m.petropoulou@astro.princeton.edu}

\author[0000-0001-6640-0179]{Maria Petropoulou}
\affil{Department of Astrophysical Sciences, Princeton University \\
4 Ivy Lane, Princeton, NJ 08544, USA}

\author{Lorenzo Sironi}
\affil{Department of Astronomy, Columbia University \\
550 W 120th St, New York, NY 10027, USA}

\author{Anatoly Spitkovsky}
\affil{Department of Astrophysical Sciences, Princeton University \\
4 Ivy Lane, Princeton, NJ 08544, USA}

\author{Dimitrios Giannios}
\affil{Department of Physics \& Astronomy, Purdue University \\
525 Northwestern Avenue, West Lafayette, IN, 47907 USA}
\affil{Department of Physics, University of Crete \\
Voutes, GR-70013, Heraklion, Greece}
\affil{Institute of Astrophysics, Foundation for Research and Technology Hellas \\
Voutes, GR-70013, Heraklion, Greece}


\begin{abstract}
Magnetic reconnection is often invoked to explain the non-thermal radiation of relativistic outflows, including jets of active galactic nuclei (AGN). Motivated by the largely unknown plasma composition of AGN jets, we study reconnection in the unexplored regime of electron-positron-proton (pair-proton) plasmas with large-scale two-dimensional  particle-in-cell simulations. We cover a wide range of pair multiplicities (lepton-to-proton number ratio $\kappa=1-199$) for different values of the all-species plasma magnetization ($\sigma=1,3$ and 10) and electron temperature ($\Theta_e\equiv kT_e/m_ec^2=0.1-100$).  We focus on the dependence of the post-reconnection  energy partition and lepton energy spectra on the hot pair plasma magnetization  $\sigeh$ (i.e., the ratio of magnetic to pair enthalpy densities). We find that the post-reconnection energy is shared roughly equally between magnetic fields, pairs, and protons for $\sigeh\gtrsim 3$. We empirically find that the mean lepton Lorentz factor in the post-reconnection region depends on $\sigma, \Theta_e$, and $\sigeh$ as $\langle \gamma_e-1\rangle \approx \sqrt{\sigma}(1+4\Theta_e)\left(1+\sigeh/30\right)$, for $\sigma\ge1$.
The high-energy part of the post-reconnection lepton energy distributions can be described by a power law, whose slope is mainly controlled by $\sigeh$ for $\kappa \gtrsim 3-6$, with harder power laws obtained for higher magnetizations.  We finally show that reconnection in pair-proton plasmas with multiplicities $\kappa \sim 1-20$, magnetizations $\sigma \sim 1-10$, and temperatures $\Theta_e \sim 1-10$ results in particle power law slopes and average electron Lorentz factors that are consistent with those inferred in leptonic models of AGN jet emission. \end{abstract}


\section{Introduction}\label{sec:intro}
A fundamental question in the physics of astrophysical relativistic outflows is 
how their energy, which is initially carried in the form of Poynting flux, is first transferred to the plasma, and then radiated away to power the observed emission. Magnetic field dissipation via reconnection has been often invoked to explain the non-thermal signatures of pulsar wind nebulae \citep[PWNe; e.g.,][see 
\citealt{sironi_17} for a recent review]{lyubarsky_kirk_01,petri_lyubarsky_07,sironi_spitkovsky_11b,cerutti_12b,philippov_14}, 
gamma-ray bursts \citep[GRBs; e.g.,][]{thompson_94,usov_94,spruit_01,drenkhahn_02a,lyutikov_03,giannios_08, beniamini_17}, and jets from active galactic nuclei \citep[AGN; e.g.,][]{romanova_92,giannios_09,giannios_10b,giannios_13,petropoulou_16, nalewajko_18,christie_19}.  

In most relativistic astrophysical outflows, reconnection proceeds in the so-called relativistic regime in which the \alf \, velocity of the plasma approaches the speed of light (or equivalently the plasma magnetization, defined as the ratio of magnetic to particle enthalpy densities, is $\sigma \gtrsim 1$). The physics of reconnection can only be captured from first principles by means of fully-kinetic particle-in-cell (PIC) simulations. Extensive numerical work on relativistic reconnection of electron-positron (pair) plasmas has been performed in two dimensions \citep[2D; e.g.,][]{zenitani_01,zenitani_07,daughton_07,cerutti_12b,ss_14,guo_14,guo_15a,liu_15,nalewajko_15,sironi_15,sironi_16,
werner_16,kagan_18,petropoulou_18, hakobyan_18} and in three dimensions \citep[3D; e.g.,][]{zenitani_05b,zenitani_08,liu_11,sironi_spitkovsky_11b, sironi_spitkovsky_12,kagan_13,cerutti_13b,ss_14,guo_15a,werner_17}, whereas the study of trans-relativistic and relativistic reconnection in 2D electron-proton plasmas became  possible more recently \citep[e.g.,][]{melzani_14,sironi_15,guo_16,rowan_17,werner_18,ball_18}. 

In contrast to other astrophysical outflows, such as PWNe, the plasma composition of astrophysical jets is largely unknown. On the one hand, there is no direct way of probing the plasma composition in jets and, on the other hand, there are large theoretical uncertainties about the jet baryon loading mechanisms \citep[for recent kinetic simulations of black-hole jet launching, see][]{parfrey_19}. As a result, any attempts to infer the jet plasma composition rely on the modeling of the emitted radiation \citep[e.g.,][]{ghisellini_12,ghisellini_14}, which, however, suffers from degeneracies that are inherent in the radiative models. For AGN jets, in particular, both pair and electron-proton compositions have been discussed in the literature. A pure pair composition in powerful AGN jets (e.g., in flat spectrum radio quasars, FSRQs) is disfavored, since bulk Comptonization of the ambient  low-energy photons by the pairs would result in luminous spectral features in X-rays that are not observed (e.g., \cite{sikora_97, sikora_00}; see however \cite{kammoun_18}). This argument does not apply to less powerful jets (such as BL Lac type sources), since the ambient radiation fields are weak or even absent, and a pure pair plasma cannot be excluded in this case. If jets are devoid of pairs, namely they are composed of electron-proton plasmas, the inferred power (which is dominated by the kinetic power of protons) is large, usually exceeding the accretion power \citep[e.g.][]{ghisellini_14, madejski_16}. A mixed composition with tens of pairs per proton may be more realistic, as it can reduce the inferred jet power by  a factor equal to the lepton-to-proton number ratio, the so-called pair multiplicity  \citep[e.g.][]{ghisellini_10,ghisellini_12,madejski_16}. The presence of pairs in the dissipation regions of jets is also expected to affect the average energy per lepton available for particle heating as well as the efficiency with which non-thermal particles are accelerated. 
 
The goal of this work is to study the general properties of relativistic reconnection in the unexplored regime of plasmas with mixed composition. We focus on  electron-positron-proton (or pair-proton) plasmas, as they bridge the gap between the pair plasma and electron-proton plasma cases that have been extensively studied in the past. We perform a suite of large-scale 2D PIC simulations using the realistic proton-to-electron mass ratio ($m_i/m_e=1836$) while varying three physical parameters, namely the plasma magnetization ($\sigma=1,3$ and 10), the plasma temperature ($\Theta_e \equiv k T_e / m_e c^2 = 0.1-100$ with equal electron and proton temperatures), and the number of pairs per proton ($\kappa=1-199$). In this study, even in cases where the pairs dominate by number, the plasma rest mass energy is governed by protons. We study, for the first time, the inflows and outflows of plasma in the reconnection region, the energy partition between pairs, protons, and magnetic fields, and the energy distributions of accelerated particles as a function of the pair multiplicity. 
 
This paper is organized as follows. In \sect{setup} we describe the setup of our simulations. In \sect{layer} we present the structure of the reconnection layer for different pair multiplicities. In \sect{flows} we focus on the inflow and outflow motions of the plasma  and in \sect{energy} we discuss the energy partition between magnetic fields and different particle species in the reconnection region. In \sect{spectra} we focus on the evolution of the particle energy spectrum, illustrating how the lepton power-law slope depends on the pair multiplicity. In \sect{discussion} we discuss the  astrophysical implications of our findings and conclude in \sect{summary} with a summary of our results. Readers interested primarily in the application of our results to jetted AGN can move directly to \sect{discussion}.

\section{Numerical setup} \label{sec:setup}
We use the 3D electromagnetic PIC code TRISTAN-MP \citep{buneman_93, spitkovsky_05} to study magnetic reconnection in pair-proton plasmas. We explore anti-parallel reconnection, i.e., we set the guide field perpendicular to the alternating fields to be zero. The reconnection layer is initialized as a Harris sheet of length $L$ with the magnetic field $\bm{B}=-B_0\,\tanh\,(2\pi y/\Delta)\,\bm{\hat{x}}$ reversing at $y=0$ over a thickness $\Delta$.  Here, we set $\Delta$=80 $c/\omega_p$, where $\omega_p$ is the all-species plasma frequency defined in \eq{omp_tot}, and choose a spatial resolution of $\comp$=3 computational cells. 

The field strength $B_0$ is defined through the (total) plasma magnetization $\sigma=B_0^2/4\pi h$, where $h$ is the enthalpy density of the unreconnected plasma including all species (see \eq{stot}). The \alf\ speed is related to the magnetization as $v_A/c=\sqrt{\sigma/(\sigma+1)}$. We focus on the regime of relativistic reconnection (i.e., $v_A/c\sim 1$) and explore cases with $\sigma=1, 3$ and 10 (see \tab{setup}). The proton and pair plasmas outside the layer are initialized with the same temperature ($T_i=T_e$). We consider cases where the pairs are initially relativistically hot ($\Theta_e \equiv k T_e / m_e c^2 = 1$, 10, and 100), but for completeness we study also a few cases with initially colder pairs ($\Theta_e=0.1$). In all simulations, the protons are non-relativistic ($\Theta_i \equiv k T_i / m_i c^2 = \Theta_e m_e / m_i \ll 1$).

\begin{deluxetable}{lccccccc}
\tablewidth{0.8\columnwidth} 
\tablecaption{Simulation parameters.\label{tab:setup}}
\tablehead{
\colhead{Run} & \colhead{$\sigma$} & \colhead{$\Theta_e$} & \colhead{$\kappa$} & \colhead{$\sigeh$\tablenotemark{a}} & \colhead{$L/\rLe$} & \colhead{$L/\rLi$} & 
\colhead{$T_{\max}$\tablenotemark{b}}
} 
\startdata
A0* & 1	& 1 & 199 & 2.9 &2706.5 & 13.6 & 1.8 \\
A1 & 1	& 1 & 66 & 6.9 &1769.0 & 26.9 & 1.8 \\
A2 & 1	& 1 & 19 & 21.4 &1004.7 & 52.9 & 1.8 \\
A3 & 1	& 1 & 6	& 69.3 &273.2 & 48.2 & 6.1\tablenotemark{c}\\
A4* & 1	& 1 & 6	& 69.3 &557.9 & 98.4 & 1.7\\	
A5 & 1	& 1 & 6	& 69.3 &1195.4 & 210.9 & 1.3\\
A6 & 1	& 1 & 3	& 130.0 & 398.9 & 133.0 & 1.7\\	
A7 & 1	& 1 & 1.2 &317.6 & 257.7 & 206.6 & 1.7\\	
A8 & 1	& 1 & 1 & 387.9 &245.2 & 245.2 & 5.5 \\
A9 & 1 & 1 & 199 & 2.9& 5799.8  & 29.1 & 1.5 \\
\hline
B0 & 1  & 10 & 199 & 1.2 &4099.7 & 20.6 & 2.0\\
B1 & 1  & 10 & 66 & 1.7 &3492.7 & 53.2 & 1.9\\
B2 & 1  & 10 & 19 & 3.4 &2466.1 & 129.8 & 1.9\\
B3 & 1  & 10 & 6 & 9.0 &1510.7 & 266.6 & 1.8 \\
B4 & 1  & 10 & 3 & 16.1 &1104.6  & 368.2 & 1.8\\
B5 & 1  & 10 & 1 & 46.4 &688.2 & 688.2 & 3.1\\
\hline
C1 & 3  & 1  & 199 & 8.8 &1562.6 & 7.85 & 1.7 \\
C2 & 3  & 1  & 66 & 20.7 &1021.3 & 15.5 & 1.6\\
C3 & 3  & 1  & 19 & 64.1 &580.1 & 30.5 & 1.5\\
C4* & 3  & 1  & 6 & 207.8 &322.1 & 56.8 & 3.1\\
C5 & 3  & 1  & 6 &  207.8 &690.2 & 121.8 & 1.3\\
C6 & 3  & 1  & 1 & 1163.8 &141.5 & 141.5 & 3.9\tablenotemark{c}\\
\hline
D1 & 3  & 10 & 66 & 5.1 &1975.4 & 30.1 & 1.2\\
D2 & 3  & 10 & 19 & 10.2 &1423.8 & 74.9 & 1.7\\
D3 & 3  & 10 & 6  & 27.1 &872.2 & 153.9 & 1.5\\
D4 & 3  & 10 & 1 & 139.3 &397.3 & 397.3 & 3.9\\
\hline
E1 & 10  & 1 & 199 & 29.4 & 838.4 & 4.2 & 1.5 \\
E2 & 10  & 1 & 19 & 213.6 &311.2 & 16.4 & 1.4\\
E3* & 10  & 1 & 1 & 3879.2 &77.5 & 77.5 & 3.9\tablenotemark{c}\\
E4 & 10 & 1 & 1 & 3879.2 & 229.8 & 229.8 & 1.0 \\
\hline
F1 & 10  & 10 & 199 & 12.9 &1296.4 & 6.5 & 1.6\\
F2 & 10  & 10 & 6 & 90.2 &468.0 & 82.6 & 1.5\\
F3 & 10  & 10 & 1 & 464.5 &217.6 & 217.6 & 3.9\\
\hline
G1 & 1 	& 0.1 & 19 & 76.4 &1230.6 & 64.8 & 1.5\\
G2 & 1  & 0.1 &  3 & 478.3 &491.7 & 163.9 & 1.3\\
\hline 
H1 & 1  & 100 & 199 & 1.0 &4376.9 & 22.0 & 1.9\\
H2 & 1  & 100 & 19 & 1.3 &3911.1 & 205.8 & 2.0\\
H3 & 1  & 100 & 3 & 2.7 &2610.6 & 870.2 & 2.0 \\
H4 & 1 & 100 & 1 & 6.2 &1758.0 & 1758.0 & 6.0\\ 
\hline
\enddata
\tablenotetext{a}{Hot pair plasma magnetization defined in \eq{seh}.}
\tablenotetext{b}{Duration of the simulation in units of $L/c$.}
\tablenotetext{c}{The reconnection rate decreases after $\sim 3L/c$ due to the formation of a large boundary island.}
\tablecomments{For simulations performed with the same physical parameters but different box sizes, we mark the default cases for display in the figures  with an asterisk (*). Simulations with $\kappa=1$ and $\kappa>1$ are performed with 4 and 32 computational particles per cell, respectively. In all cases, the plasma skin depth $c/\omega_p$ is resolved with 3 computational cells and the typical domain size is $L/(c/\omega_p) \simeq5200-11200$.}
\end{deluxetable} 

Let $N_{ppc}$ denote the total number of computational particles per cell, which is equally partitioned between negatively and positively charged particles. If $q=2/(\kappa+1)$ denotes the physical number ratio of protons to electrons in a plasma with pair multiplicity $\kappa$, then the number of computational protons and positrons per cell is given, respectively, by $(q/2)N_{ppc}$ and $[(1-q)/2]N_{ppc}$\footnote{We fill the cells with particles by performing two cycles of injection. We first inject protons and electrons at equal numbers (i.e., $N_1=(q/2)N_{ppc}$ per cell) and then inject positrons and electrons, with a number of $N_2=[(1-q)/2]N_{ppc}$ per cell for each component. The injection is not done on a cell-by-cell basis, but in slabs partitioned along the $y$ direction with $N_{cells}$ each, which are handled by different computer cores. When either $N_1\times N_{cells}$ or $N_2\times N_{cells}$ is $<10$, the actual number of particles injected is randomly drawn from a Poisson distribution with mean value equal to $N_1\times N_{cells}$ or $N_2\times N_{cells}$, respectively.}. We varied $N_{ppc}$ from 4 to 64 and checked the convergence of our results in regard to the reconnection rate, outflow four-velocity, and particle energy distributions. For pair-proton simulations with high pair multiplicity (e.g., $\kappa>10$), we need to use $N_{ppc}>16$ to achieve convergence (within a few percent in inflow rate and outflow four-velocity), whereas for electron-proton simulations we find that 4 particles per cell are sufficient. 
For cases with high $\kappa$ (low $q$) there is a low probability of proton ``injection'' in a given cell due to the small (physical) fraction of protons per electron. This introduces an appreciable level of shot noise in fluid quantities that are computed from (or governed by) the protons (e.g., outflow four-velocity), which can be mitigated by increasing the number of computational particles per cell.

The magnetic pressure outside the current sheet is balanced by the particle pressure in the sheet. This is achieved by adding a component of hot plasma with the same composition as in the upstream region and over-density $\eta=3$ relative to the all-species number density outside the layer. We exclude the hot particles initialized in the current sheet from the particle energy spectra and from all thermodynamical quantities (except the plasma number density), as their properties depend on our choice of the sheet initialization.

Our simulations are performed in a 2D domain, but all three components of the velocity and of the electromagnetic fields are tracked. We adopt periodic boundary conditions in the $x$ direction of the reconnection outflow and we employ an expanding simulation box in the $y$ direction (i.e., the direction of the reconnection inflow). We also use two moving injectors receding from $y=0$ along $\pm {\bm{\hat{y}}}$, which constantly introduce fresh magnetized plasma into the simulation domain \citep[for details, see][]{sironi_spitkovsky_11a, ss_14, sironi_16, rowan_17, ball_18}. In all cases, the box size along the $y$ direction increases over time and by the end of the simulation it is comparable or larger than the $x$ extent.  

We trigger reconnection at the center of the simulation domain by instantaneously removing the pressure of hot particles that were initialized in the sheet \citep{sironi_16, ball_18}. This causes a central collapse of the current sheet and the formation of two ``reconnection fronts'' that are are pulled along the layer towards the edges of the box by the magnetic tension force and reach the boundaries at $t \sim L/2 v_A$. The main advantage of this simulation setup is that the results are independent of the initialization of the sheet (i.e., overdensity, temperature, and thickness\footnote{This is true if the sheet is thick enough so that it does not become spontaneously tearing unstable at locations that have not been swept up yet by the receding reconnection fronts.}) in contrast to the untriggered cases where the influence of the initial conditions may affect the temporal evolution of the reconnection rate and the particle energy distributions at early times \citep[see, e.g., Fig.~4 of][]{petropoulou_18}.  

We choose as our typical unit of length the Larmor radius of electrons ($\rLe$) with Lorentz factor equal to the cold pair plasma magnetization ($\sigec$), namely $\rLe=\sigec m_e c^2/e B_0$, implicitly assuming  that reconnection transfers all the magnetic energy to relativistic pairs (for definitions, see \eq{rLe} and \eq{sec}). The proton Larmor radius is defined in a similar way, i.e., $\rLi=\sigic m_i c^2/e B_0$, where $\sigic$ is the cold proton plasma magnetization (see \eq{sic}). The size of the computational domain along the reconnection layer $L$ ranges from hundreds to thousands of $\rLe$ and tens to hundreds of $\rLi$ (see \tab{setup}). The fact that the Larmor radii change as a function of pair multiplicity is a direct result of our choice to fix the total $\sigma$ and electron thermal spread $\Theta_e$, as shown in \fign{rho_ionfrac} of Appendix~\ref{sec:app1}. 

A key parameter in our study, as it will become clear in the following sections, is the hot pair plasma magnetization. This is defined as $\sigeh\equiv B_0^2/4\pi h_{\pm}$, where $h_{\pm}$ is the enthalpy density of the upstream pair plasma, and it relates to the total $\sigma$ as:
\eqb
\frac{\sigeh}{\sigma} =\frac{q \left(\frac{m_i}{m_e} + \Theta_e\frac{\hat{\gamma}_i}{\hat{\gamma}_i-1}\right) +(2-q)\left(1+\Theta_e\frac{\hat{\gamma}_e}{\hat{\gamma}_e-1} \right)}{(2-q) \left(1+\Theta_e\frac{\hat{\gamma}_e}{\hat{\gamma}_e-1}\right)},
\label{eq:seh}
\eqe  
where $q=2/(\kappa+1)$ is the ratio of proton-to-electron number densities and $\hat{\gamma}_{i,e}$ are the adiabatic indices of protons and leptons.
\eq{seh} can be simplified in the following asymptotic regimes:
\begin{itemize}
    \item relativistically cold electrons ($\Theta_e \ll 1$). Here, $\sigeh \approx \sigma [m_i/m_e + \kappa]/\kappa$. For electron-proton plasmas (or in general, if $\kappa \ll m_i/m_e$) this reduces to the well-known result $\sigeh \approx \sigma m_i/m_e$, whereas for pair-dominated plasmas with  $\kappa \gg m_i/m_e$, we find $\sigeh \approx \sigma$. Although pairs are cold, if their number density is sufficiently high, like in the latter case, their pressure  (which is $\propto \kappa \Theta_e$) can be more important than the proton rest-mass energy density. 
    \item relativistically hot electrons ($1 < \Theta_e < m_i/m_e$). Here,  $\sigeh \approx \sigma [m_i/m_e + 4\Theta_e \kappa]/4\Theta_e\kappa$. This reduces to $\sigeh \approx \sigma m_i/ (4 m_e \Theta_e)$ for $\kappa=1$, while for $\kappa \gg (m_i/m_e)/4\Theta_e$ we find $\sigeh \approx \sigma$. In the latter case, the pressure of the hot pairs is large enough to dominate over the rest-mass energy density of protons. Note that the critical pair multiplicity here is lower by a factor of $\sim 4\Theta_e$ compared to the cold electron case (see first bullet point).
    \item relativistically hot protons ($\Theta_e \gg m_i/m_e$). In this ultra-relativistic regime, all fundamental plasma scales (e.g., the plasma frequencies and skin depths) become independent of the particle rest mass. They depend only on the average particle
energy which, in this regime, is similar for protons and pairs. Here, $\sigeh \approx \sigma$ independent of $\kappa$.
\end{itemize}
In this study, we focus on cases where the protons are non-relativistic and dominate the mass density. We refer the reader to Appendix~\ref{sec:app1}, for the full list of parameters and their definitions.

\begin{figure}
 \centering
 \includegraphics[width=0.49\textwidth, trim= 0 0 20 0]{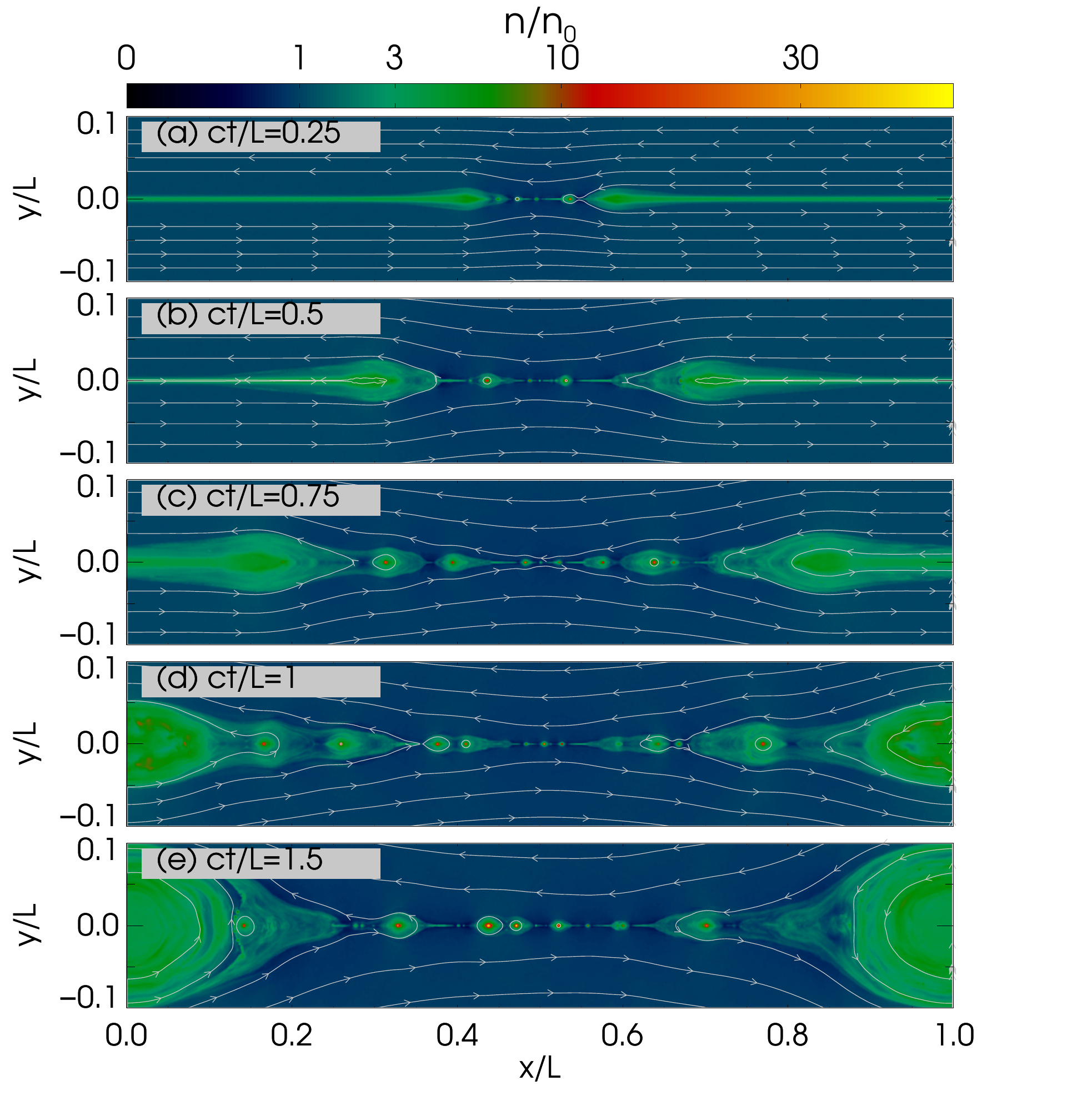}
 \caption{2D structure of the all-species particle number density $n$ (normalized to the number density $n_0$ far from the reconnection layer), from a simulation with $\sigma=1, \Theta_e=1$, and $\kappa=19$ (A2 in \tab{setup}). We show only the region $|y|/L<0.1$ to emphasize the small-scale structures in the reconnection layer (the extent of the computational box along $y$ increases over time, as described in \sect{setup}). The 2D density structure at different times (as marked on the plots) is shown in the panels from top to  bottom, with overplotted magnetic field lines (solid white lines). A movie showing the temporal evolution of the 2D structure of the number density of each particle species can be found at \url{https://bit.ly/2HmZR7j}.
 } 
 \label{fig:fluidtime}
\end{figure}

\begin{figure*}
 \centering
 \includegraphics[width=0.49\textwidth, trim=0  0 0 0]{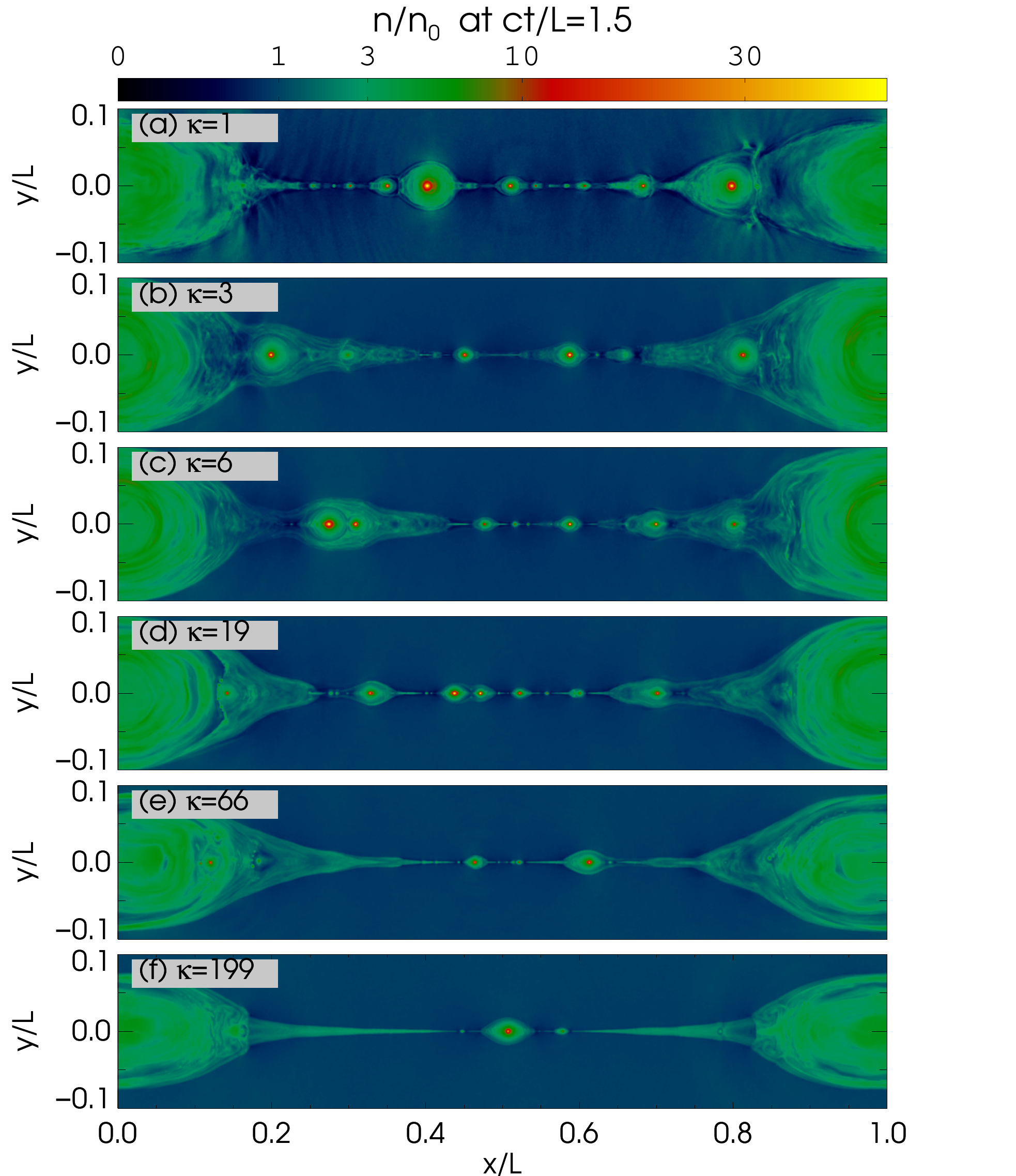}
 \hfill
 \includegraphics[width=0.49\textwidth, trim=0  0 0 0]{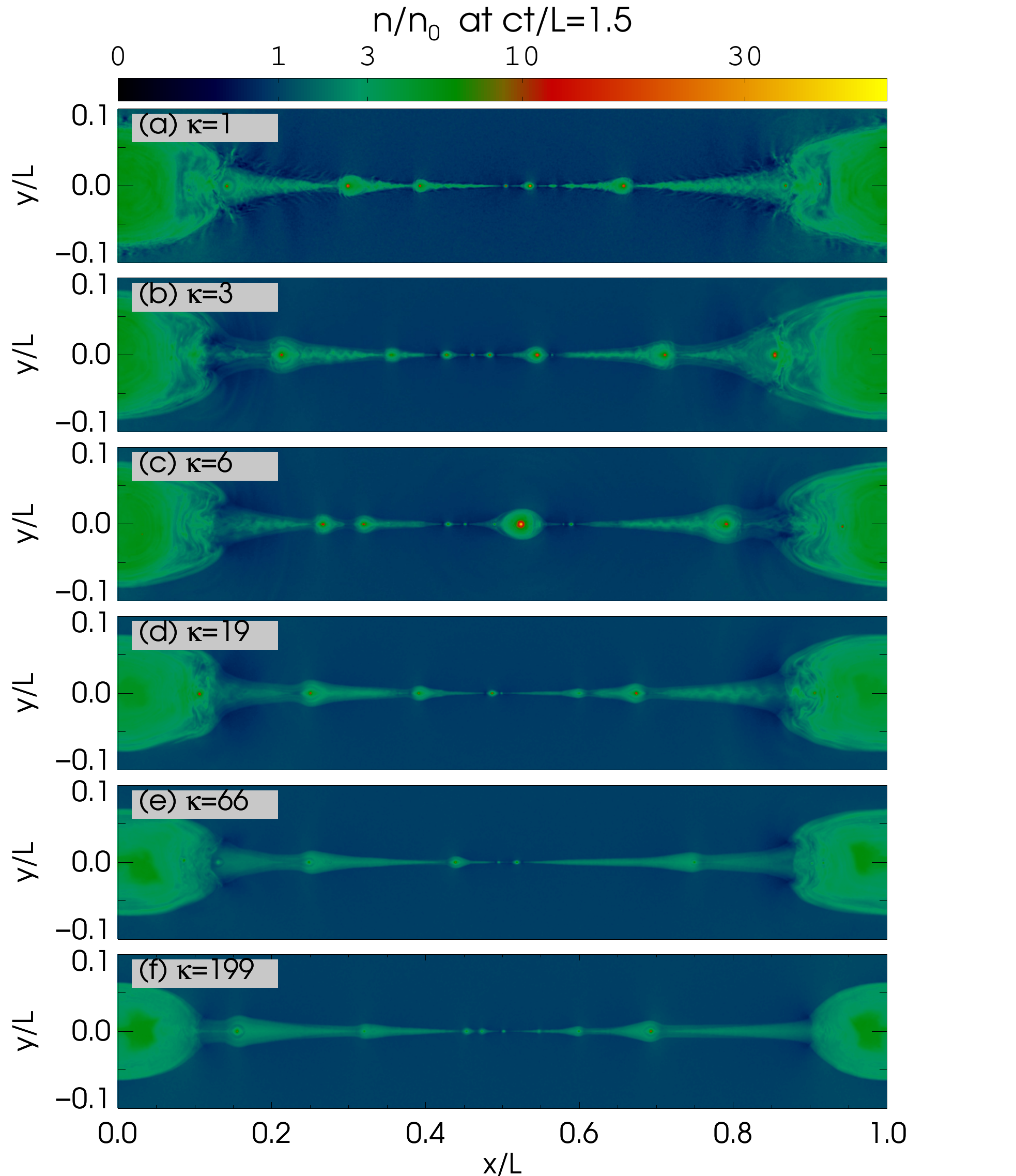}
 \caption{2D snapshots of the all-species particle number density $n$ (normalized to the number density $n_0$ far from the reconnection layer), including the particles initially present in the sheet. Results are displayed at $t=1.5 L/c$ for different values of the pair multiplicity $\kappa$, as marked on each panel. The simulations were performed for plasmas with $\sigma=1$ and $\Theta_e=1$ (left panel) and  $\sigma=1$ and $\Theta_e=10$ (right panel); for reference see cases A0-A2, A4, A6, A8, and B0-B5 in \tab{setup}. The appearance of the layer is similar for cases with similar $\sigeh$ values, as exemplified by panels (e) on the left and (c) on the right sides of the figure (see also \tab{index-gammae}). Movies showing the temporal evolution of the layer structure for different pair multiplicities can be found at \url{https://bit.ly/2HmZR7j}.} 
\label{fig:fluidtime-ionfrac}
\end{figure*}
\section{Structure of the reconnection layer} \label{sec:layer}
\subsection{Temporal evolution}\label{sec:layer-evolution}
To illustrate the temporal evolution of the reconnection region we show in \fign{fluidtime} snapshots of the 2D structure of the particle number density from one of our simulations in a $\sigma=1, \Theta_e=1$ pair-proton plasma with pair multiplicity $\kappa=19$ (A2 in \tab{setup}). The localized (at the center) removal of pressure from the hot particle population initialized in the sheet (see \sect{setup}) causes its collapse, thus leading to the formation of a central  (or primary) X-point. Two reconnection fronts form on opposite sides of the primary X-point and move outwards due to the tension of the magnetic field lines. Plasmoid and secondary X-point formation takes place in the low-density region between the moving fronts, as shown in panels (a) and (b). The fronts reach the boundaries of the simulation domain at $t \approx L/2 v_A \approx 0.7 L/c$ and form the so-called boundary island, whose size eventually becomes a significant fraction of the layer length (here, $\sim 0.4\,L$ as shown in panels d and e). The formation of such a large plasmoid, which is the result of  periodic boundary conditions, will eventually inhibit the inflow of fresh plasma into the layer, thus shutting off the reconnection process. We verified that the reconnection process remains active\footnote{We characterize the reconnection process as active, as long as the inflow rate of plasma into the reconnection region does not show a monotonically decreasing trend with time and remains $\gtrsim 0.01 v_A$ at all times.} for the entire duration of all simulations listed in \tab{setup} except A3, C6, and E3.

\subsection{Dependence on pair multiplicity}\label{sec:layer-ionfrac}
The effect that the pair multiplicity $\kappa$ has on the appearance of the reconnection region is illustrated in \fign{fluidtime-ionfrac}, where 2D snapshots of the all-species particle density, including particles initially present in the sheet, are plotted for increasing values of $\kappa$ (top to bottom) in plasmas with $\sigma=1, \Theta_e=1$ (left panel) and $\sigma=1, \Theta_e=10$ (right panel). In cases with fixed $\sigma$ and $\Theta_e$ but increasing $\kappa$ we find that the plasma outflows along the layer become more uniform (i.e., fewer X-points and plasmoids form in the layer) and the typical size of the plasmoids decreases. For fixed $\sigma$ and $\kappa$, an increasing upstream plasma temperature also leads to smaller plasmoids and less fragmentation in the reconnection region (compare left and right panels in \fign{fluidtime-ionfrac}). 

One might argue that the differences in the appearance of the layer as a function of $\kappa$ are merely a result of the different box sizes in terms of the proton skin depth or alternatively $\rLi$ (see \tab{setup}). To check this possibility, we compare  cases with different physical conditions, but similar box sizes in terms of $\rLi$. We find that the plasma conditions have a major effect on the appearance of the layer (for details, see Appendix~\ref{sec:layer-boxsize}) and that the differences seen in \fign{fluidtime-ionfrac} are not just a numerical artifact.

Empirically, we find that the most important parameter controlling the appearance of the layer turns out to be $\sigeh$. We find that the layer structure is similar for different values of the pair multiplicity and temperature, as long as $\sigeh$ is nearly the same. For example, compare panel (e) on the left side to panel (c) on the right side of \fign{fluidtime-ionfrac}. Typically, the density profile is smoother and the plasmoid sizes are smaller for lower $\sigeh$ values\footnote{ The apparent correlation of the plasmoid size with $\sigeh$ is likely related to the dependence of the electron Larmor radius on  $\sigeh$ (i.e., $\rLe \propto \sigeh^{1/2}$).} (e.g., compare panels (a) and (f) on the left side of \fign{fluidtime-ionfrac}).

Similar results have been presented by \cite{ball_18} (see Fig.~4 therein) for trans-relativistic electron-proton reconnection and an increasing electron plasma $\beta_e$, defined as the ratio of upstream electron plasma pressure and magnetic pressure (see \eq{betae}). 
The similarity of our findings is not unexpected and can be understood as follows. The increasing pair multiplicity corresponds to a decreasing hot pair plasma magnetization $\sigeh$ (see \eq{seh} and \fign{sigma_ionfrac}), which in turn is inversely proportional to $\beta_e$ in the limit of $\kappa \gg 1$ (see \eq{betae}). 

Henceforth, we choose $\sigeh$ over $\beta_e$ to perform our parameter study, since the relative contribution of the rest-mass and internal energy densities to the enthalpy density of the upstream plasma varies among our simulations. In Sections~\ref{sec:energy} and \ref{sec:spectra} we will also demonstrate that $\sigeh$ is the main parameter that regulates the energy partition and the power-law slope of the lepton energy spectrum. 


\section{Inflows and outflows} \label{sec:flows}
To compute the reconnection rate in our simulations, we average at each time 
the inflow speed over a slab centered at $x=0.5 \, L$ with width $0.2\, L$ across the layer (i.e., along the $y$ direction) and  length $0.5 \, L$ (along the $x$ direction). Our results are nearly insensitive to the choice of the slab dimensions as long as the region occupied by the boundary island, where the inflow rate is inhibited, is excluded from the averaging process. The spatially averaged inflow rate is then averaged over time for $t > L/2 v_A$, i.e., excluding times when the reconnection fronts are still in the slab. 
\begin{figure}
\begin{center}
 \includegraphics[width=0.45\textwidth]{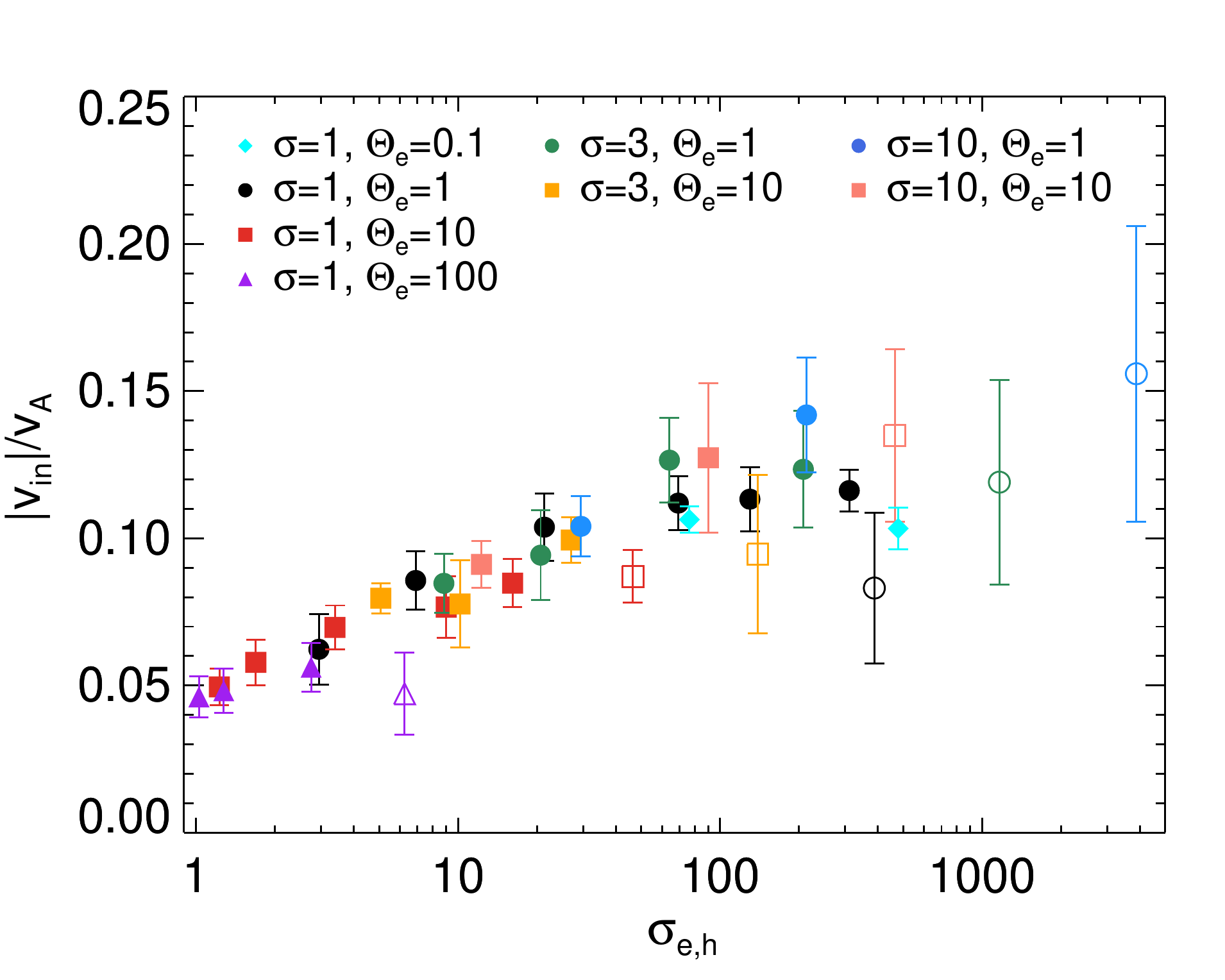}
 \caption{Average inflow rate (in units of the \alf \, speed) as a function of $\sigeh$ for all the simulations presented in \tab{setup} in which the reconnection process is not inhibited by the boundary island. Filled and open symbols are used for simulations in pair-proton and electron-proton plasmas, respectively. Error bars indicate the standard deviation of the spatially averaged inflow rate during the course of the simulation. Results from the box-size scaling simulations are not included here.}
 \end{center}
 \label{fig:inflow}
\end{figure}

Our results for simulations with different $\sigma,\, \Theta_e$, and $\kappa$ (\tab{setup}) are presented in \fign{inflow}, where the average inflow speed $v_{in}$ (normalized to $v_A$) is plotted as a function of the hot pair plasma magnetization $\sigeh$. Results for pair-proton and electron-proton cases are indicated with filled and open symbols, respectively. The error bars, which indicate the standard deviation of the reconnection rate over the duration of the simulation, become typically larger with increasing $\sigeh$ and fixed $\sigma, \Theta_e$. This suggests that the layer becomes accordingly more structured (see also \fign{fluidtime-ionfrac}), since the temporal variations of the reconnection rate about its average value relate to the motion and coalescence of plasmoids \citep[see also][]{petropoulou_18}. We find a weak dependence of the average reconnection rate on $\sigeh$, as this changes only by a factor of $\sim 3$ ($\sim 0.05-0.15$) over more than three orders of magnitude in $\sigeh$. Despite this weak dependence, our results reveal a clear trend of lower reconnection rates at lower $\sigeh$ (i.e., at higher $\beta_e$), in agreement with the findings of \cite{ball_18}. 

\begin{figure}
 \centering 
 \includegraphics[width=0.45\textwidth]{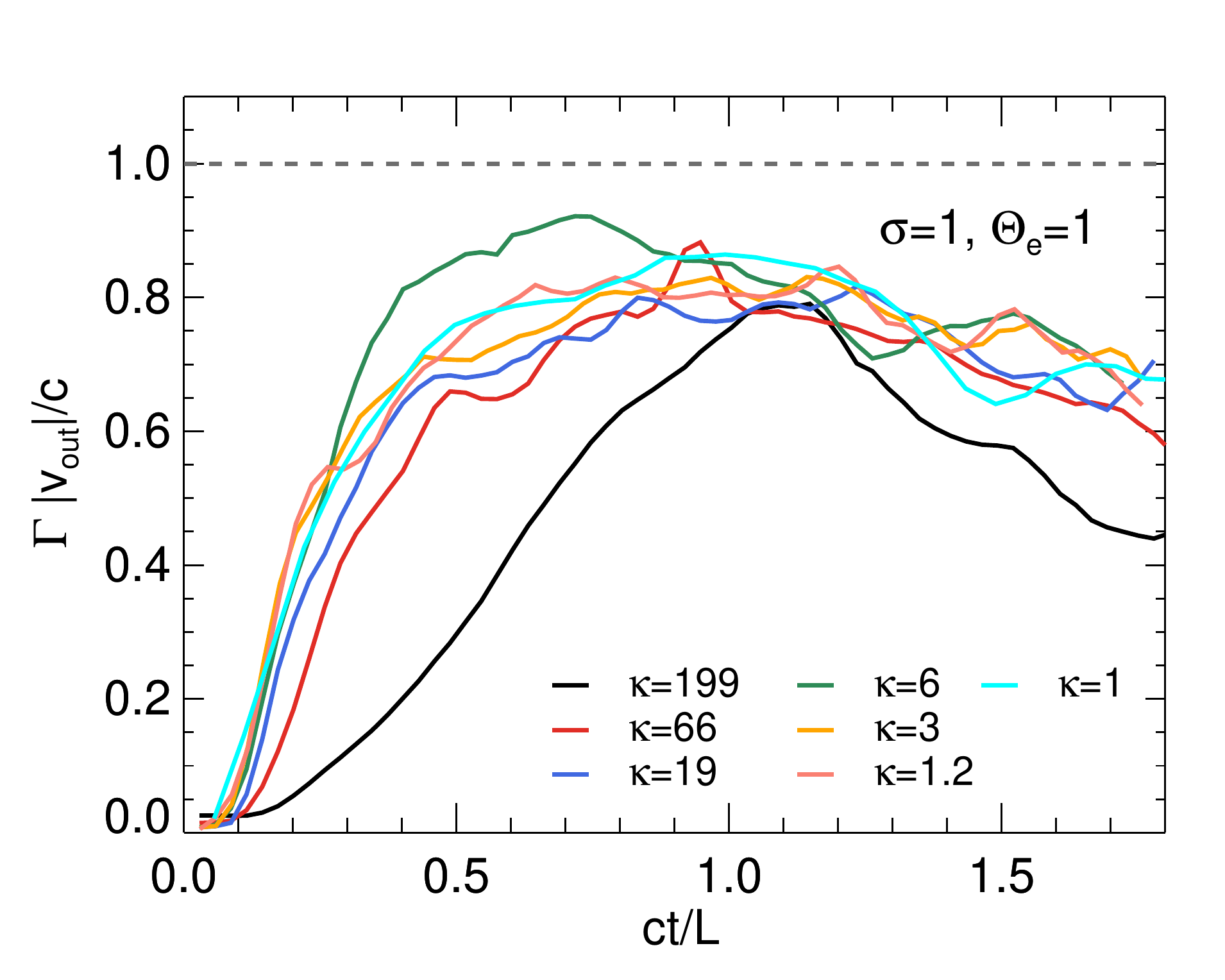}
 \caption{Temporal evolution of the maximum outflow four-velocity (in units of the speed of light) for reconnection in a pair-proton plasma with $\sigma=1, \Theta_e=1$, and different pair multiplicities marked on the plot (see A0-A2, A4, A6-A8 in \tab{setup}). At each time, we take a slice at $y = 0$ and use the 95th percentile of all values measured along the layer as a proxy of the maximum four-velocity. The horizontal dashed grey line marks the Alfv{\'e}n four-velocity. Time is normalized to the light crossing time of the layer.}
 \label{fig:outflow}
\end{figure}

The four-velocity of the plasma outflows in the reconnection region along the $x$ direction, $\Gamma v_{out}/c$, is computed using all particle species, although it is controlled by the protons that contribute most to the plasma inertia. The Lorentz factor $\Gamma$ takes into account the motion in all three directions, but the bulk motion along $x$ dominates.  
To estimate the maximum four-velocity we compute at each time the 95th percentile\footnote{We compute the absolute values of the four-velocity measured at different locations along the layer at $y=0$, sort them in descending order, and determine the value below which 95\% of the measurements falls.} of all values of $\Gamma v_{out}/c$ at $y=0$, and show  in \fign{outflow} its temporal evolution from simulations with $\sigma=1, \Theta_e=1$, and different pair multiplicities. The outflowing plasma accelerates soon after the onset of reconnection, its motion becomes relativistic, and its maximum four-velocity approaches the asymptotic value $\sqrt{\sigma}$ \citep{lyubarsky_05}.  We note that the 95th percentile  of $\Gamma v_{out}/c$ values in the layer provides a more conservative estimate of the maximum outflow four-velocity than the one derived using, for example, the fifth (or tenth) largest value \citep[see e.g.][]{sironi_16}. We verified that with the latter method the peak four-velocity is even closer to $\sqrt{\sigma}$. We find no systematic dependence of the maximum outflow four-velocity on the pair multiplicity, apart from the fact that the bulk acceleration is more gradual in plasmas with $\kappa=199$ (see black line in \fign{outflow}); this is also true for other values of $\Theta_e$ and $\sigma=1-3$.

\section{Energy partition in the reconnection region} \label{sec:energy}
The question of how the available energy is shared between particles and magnetic fields in the region where plasma has undergone reconnection (henceforth, the reconnection region) is of particular astrophysical importance, since it is related to the intensity and spectrum of the associated electromagnetic radiation. Here, we study the energy partition in pair-proton plasmas post reconnection, as a function of pair multiplicity, magnetization, and temperature of the unreconnected plasma. 

\begin{figure}
 \centering 
 \includegraphics[width=0.49\textwidth]{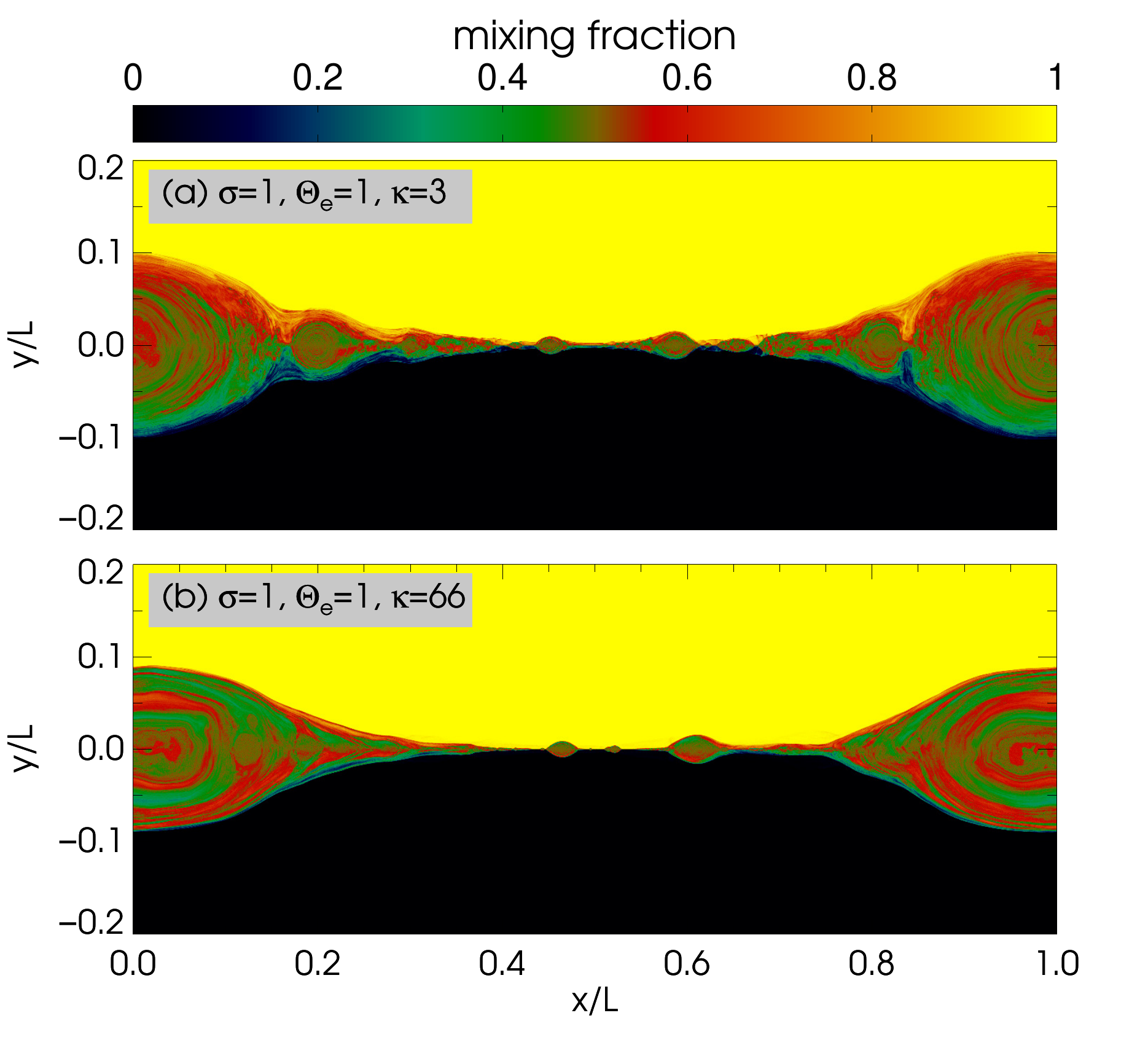}
 \caption{2D snapshots of the mixing fraction computed at $t=1.5 L/c$ for two pair-proton simulations with $\kappa=3$ and $\kappa=66$ (see A1 and A6 in \tab{setup}).}
 \label{fig:mixing}
\end{figure}

To identify the reconnection region we use a mixing criterion, as proposed by \citet{daughton_14}. Particles are tagged with an identifier (0 or 1) based on their initial location (below or above) with respect to the current sheet. Particles from these two regions get mixed in the course of the reconnection process. We identify the reconnection region by the ensemble of computational cells with mixing fraction above a certain threshold $\epsilon$ and below $1-\epsilon$; here, we employed $\epsilon=0.01$\footnote{We verified that our results are insensitive to the exact value, except for very early times (i.e., $\lesssim 0.15L/c$) where the small size of the reconnection region makes the computation of quantities therein sensitive to the choice of $\epsilon$.} \cite[for more details, we refer the reader to][]{rowan_17, ball_18}. 2D snapshots of the mixing fraction from two indicative simulations (see A1 and A6 in \tab{setup}) are presented in \fign{mixing}, where the reconnection region is identified by the mixed colors (green and red).

\begin{figure}
 \centering 
 \includegraphics[width=0.47\textwidth]{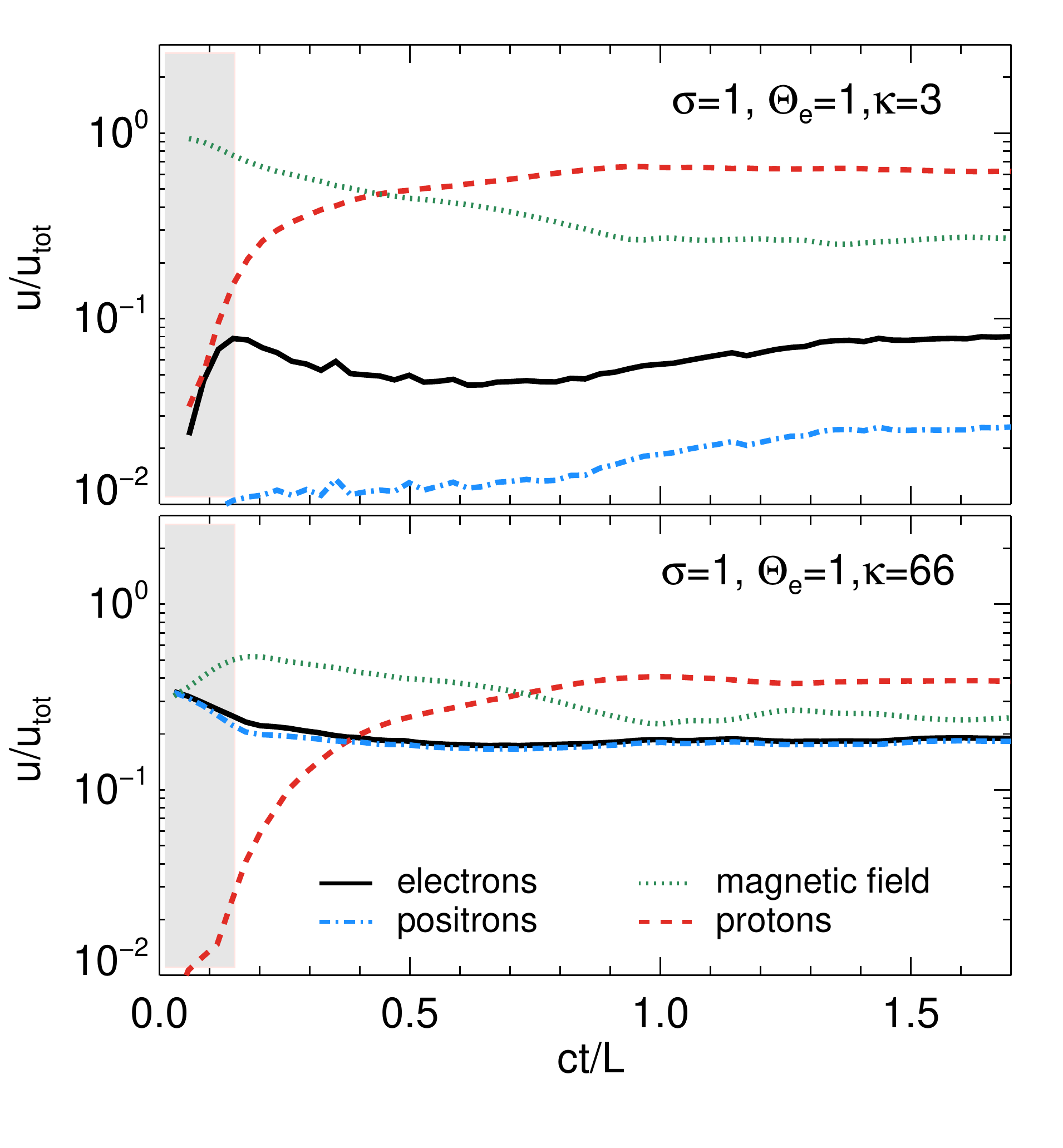}
 \caption{Temporal evolution of the energy stored in magnetic field (dotted green line), protons (dashed red line), electrons (solid black line), and positrons (dash-dotted blue line) in the reconnection region. The energies of all components are normalized to the total (particle and magnetic) energy at each time. Results for $\kappa=3$ and $\kappa=66$ are shown in the top and bottom panels. Snapshots of the mixing fraction used to identify the reconnection region are shown in \fign{mixing}. The early-time evolution of the energy ratios (grey-colored region) is sensitive to the choice of the mixing threshold.
 }
 \label{fig:enratio-time}
\end{figure}

We compute the kinetic energy of each particle species by summing up the contributions from all computational cells that define the reconnection region, namely $u_j=m_j c^2\sum\limits_{cells} n_j (\gamma_j-1)$, where $\gamma_j$ is the average Lorentz factor of particles of species $j$ in a computational cell. We then normalize $u_j$ to the total energy  $u_{tot}=u_B+\!\!\!\sum\limits_{j=i,e^\pm} \!\!\! u_j$, where $u_B=\sum\limits_{cells}B^2/8\pi$. In \fign{enratio-time} we show the temporal evolution of $u_j/u_{tot}$ for the same cases as those shown in \fign{mixing}. At very early times, when the reconnection region is small (see grey-colored region in \fign{enratio-time}), the plasma properties therein depend on how exactly the reconnection region is identified. Yet, neither the time-averaged properties nor their late-time evolution are sensitive to the definition of the reconnection region. Given that there might be also other factors affecting the early time evolution (e.g., initial setup), we henceforth ignore this transitional early period. At later times, the ratio of post-reconnection magnetic energy to the total energy decreases gradually with time, whereas the pair energy density ratio reaches an almost constant value very soon after the onset of reconnection (i.e., already at $0.4 L/c$). The proton energy ratio asymptotes to a constant value typically at later times compared to the pairs, but our simulations are long enough to capture the steady-state values of all energy ratios. We find similar temporal trends for other cases as well.

\begin{figure}
 \centering 
 \includegraphics[width=0.47\textwidth]{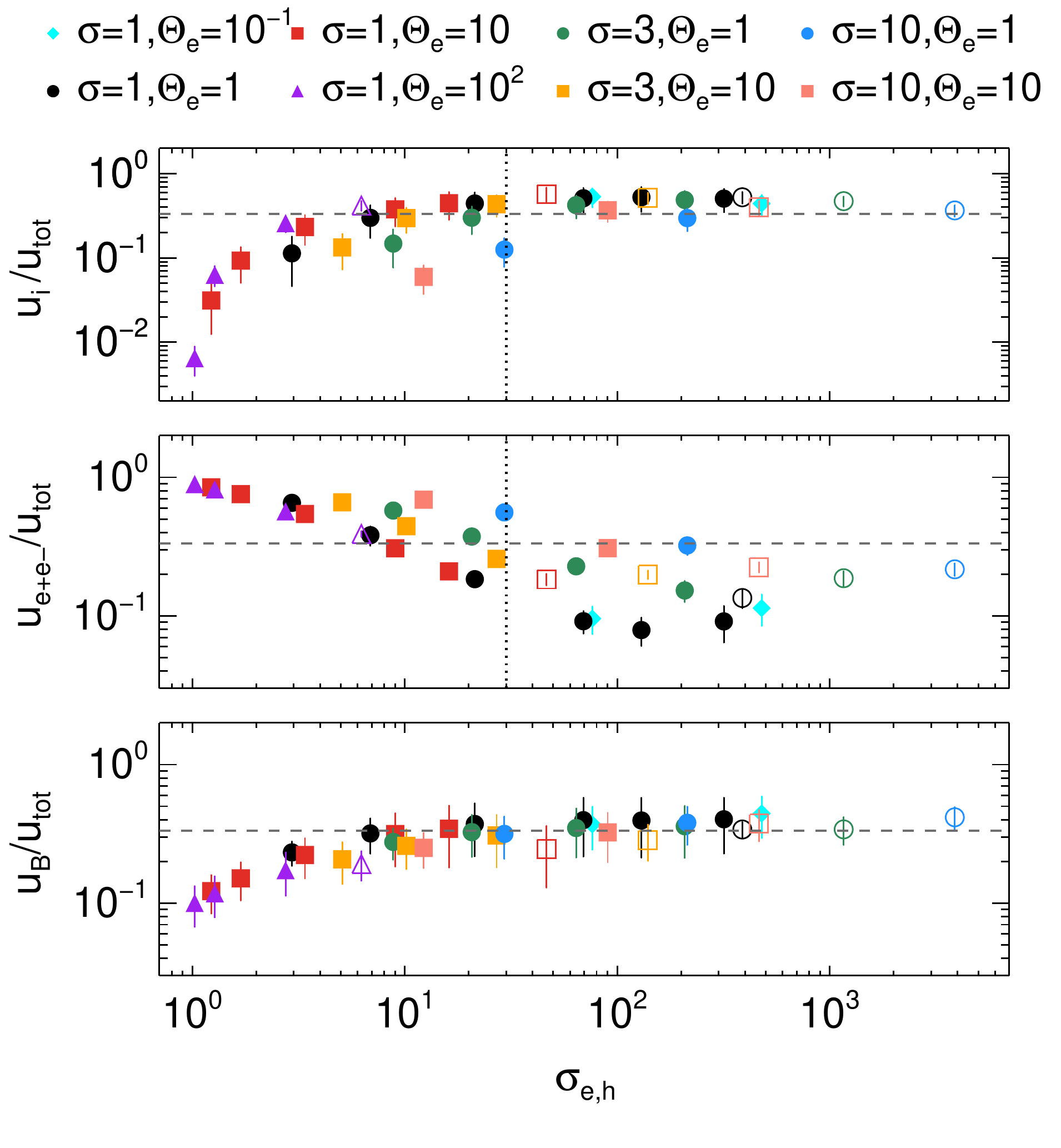}
 \caption{Time-averaged energy ratios of protons (top panel), pairs (middle panel), and magnetic field (bottom panel) in the reconnection region plotted against $\sigeh$ for our complete set of simulations with different physical parameters marked on the plot (same color coding used as in \fign{inflow}). Results from the size-scaling simulations are not included here. Filled and open symbols are used for simulations in pair-proton and electron-proton reconnection, respectively. Error bars indicate the standard deviation of the energy ratios during the course of the simulation. In all panels, the horizontal dashed line marks the equipartition value of 1/3. The dependence of the particle energy ratios on $\sigeh$ changes at $\sigeh \sim 30$, as noted by the dotted vertical line in the upper two panels.}
 \label{fig:enratio-seh}
\end{figure}

The time-averaged energy ratios of protons, pairs, and magnetic fields in the reconnection region are presented in \fign{enratio-seh}. The leftmost point in each series with a given color corresponds to pair-proton plasmas with $\kappa=199$ and the rightmost point corresponds to the pure electron-proton case with $\kappa=1$ (open symbols). 
The fraction of energy that remains in the post-reconnection magnetic field is $\sim 1/3$ and is approximately constant for a wide range of $\sigeh$ values, spanning almost three orders of magnitude (bottom panel). Only for $\sigeh < 3$, we find sub-equipartition values, i.e., $u_{B}/u_{tot} <1/3$. 
In this parameter regime, the pairs in the plasma carry most of the upstream total energy. Upon entering the reconnection region,
the pair kinetic energy increases even further at the expense of magnetic energy due to field dissipation. 
As a result, the post-reconnection magnetic energy for $\sigeh < 3$ is only a small fraction of the total energy ($u_B/u_{tot} \sim 0.1-0.2$). 

One can empirically define two regimes of interest for the particle energy ratios: a low-$\sigeh$ regime ($\sigeh\lesssim 30$), where  $u_{i}/u_{tot}\propto \sigeh$ and  $u_{e^\pm}/u_{tot} \propto \sigeh^{-1/2}$, and a high-$\sigeh$ regime ($\sigeh> 30$), where both ratios are almost independent of the hot pair plasma magnetization. In both regimes, there is no dependence of the particle energy ratios on $\Theta_e$, but a weak dependence on the total plasma magnetization $\sigma$ is evident. This can be more clearly seen in the middle panel of \fign{enratio-seh}, where points with the lowest $\sigma$ (black and cyan symbols) systematically lie below points with higher $\sigma$. Finally, energy equipartition between magnetic fields, protons, and pairs is asymptotically achieved for $\sigma \gg 1$ and $\sigeh \gtrsim 30$, with each component carrying $\sim 1/3$ of the total energy. 

\begin{figure*}
 \centering 
 \includegraphics[width=0.49\textwidth]{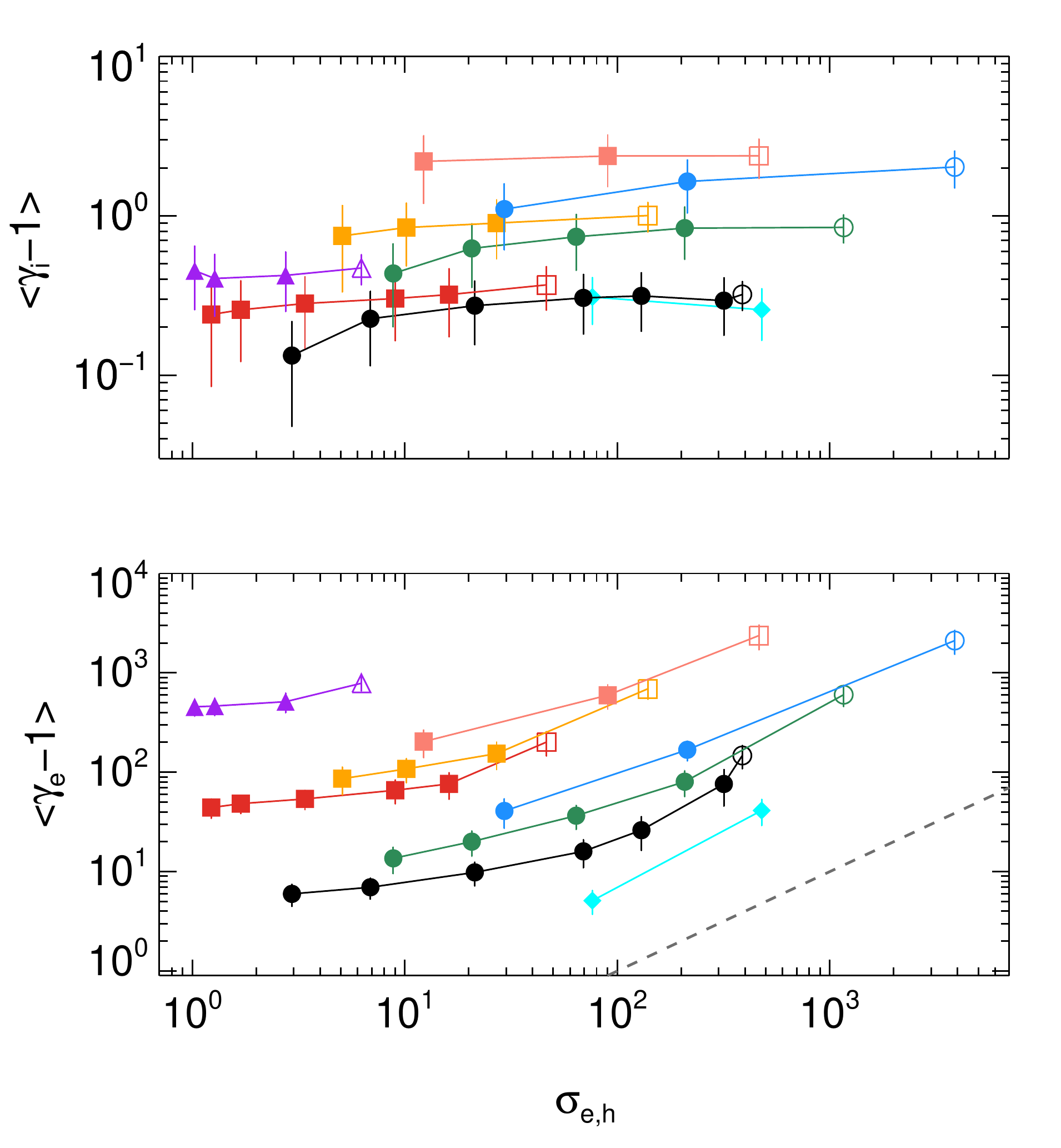}
 \includegraphics[width=0.49\textwidth]{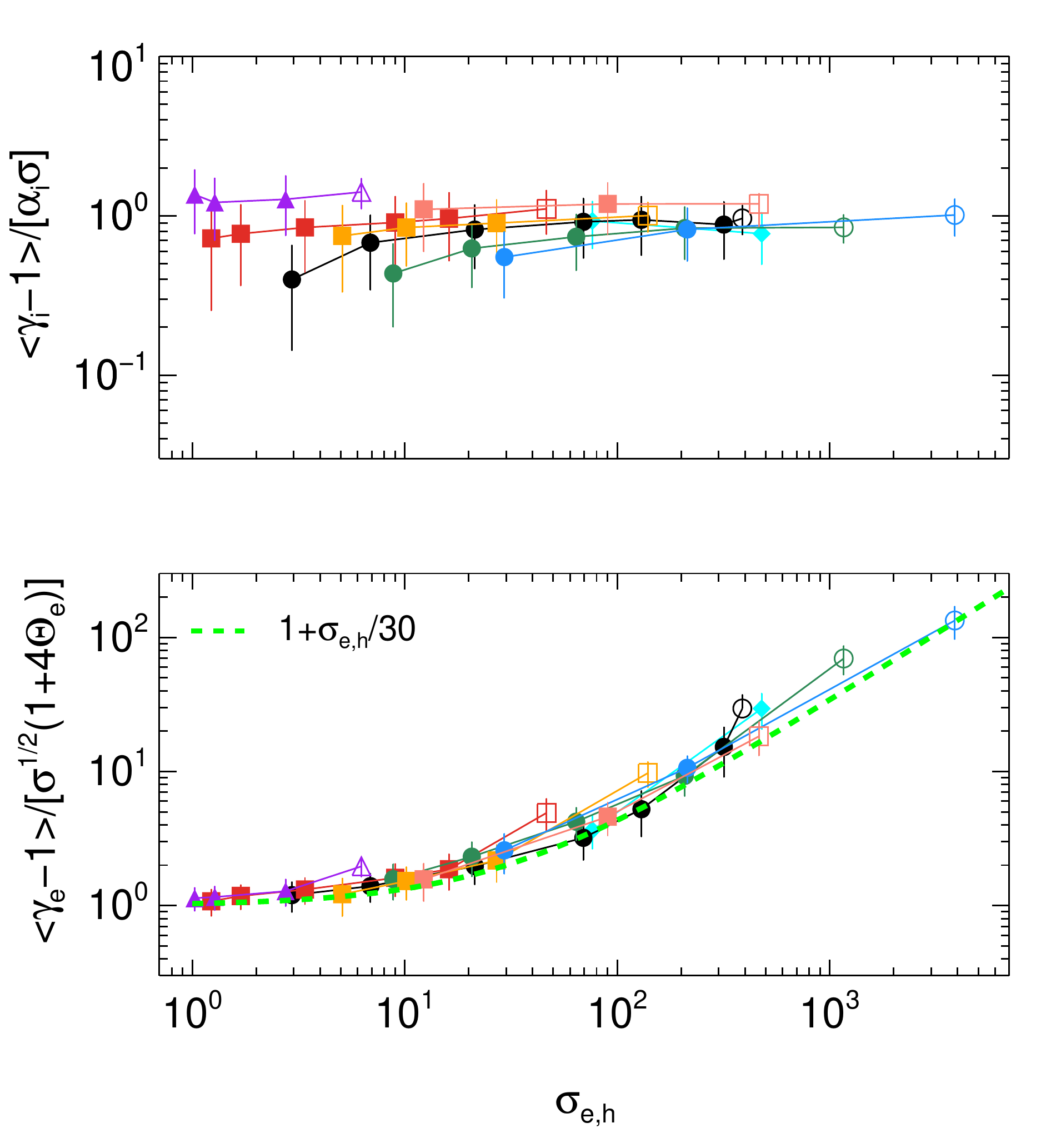}
 \caption{Left panel: Time-averaged ratios of the energy density to the rest mass energy density of protons (top panel) and pairs (bottom panel), which serve as a proxy of $\langle \gamma_j-1\rangle$.  A dashed line with slope unity is also plotted in the bottom panel to show the linear asymptotic dependence of the  mean lepton Lorentz factor on $\sigeh$. All symbols have the same meaning as in \fign{enratio-seh}. Right panel: Proxy of the post-reconnection particle Lorentz factor normalized to $\alpha_i\sigma$ for protons (with $\alpha_i=1/3$ for $\sigma=1,3$ and $\alpha_i=1/5$ for $\sigma=10$) and $\sqrt{\sigma}(1+4\Theta_e)$ for leptons.} 
 \label{fig:ratio-seh}
\end{figure*}

The dependence of the particle energy densities on $\sigeh$ could originate from either changes in the number density or in the mean particle Lorentz factor, or both. A proxy of the average post-reconnection particle Lorentz factor, $\langle\gamma_j-1\rangle=\sum\limits_{cells} u_j /\sum\limits_{cells} n_j m_j c^2$, is plotted against $\sigeh$ in the left panel \fign{ratio-seh} for protons (top panel) and pairs (bottom panel). In all cases, we find that the post-reconnection mean proton Lorentz factor is almost independent of $\sigeh$ and $\Theta_e$, but has a dependence on $\sigma$, with larger values leading to higher mean proton Lorentz factors. Indeed, when $\langle\gamma_i-1\rangle$ is normalized to $\alpha_i\sigma$ (with $\alpha_i=1/3$ for $\sigma=1,3$ and $\alpha_i=1/5$ for $\sigma=10$) all curves coincide, as shown in the right panel of \fign{ratio-seh}. In contrast to the protons, the mean lepton Lorentz factor depends on $\Theta_e$, $\sigma$, and $\sigeh$, as shown in the left panel of \fign{ratio-seh}. We empirically find for $\sigma\ge1$ that the mean lepton Lorentz factor can be approximated as (see also right panel in \fign{ratio-seh}): 
\eqb 
\langle\gamma_e-1\rangle \approx \sqrt{\sigma}
\left(1+4\Theta_e\right)\left(1+\frac{\sigeh}{30}\right)\cdot
\label{eq:gammae}
\eqe
The asymptotic value of the mean lepton Lorentz factor for $\sigeh \ll 30$ implies that, in this regime, the pairs in the reconnection region still bear memory of their initial (pre-reconnection) conditions (and, in particular, of $\Theta_e$), in agreement with the discussion on \fign{enratio-seh}. 
In the high-$\sigeh$ regime, the  mean lepton Lorentz factor scales almost linearly with $\sigeh$. This asymptotic behavior of $\langle \gamma_e-1\rangle$ can be understood as follows. For fixed $\sigma$ and $\Theta_e$ (i.e., fixed amount of post-reconnection energy available for the particles), the energy per lepton increases as the number of leptons per proton decreases, or equivalently, as $\sigeh$ increases (see also \eq{seh}). We refer the reader to Appendix~\ref{sec:app3}, for a quantitative discussion on the dependence of  the mean lepton Lorentz factor on the physical parameters $\sigma, \Theta_e$, and $\kappa$ of the upstream plasma.
\vspace{0.05in}
\section{Particle energy distributions} \label{sec:spectra}
After having discussed the general properties of reconnection in pair-proton plasmas (see sections \ref{sec:layer}-\ref{sec:energy}), we continue our study by examining the particle energy distributions and their dependence on physical parameters, most notably on $\sigeh$.

\subsection{Temporal evolution of particle energy spectra}\label{sec:spectra-time}
The energy distribution of each particle species is defined as $f_j(E)\equiv dN_j/dE$, where $E$ is the particle kinetic energy and $j=i, e^-, e^+$. Henceforth, all particle energies are kinetic (i.e., excluding rest mass), unless stated otherwise. 

\begin{figure}
 \centering 
 \includegraphics[width=0.47\textwidth, trim=0 0 0 0]{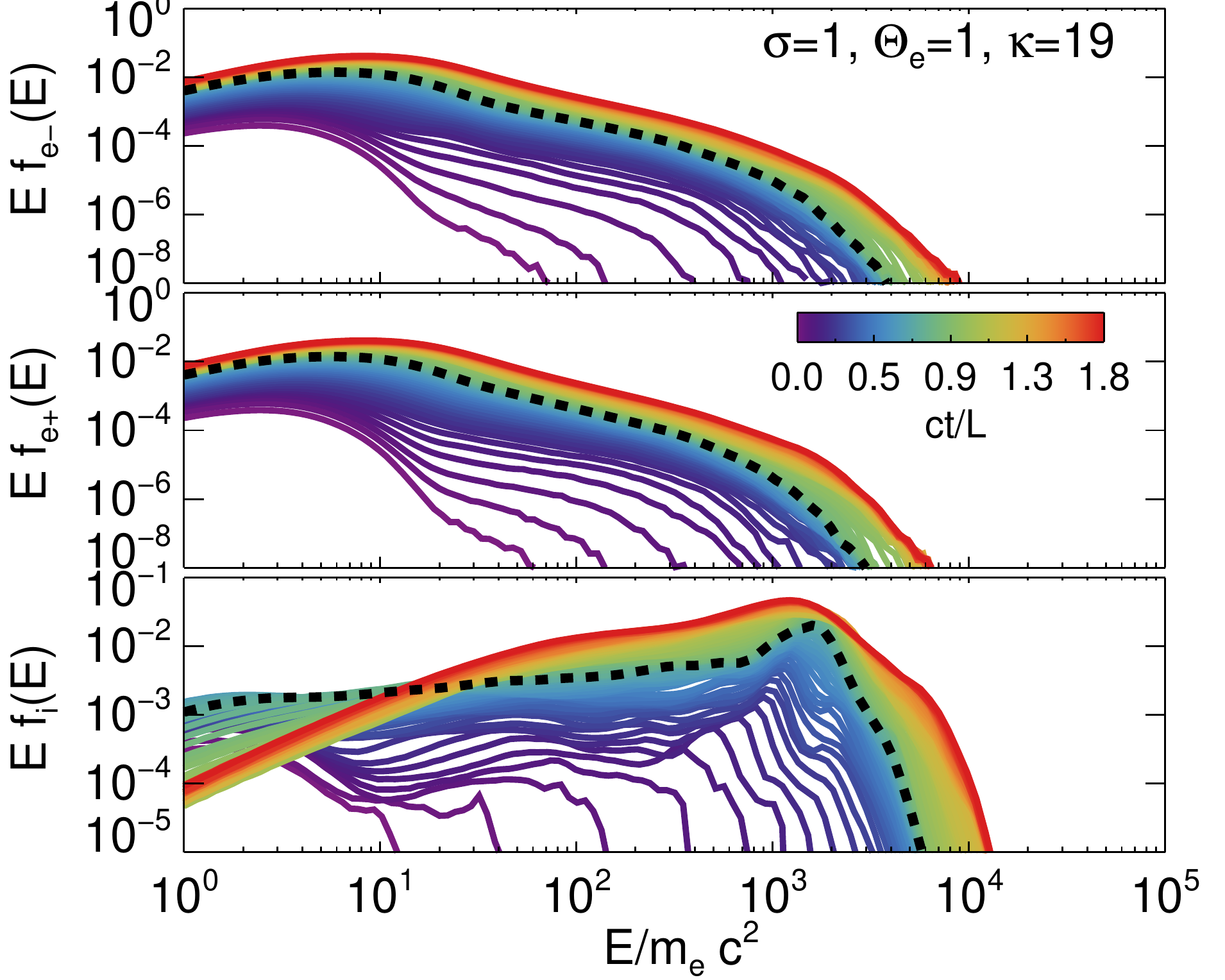}
 \caption{Temporal evolution (see inset color bar) of the electron, positron, and proton energy distributions (from top to bottom) extracted from the reconnection region of a simulation with $\sigma=1, \Theta_e=1$, and $\kappa=19$ --  see also \fign{fluidtime}, for a depiction of the layer structure. The spectrum obtained at the time the reconnection fronts reach the boundaries (i.e., $t=L/2 v_A$) is highlighted with a dashed black line. The energy distributions of each particle species are normalized to the total number of particles of that species within the reconnection region at the end of the simulation.}
 \label{fig:spec-time}
\end{figure}

As a representative example, we present in \fign{spec-time} the temporal evolution of the electron, positron, and proton energy distributions (from top to bottom) from a simulation with $\sigma=1, \Theta_e=1$, and $\kappa=19$ (see also \fign{fluidtime}, for a depiction of the layer structure). The energy distributions of each particle species are normalized to the total number of particles of that species in the reconnection region at the end of the simulation. The displayed spectra  exclude the particle population that was initialized in the current sheet. For reference, the spectrum obtained at the time the reconnection fronts reach the boundaries (i.e., $t=L/2 v_A$) is shown with a dashed black line. 

Soon after the onset of reconnection, the electron and positron energy spectra in the reconnection region begin to deviate from their initial Maxwell-J{\"u}ttner distributions. They develop a non-thermal component even before the time the reconnection fronts reach the boundaries of the layer (i.e., at $ct/L\sim 0.7$). The non-thermal part of the spectrum of pairs can be described by a power law above a characteristic energy 
where the post-reconnection energy spectrum $Ef_j(E)$ obtains its peak value. For the adopted parameters, we find $E_{pk,e}/m_e c^2\sim 10$ in agreement with the value of the mean post-reconnection Lorentz factor that we derived in \sect{energy} (see third black symbol from the left in bottom panel of \fign{ratio-seh}). 

There is a clear difference between the temporal evolution of the lepton and proton energy distributions. More specifically, the non-thermal component of the proton spectrum begins to emerge only at $t>L/2 v_A$, after the fronts have reached the boundaries. At earlier times, the proton energy spectrum shows a narrow peak that evolves with time. We interpret this early-time spectral feature as a result of heating and bulk motion of the proton plasma, whose outflow four-velocity evolves strongly for $t<0.5 L/c$ (see blue curve in the top panel of \fign{outflow}). Similar results were obtained by \citet{ball_18} for trans-relativistic reconnection in electron-proton plasmas (see Fig.~3 therein). 

The late-time development of the power law in the proton distribution can be understood in terms of the interactions of particles with various structures in the layer. X-points are typically smaller than the proton Larmor radius, so direct proton acceleration by the non-ideal reconnection electric field is not very efficient. We find evidence of proton acceleration only when the boundary island, which is the biggest structure in the layer, begins to form. We argue that  in a much larger simulation domain, where bigger secondary plasmoids could  form, protons should show signs of acceleration even before the reconnection fronts interact with the boundaries.

\begin{figure*}
 \centering 
 \includegraphics[width=0.48\textwidth]{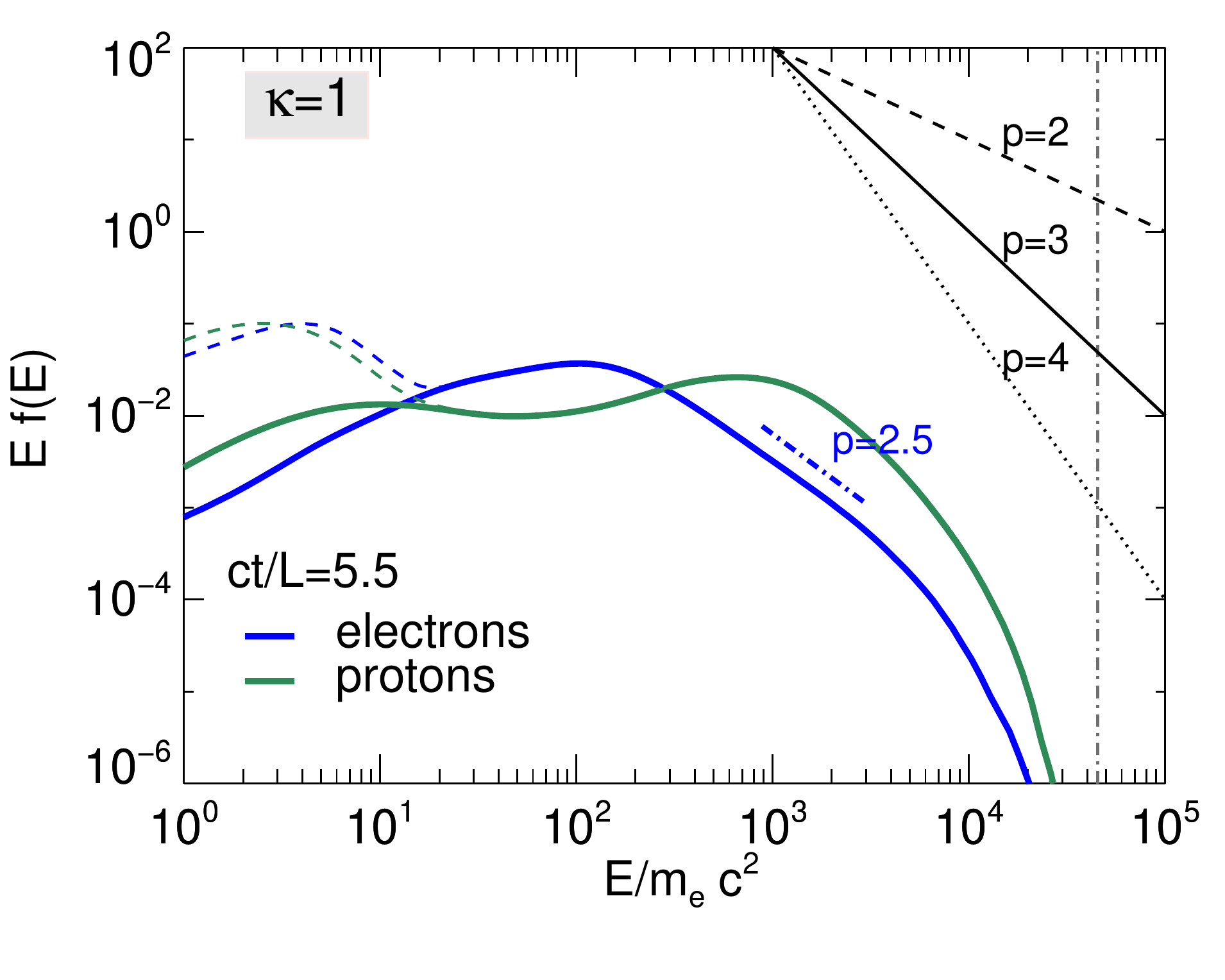}
 \includegraphics[width=0.48\textwidth]{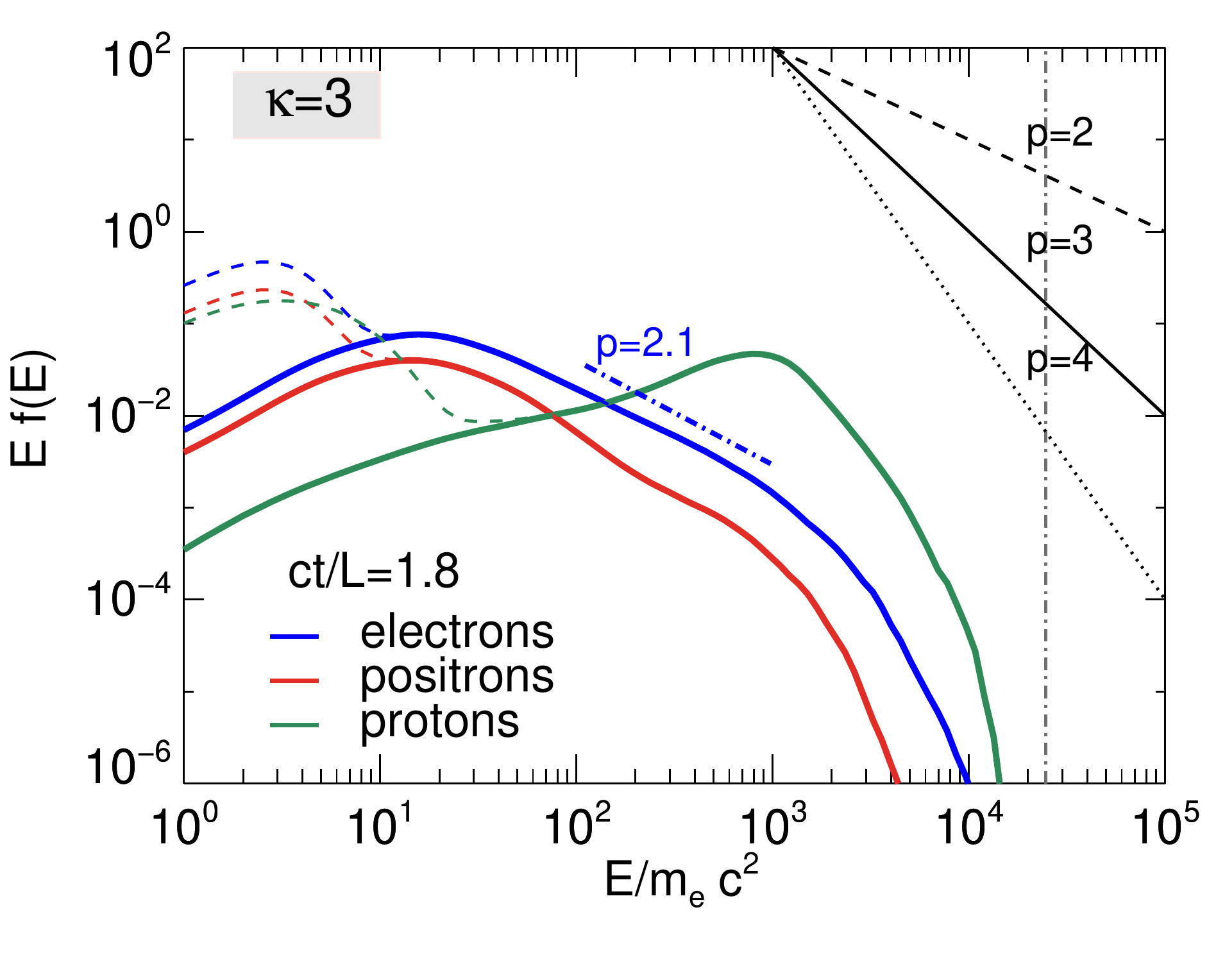}
 \includegraphics[width=0.48\textwidth]{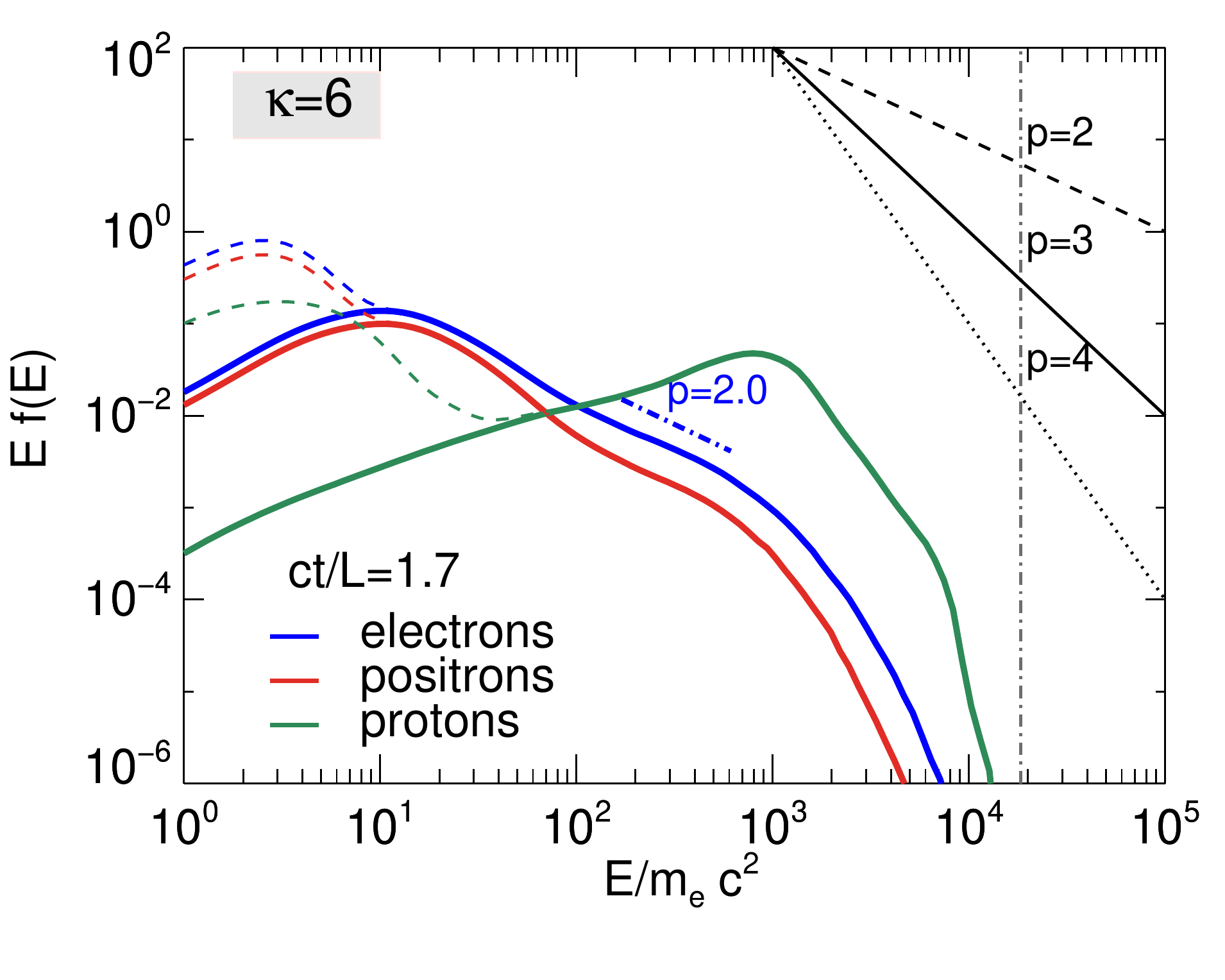}
 \includegraphics[width=0.48\textwidth]{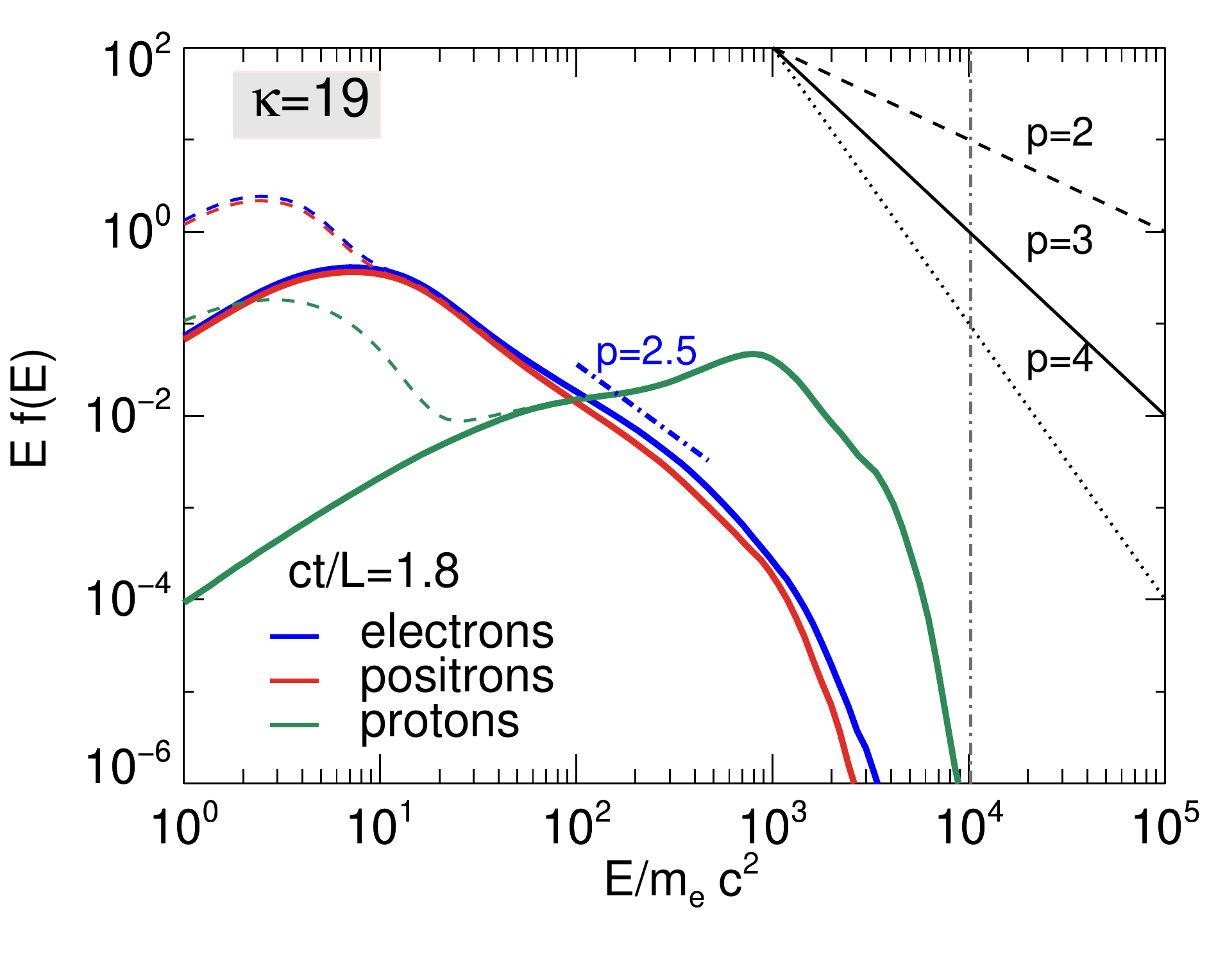}
 \includegraphics[width=0.48\textwidth]{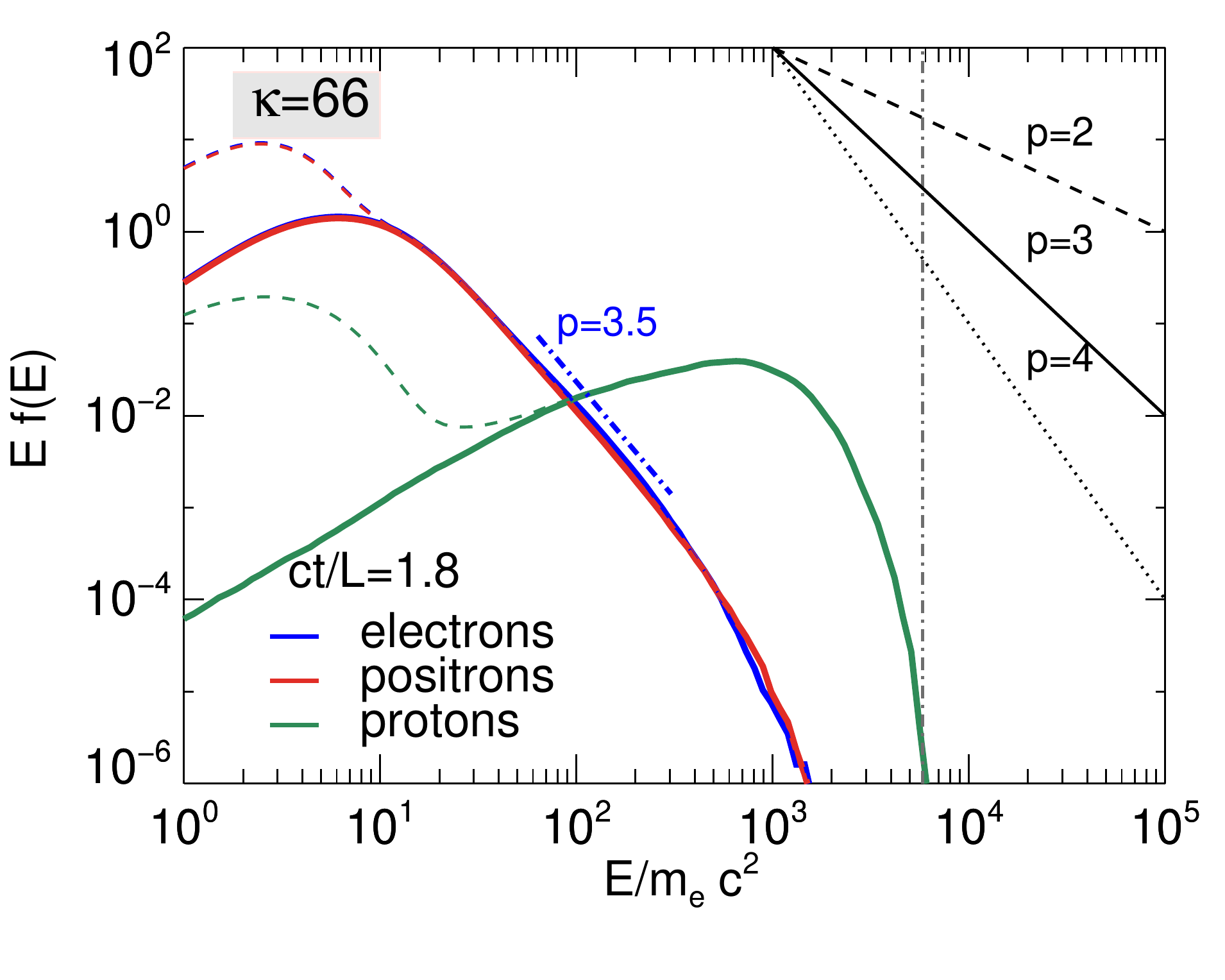}
 \includegraphics[width=0.48\textwidth]{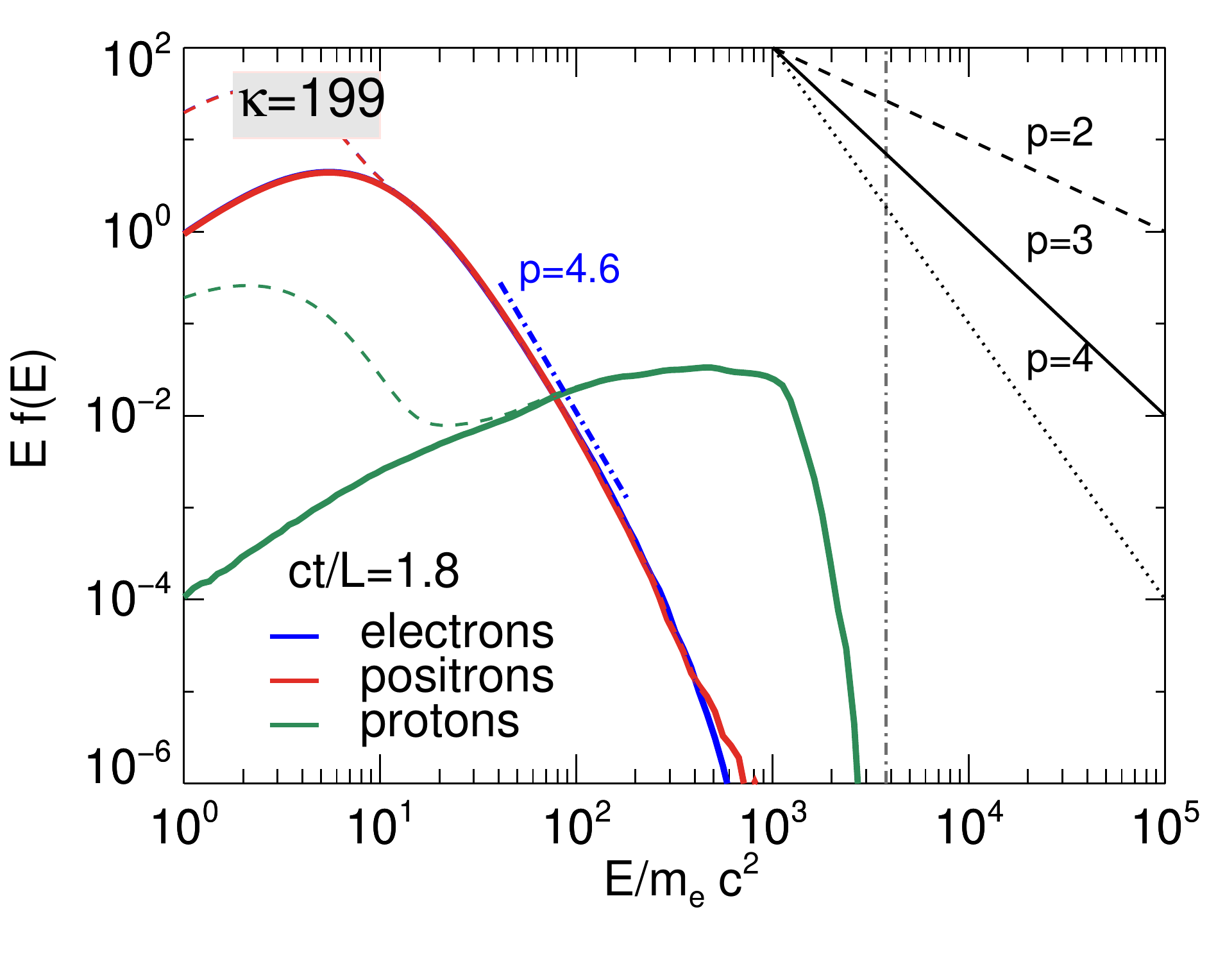}
 \caption{Electron (blue lines), positron (red lines), and proton (green lines) energy distributions computed from a set of simulations with $\sigma=1$, $\Theta_e=1$ and different pair multiplicities marked on the plots (see runs A0-A2, A4, A6 and A8 in \tab{setup}). The spectra are computed at the end of each simulation and are normalized to the total number of protons within the reconnection region at that time. Thick solid and thin dashed lines show the spectra from the reconnection region and the whole simulation domain, respectively. The power-law segment of the electron distributions used to measure the slope is indicated with dash-dotted blue lines. The black lines in the upper right corner of each panel have slopes of $-p+1$ and are plotted for three values of $p$ in order to facilitate the comparison with the power-law segments of the particle distributions. The vertical dash-dotted line in each panel marks the energy of relativistic  particles with Larmor radius $0.1 \, L$.}
 \label{fig:spec-comp}
\end{figure*}

\begin{figure}
 \centering 
 \includegraphics[width=0.45\textwidth, trim=0 0 0 0]{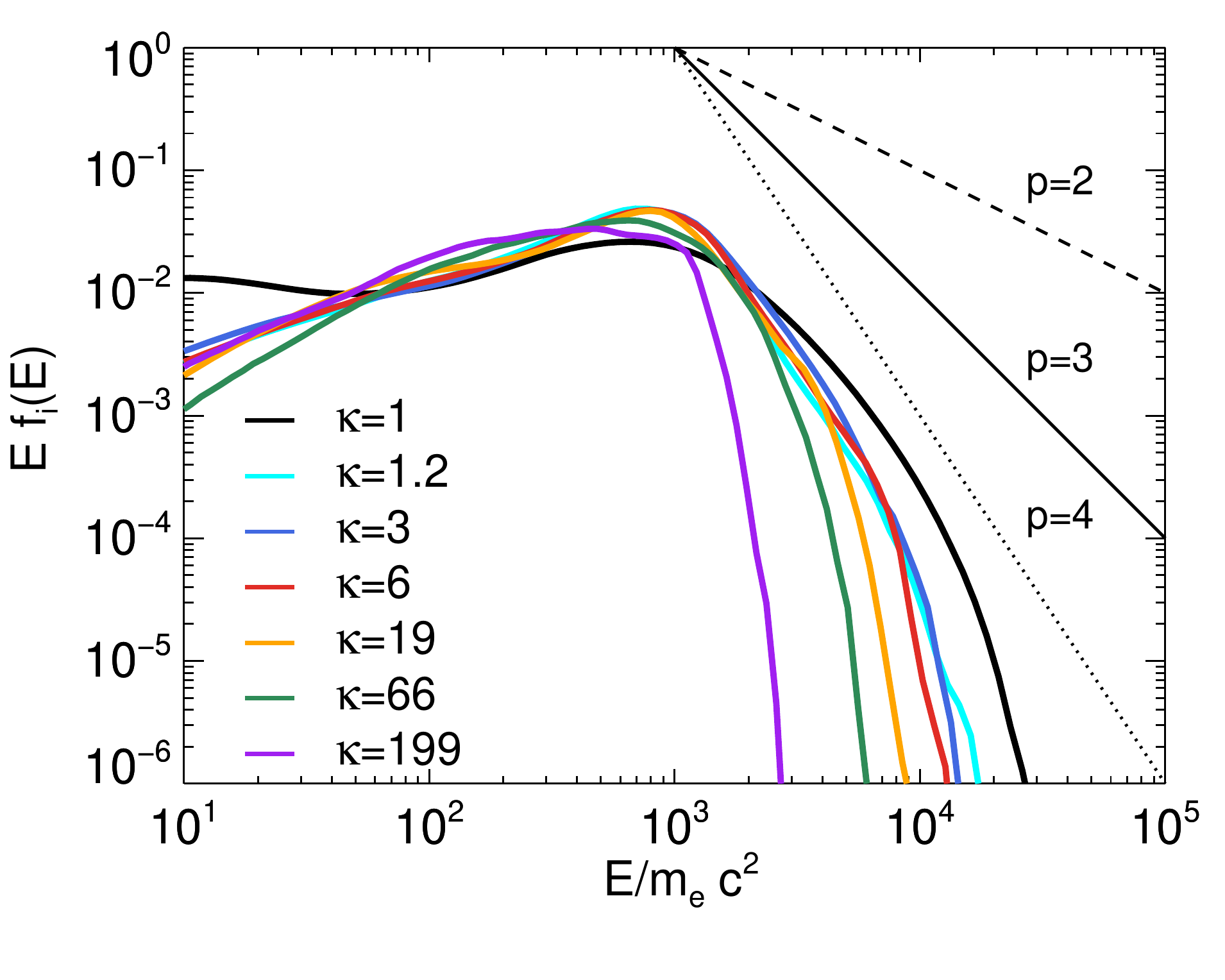}
  \includegraphics[width=0.45\textwidth, trim=0 0 0 0]{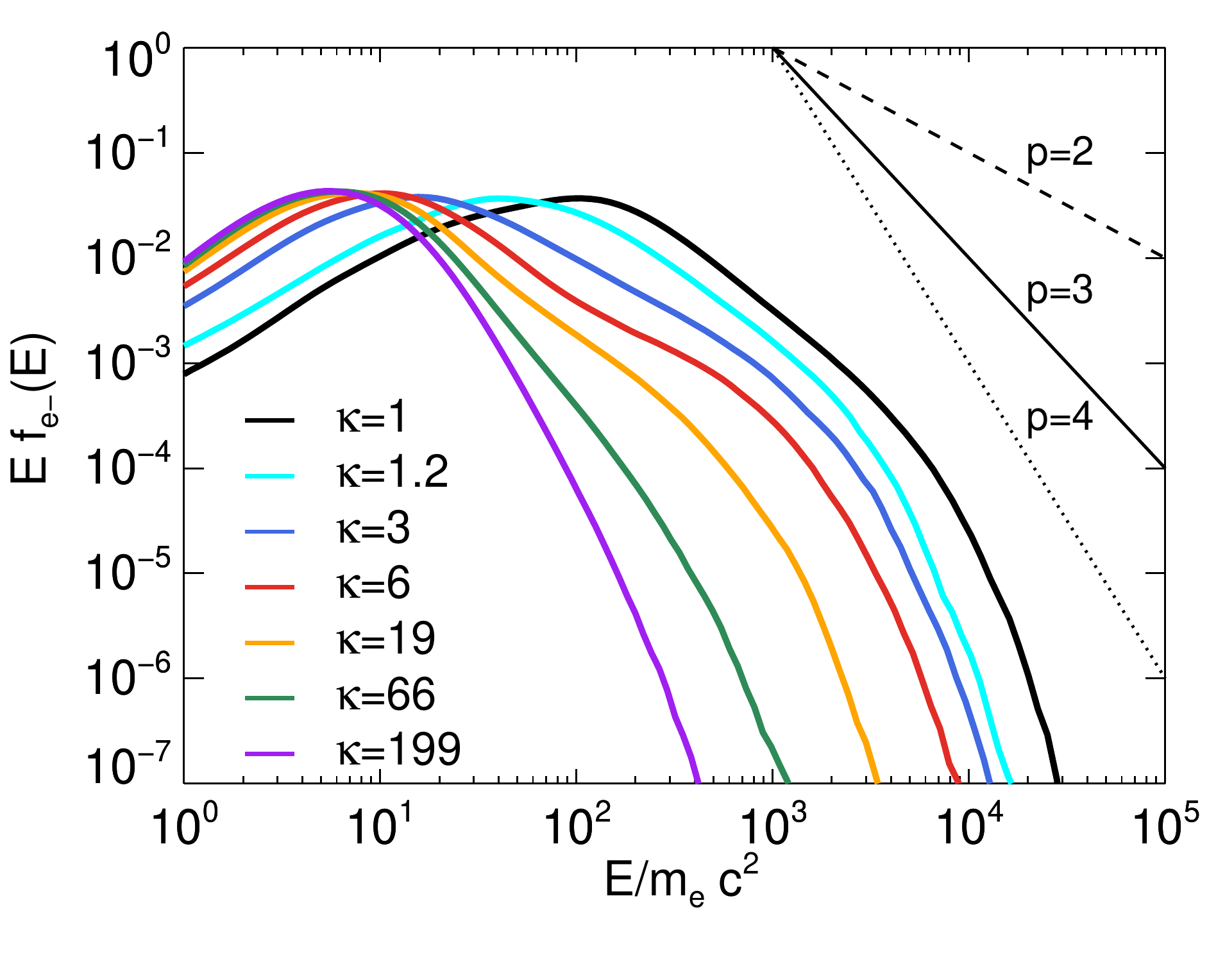} 
    \includegraphics[width=0.45\textwidth, trim=0 0 0 0]{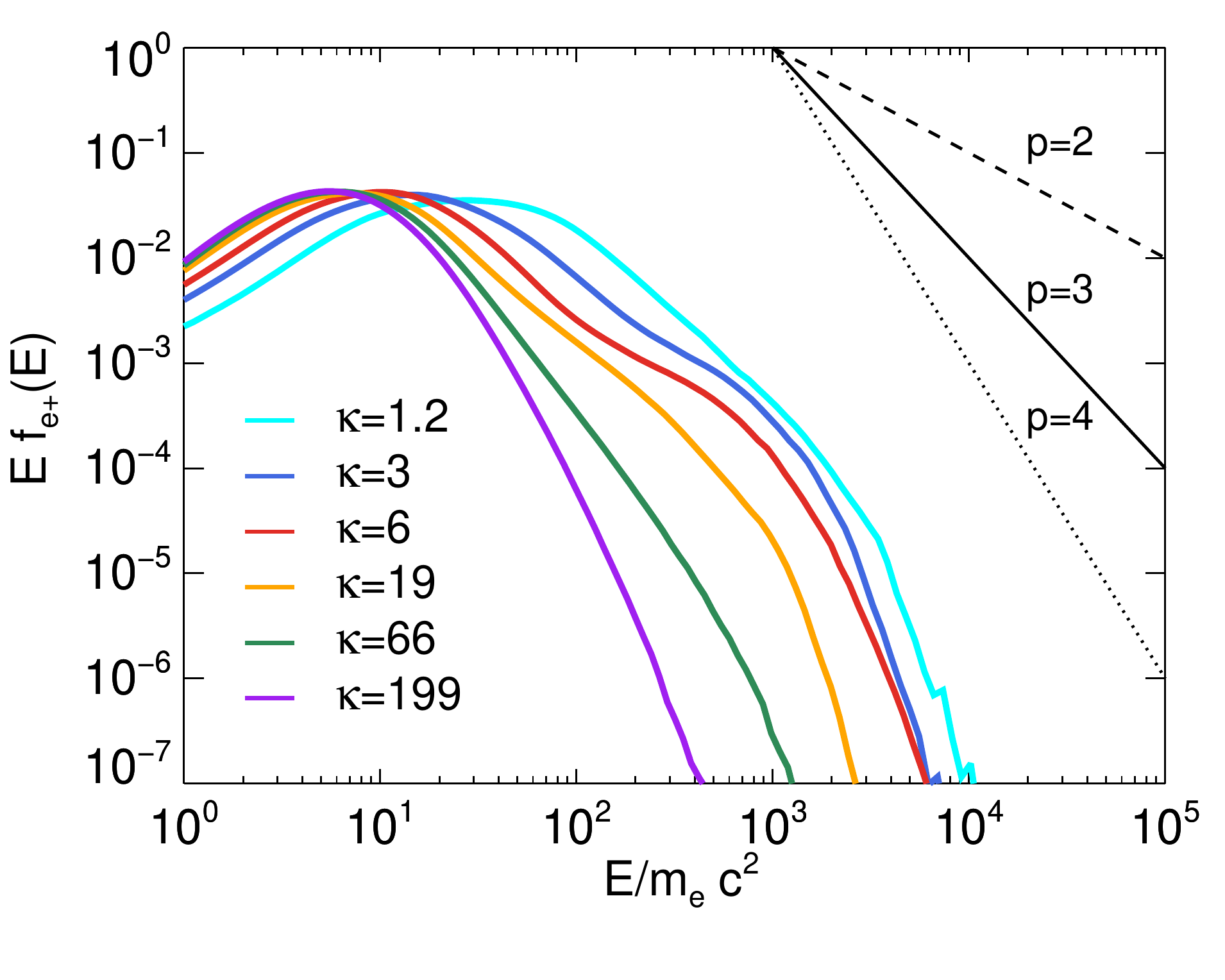}
 \caption{Post-reconnection energy distributions of protons, electrons, and positrons (from top to bottom) from simulations with $\sigma=1$, $\Theta_e=1$, and different pair multiplicities marked on each plot (see runs A0-A2, A4, A6-A8 in \tab{setup}). The spectrum of each particle species is computed at the end of each simulation and is normalized to the total particle number of that species in the reconnection region. The black lines in the upper right corner of each panel have slopes of $-p+1$ and are plotted for three values of $p$ to facilitate the comparison with the power-law segments of the particle distributions.}
 \label{fig:spec-ionfrac}
\end{figure} 

\subsection{Effects of pair multiplicity}\label{sec:spectra-kappa}
To illustrate the dependence of the particle energy distributions on pair multiplicity, we show in \fign{spec-comp} the energy spectra from a set of simulations with $\sigma=1$, $\Theta_e=1$ and different values of $\kappa$ marked on the plots.  Thick solid and thin dashed lines show the spectra from the reconnection region and the whole simulation domain, respectively. The spectra are computed at the end of each simulation and are normalized to the total number of protons within the reconnection region. The vertical dash-dotted line in each panel marks the energy of particles with Larmor radius\footnote{The Larmor radius is computed using the upstream magnetic field strength.} $0.1 \, L$, i.e., comparable to the size of the largest plasmoids in the layer. 

The peak energy of the pair energy distributions depends strongly on the pair multiplicity for $\kappa <6$, and becomes approximately constant (here, $E_{pk,e}/ m_e c^2\sim 10$) for higher pair multiplicities. On the contrary, the peak proton energy is approximately constant for all $\kappa$ values we explored. The dependence of the peak particle energy on $\kappa$ is more clearly illustrated in  \fign{spec-ionfrac}, where the energy distributions of each particle species are plotted for different values of $\kappa$. These findings are in agreement with those presented in \fign{ratio-seh}  for the mean particle Lorentz factor. The fact that the mean and the peak lepton energies are comparable is not surprising; most of the energy is expected to reside at the peak of the energy distribution, given that the power-law slopes of the lepton energy spectra are typically $\gtrsim2$ (see below and  \sect{spectra-slope}).

Above the peak energy $E_{pk,e}$, the pair energy spectra can be approximated by a power law with slope $p$ (i.e., $f(E)\propto E^{-p}$) followed by a cutoff. The power-law segment used for the estimation of the slope (see \sect{spectra-slope}) is overplotted (dash-dotted blue lines) for guiding the eye.  Inspection of the figure (see also \fign{spec-ionfrac}) shows that the power law of the pair distributions becomes steeper (i.e., larger $p$ values) as the pair multiplicity increases (for details, see \sect{spectra-slope}). The power law of the pair distributions extends well beyond their peak energy for all the cases we explored, except for the cases with the highest $\Theta_e$, which are discussed in Appendix~\ref{sec:app4}. For protons a well-developed power law forms only for small pair multiplicities (here, for $\kappa <19$), while their energy distribution shows a steep drop above the peak energy $E_{pk,i}/m_e c^2\sim 10^3$ for $\kappa=66$ and 199. This should not be mistakenly interpreted as a limitation of reconnection in accelerating protons in plasmas with high pair multiplicities. It is merely a result of the limited size of the computational domain in terms of the proton Larmor radius: $L/\rLi$ drops by a factor of ten between the simulations with $\kappa=3$ and $\kappa=199$, as shown in \tab{setup} (the dependence of the particle energy distributions on the box size is discussed in Appendix~\ref{sec:app2}). For these reasons, we do not attempt to study the spectral properties of the proton energy distributions and, in what follows, we focus  on the energy distributions of pairs. 

\subsection{Power-law slope of pair energy spectra}\label{sec:spectra-slope}
We compute the slope of the power-law segment of the pair energy distributions and explore its dependence on the physical parameters. Due to the similarity between the energy distributions of positrons and electrons (see \fign{spec-comp}) it is sufficient to use one of the two for computing the slope. Henceforth, we use for this purpose the electron energy spectrum obtained at the end of each simulation. 

The electron energy distribution can be generally described by two components: a low-energy broad component that forms due to heating and a high-energy component, which can be described as a truncated power law at low energies with an exponential cutoff at higher energies (see e.g., panels in middle row of \fign{spec-comp}). A detailed fit to the simulation data is very challenging due to the degeneracy in the model parameters describing the two components. For example, the choice of the low-energy end of the power law affects the broadness and normalization of the low-energy component and  vice versa. The slope inferred from the two-component fit to the data can vary at most by $0.2$ depending on the other model parameters. Given the inherit uncertainties in the fitting procedure, in what follows, we identify the power-law segment by eye and fit it with a single power law (see dash-dotted blue lines in \fign{spec-comp}).

The extent of the power law is, in most cases, sufficient to allow a reliable estimation of its slope.  We assign a systematic error of $\pm 0.2$ to the derived slope, which dominates the statistical error from the fits, to account for the subjective choice of the fitting energy range. For simulations with duration much larger than all others (see electron-proton cases in \tab{setup}), we computed the slope also at earlier times (i.e., comparable to the duration of all other cases) and found no difference in the inferred $p$ value within the systematic error. Although a hard power law can be safely distinguished from the thermal part of the energy distribution, for very steep power laws with $p\gtrsim 4$, we cannot exclude the possibility that what we are identifying as a power law is in fact the tail of a thermal-like distribution or a multi-temperature distribution (see e.g., bottom right panel in \fign{spec-comp}). 
Detailed modeling of the energy distributions which is important for determining the temporal evolution of the cutoff energy or the shape of the exponential cutoff \citep{werner_16, kagan_18, petropoulou_18} lies beyond the scope of this paper.

\begin{figure}
 \centering 
 \includegraphics[width=0.47\textwidth]{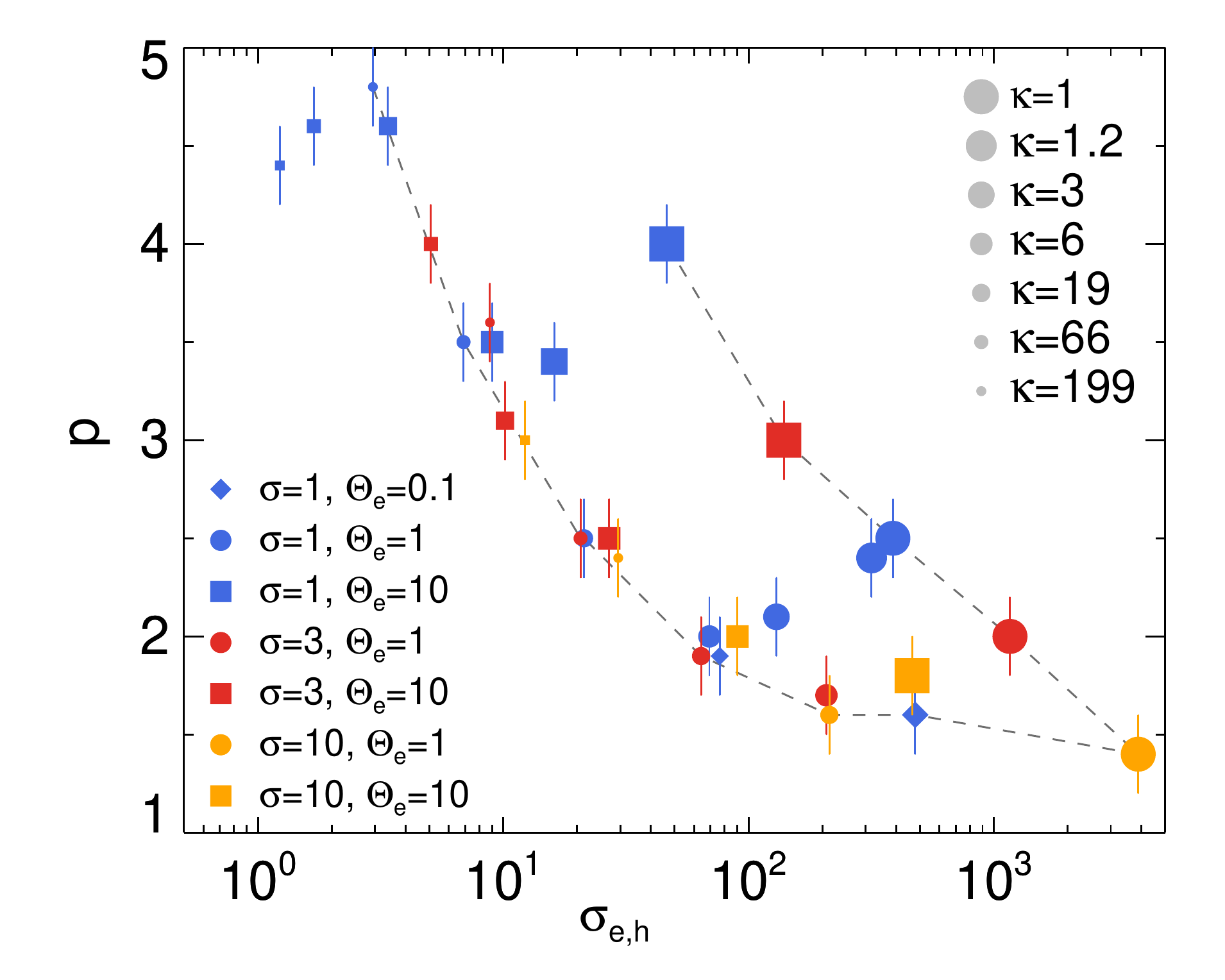} 
 \caption{Power-law index of the electron energy distribution $p$ as a function of $\sigeh$ from our simulations presented in \tab{setup} (results from the box-size scaling runs are not included). Different symbols, colors, and symbol sizes are used to indicate simulations with different values of $\Theta_e$, $\sigma$, and $\kappa$, respectively (see inset legends). Dashed grey lines indicate the two branches discussed in text. A systematic error of $\pm 0.2$ applies to all $p$ values (for details, see \sect{spectra-slope}).}
 \label{fig:index}
\end{figure}

\startlongtable
\begin{table}
\centering
\caption{Summary table of the power-law index of the pair energy distribution $p$, mean electron Lorentz factor $\langle\gamma_e-1\rangle$,  hot pair plasma magnetization $\sigeh$ (\eq{seh}), and electron plasma parameter $\beta_e$ (\eq{betae}) from our simulations of reconnection in pair-proton plasmas (see also Figs.~\ref{fig:index} and \ref{fig:index2}). A systematic error of $\pm 0.2$ applies to all $p$ values. Results from the box-size scaling simulations are not included here. \label{tab:index-gammae}}
\begin{tabular}{cccccccc}
\hline
Run & $\sigma$ & $\Theta_e$ & $\kappa$ & $\sigeh$ & $\beta_e$ &  $p$ & $\langle\gamma_e-1\rangle$ \\
\hline 
A0 & 1	& 1 & 199 & 2.9 &  0.072 & 4.8 & 6.0\\
A1 & 1	& 1 & 66 & 6.9 & 0.031 & 3.5 & 7.0\\
A2 & 1	& 1 & 19 &21.4 & 0.010 & 2.5 & 9.8\\
A4 & 1	& 1 & 6	&69.3 & 0.003 & 2.0 & 15.8\\	
A6 & 1	& 1 & 3	& 130.0& 0.002 & 2.1 & 25.8  \\	
A7 & 1	& 1 & 1.2 &317.6 & 0.001 & 2.4 & 75.0\\	
A8 & 1	& 1 & 1 &387.9 & 0.001 & 2.5 & 147.5 \\
\hline
B0 & 1  & 10 & 199 &1.2 & 0.199 & 4.4 & 44.1\\
B1 & 1  & 10 & 66 & 1.7 & 0.146 & 4.6 & 48.2\\
B2 & 1  & 10 & 19 & 3.4 & 0.076 & 4.6 & 53.9\\
B3 & 1  & 10 & 6 & 9.0 & 0.032 & 3.5 & 65.9\\
B4 & 1  & 10 & 3 & 16.1 & 0.020 & 3.4 & 76.3 \\
B5 & 1  & 10 & 1 & 46.4 & 0.010 & 4.0 & 201.3\\
\hline
C1 & 3  & 1  & 199 & 8.8 & 0.024 & 3.6 & 13.6\\
C2 & 3  & 1  & 66 & 20.7 & 0.010 & 2.5 & 20.0\\
C3 & 3  & 1  & 19 & 64.1 & 0.003 & 1.9 & 36.3\\
C4 & 3  & 1  & 6 & 207.8 & 0.001 & 1.7 & 79.8\\ 
C6 & 3  & 1  & 1 & 1163.8 & 0.0004 & 2.0 & 602.9\\
\hline
D1 & 3  & 10 & 66 & 5.1 & 0.049 & 4.0 & 86.0\\
D2 & 3  & 10 & 19 & 10.2 & 0.025 & 3.1 & 107.9\\
D3 & 3  & 10 & 6  & 27.1 & 0.011 & 2.5 & 153.7\\
D4 & 3  & 10 & 1 & 139.3 & 0.003 & 3.0 & 691.2\\
\hline
E1 & 10  & 1 & 199 & 29.4 & 0.007 & 2.4 & 40.6\\
E2 & 10  & 1 & 19 & 213.6 & 0.001 & 1.6 & 166.3\\
E3 & 10  & 1 & 1 &3879.2 & 0.0001 & 1.4\tablenotemark{a} & 2114.7\\
 \hline
 F1 & 10  & 10 & 199 &12.9 & 0.020 &  3.0 & 203.1\\
 F2 & 10  & 10 & 6 & 90.2 & 0.003 &  2.0 & 592.4\\
 F3 & 10  & 10 & 1 & 464.5 & 0.001 & 1.8 & 2371.3\\
 \hline 
 G1 & 1  & 0.1 & 19 & 76.4 & 0.001 &  1.9 & 0.31  \\
 G2 & 1  & 0.1 &  3 & 478.3  & 0.0002 & 1.6 & 0.25  \\
 \hline 
 H1 & 1 & 100 & 199 & 1.0 & 0.244 & 3.2 & 454.0\\
 H2 & 1 & 100 & 19  & 1.3 & 0.205 & 3.3 & 463.1\\
 H3 & 1 & 100 & 3 & 2.7 & 0.121 & 3.6 & 512.2\\
 \hline
\end{tabular}
\tablenotetext{$a$}{The power law might not have reached saturation, because this is the smallest box-size simulation in terms of $\rLe$ (see \tab{setup}) and the power laws tend to become steeper with increasing box size \citep{petropoulou_18, ball_18}. To check this, we ran a simulation with a three times larger box (E4 in \tab{setup}) and found a slope of 1.6, which is comparable to the reported value within the systematic uncertainties.}
\end{table}
Our results are summarized  in \fign{index} and \tab{index-gammae}, where the slope of the electron energy distribution $p$ is plotted as a function of $\sigeh$.  We do not include the results from runs H1-H3  with the highest plasma temperature (see \tab{setup}), since the energy spectra are qualitatively different from all other cases (for details, see Appendix~\ref{sec:app4} and \cite{ball_18}). The inferred power-law slopes fall onto two branches (dashed grey lines) that track each other for $\sigeh \sim 30-300$, but merge in the asymptotic regime of $\sigeh \gtrsim 10^3$, where both protons and pairs start to behave as one particle species (i.e., their Larmor radii become similar). The upper branch (i.e., larger $p$ values) is composed of results from $\kappa=1$ simulations, whereas results for larger multiplicities ($\kappa \gtrsim 6$) fall onto the lower branch (i.e., smaller $p$ values). For a fixed pair of $\Theta_e$ and $\sigma$ values, a transition from the lower to the upper branch, which is accompanied by a steepening of the power law, occurs at $\kappa\sim 3-6$. No transition is found for $\sigma=10$. The power-law slopes derived for the majority of the simulations lie on the lower branch for a wide range of $\sigeh$ values, spanning more than three orders in magnitude, despite the differences in the total plasma magnetization, temperature, and pair multiplicity. This suggests that $\sigeh$ is a key physical parameter in regard to the pair energy distribution. 

In general, higher $\sigeh$ lead to the production of harder power laws (i.e., smaller $p$ values),
which is similar to the trend reported by \cite{ball_18} for a decreasing electron plasma $\beta_e$ in electron-proton reconnection (see Fig.~13 therein). By tracking a large number of particles, \cite{ball_18} showed that at low $\beta_e$ particles primarily accelerate by the non-ideal electric field at X-points. Since their number was found to decrease with increasing $\beta_e$, the authors argued that lower acceleration efficiencies and steeper power laws are expected at high $\beta_e$. The dependence of our derived power-law slopes on $\sigeh$ can be qualitatively understood in the same context, since at high $\sigeh$ (or equivalently low $\beta_e$) more X-points and secondary plasmoids are formed (see \sect{layer-ionfrac}). A quantitative description of our results requires a detailed study of the electron acceleration, which is beyond the scope of this paper. 

\section{Astrophysical implications}\label{sec:discussion}
In this section, we discuss the findings of our simulations  in the context of AGN jets. We focus on blazars, the most extreme subclass of AGN, with jets closely aligned to our line of sight. The blazar jet emission has a characteristic double-humped shape with a broad low-energy component extending from radio wavelengths up to UV or X-ray energies, and a high-energy component extending across the X-ray and $\gamma$-ray bands  \citep[][]{ulrichetal_97, fossatietal_98, costamante_01}. The low-energy hump is believed to be produced by synchrotron emission of relativistic pairs with a power-law (or broken power-law) energy distribution \citep[e.g.][]{celotti_08}, which is suggestive of non-thermal particle acceleration. The synchrotron-emitting pairs can also inverse Compton scatter low-energy photons to $\gamma$-ray energies, which can explain the high-energy component of the blazar spectrum\footnote{This is true in leptonic scenarios where the broadband jet emission is attributed to relativistic pairs. This is our working hypothesis and our results should be interpreted in this framework.}. In blazars with TeV $\gamma$-ray emission, electrons should accelerate up to Lorentz factors $10^5-10^6$ to explain the highest photon energies \citep[e.g.][]{aleksic_12, ahnen_18}. 

\subsection{Properties of radiating particles}
A key parameter in blazar emission models is the shape of the non-thermal pair distribution (e.g., power law, broken power law,  log-parabolic, and others). The assumed distribution in most cases is phenomenological, as it is not derived from a physical scenario. Upon adopting a specific model for the energy distribution of accelerated pairs, its properties (e.g., power-law slope, minimum, and maximum Lorentz factors) are inferred by modeling the  broadband blazar photon spectrum \citep[e.g.,][]{celotti_08, ghisellini_14}. However, not all the model parameters can be uniquely determined due to degeneracies that are inherent in the radiative models \citep[e.g.,][]{cerruti_13}. 

Bearing in mind the aforementioned caveats, we continue with a tentative comparison of our results (see Sect.~\ref{sec:energy}-\ref{sec:spectra}) with those inferred by radiative leptonic models. As an indicative example, we use the results of \cite{celotti_08}. The accelerated lepton distribution that was used for the modeling was assumed to be a broken power-law:
\eqb 
f(\gamma) \propto \left \{ \begin{array}{cc}
                         \gamma^{-s_1}, &  \gamma \le  \gamma_{inj} \\
                         \gamma^{-s_2}, &  \gamma >  \gamma_{inj}
                         \end{array}
                         \label{eq:dist}
                   \right.     
                        \eqe 
where $s_1=1$, and $s_2$, $\gamma_{inj}$ were determined by the fit to the data. There is some degeneracy in the low-energy index, since distributions with even flatter spectra than the one above (i.e., $s_1<1$) cannot be usually distinguished by the data \citep[see also][]{ghisellini_14}. 

In our simulations, we find that the post-reconnection pair energy distributions exhibit a power law extending well beyond a broad thermal-like component that peaks at $E_{pk, e}$ (see e.g., Figs.~\ref{fig:spec-comp} and \ref{fig:spec-ionfrac}). At $E<E_{pk,e}$, the pair spectra in the reconnection region generally follow the low-energy tail of a Maxwell-J{\"u}ttner distribution (see e.g., \fign{spec-time-hot}), which can be modeled by an inverted power law (i.e., $s_1<0$). For the purposes of making a general comparison to the modeling results, we can phenomenologically describe the lepton energy spectra from our simulations by \eq{dist}, with $s_1 < 1$, $s_2 = p$, and a peak Lorentz factor $\gamma_{inj}=1+E_{pk, e}/m_e c^2$, which depends on the total magnetization and temperature of the plasma (see e.g., \fign{spec-comp} and \fign{spec-time-hot}). Using the fitting results of \cite{celotti_08} (see Table A1 therein), we compute the mean Lorentz factor of the accelerated distribution (i.e., without radiative cooling) and compare it against the one determined by our simulations (see e.g., \fign{ratio-seh}). 

\begin{figure}
 \centering  
 \includegraphics[width=0.47\textwidth]{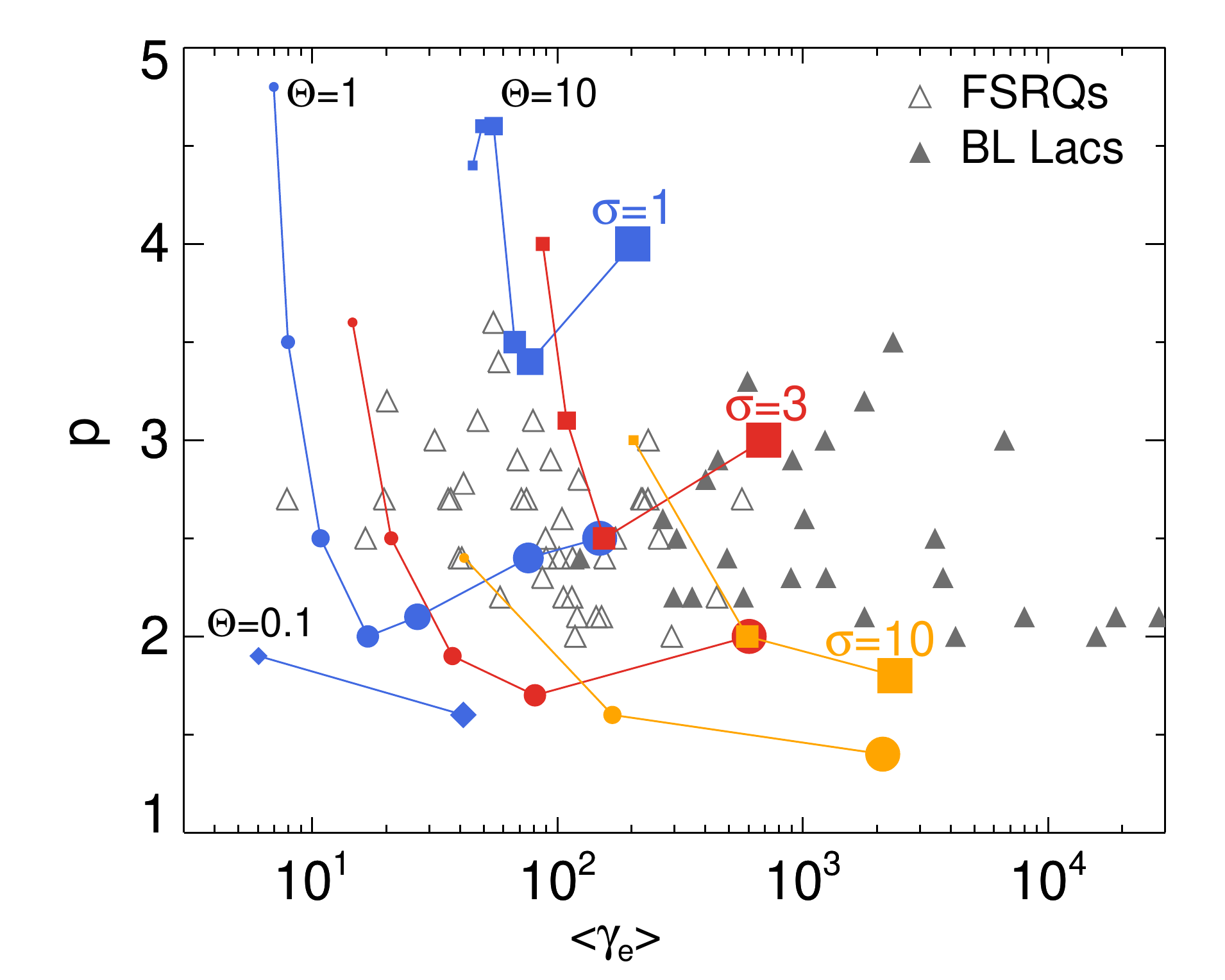}
 \caption{Power-law index of the electron distribution $p$ plotted against the mean electron Lorentz factor $\langle\gamma_e\rangle$ from our simulations. Different colors and symbols are used to indicate the total plasma magnetization $\sigma$ and temperature $\Theta_e$, respectively. The pair multiplicity $\kappa$ is indicated by the symbol size, as in \fign{index}. A systematic error of $\pm 0.2$ applies to all $p$ values, but is not plotted for clarity. Overplotted with filled and open triangles are the values inferred by leptonic modeling of blazar broadband spectra \citep{celotti_08} for different blazar types (see inset legend).}
 \label{fig:index2}
\end{figure}

Our results are summarized in \fign{index2}, where the power-law index $p$ above the peak Lorentz factor of the distribution is plotted against the mean Lorentz factor $\langle \gamma_e\rangle$ of the distribution (for a tabulated list of our results, see \tab{index-gammae}). Open and filled triangles indicate the values from the leptonic modeling of \cite{celotti_08} for FSRQs and BL Lac objects, respectively. The predictions of reconnection are shown with colored symbols (for details, see figure caption). The degeneracy of the power-law index $p$ on the physical parameters, such as $\sigma$ and $\Theta_e$ shown in \fign{index}, is lifted when $p$ is plotted against the mean lepton Lorentz factor. This is illustrated in \fign{index2}, where, for fixed $\sigma$, curves corresponding to higher $\Theta_e$ values are shifted towards larger $\langle\gamma_e\rangle$ and $p$ values (upper right corner of the plot). For fixed plasma temperature but increasing $\sigma$, the curves are shifted towards lower $p$ values (i.e., harder power laws) and larger mean particle energies, regardless of the pair multiplicity. 

Interestingly, the values from our simulations fall in the same range with those inferred by leptonic radiation models. More specifically, the  numerically obtained curves for $\Theta_e=1$ and 10 enclose most of the results for FSRQs (open triangles). One can envision different families of curves that pass through the data points for FSRQs, which can be obtained by simply changing the temperature of the upstream plasma from $\Theta_e=1$ to $10$. For example, some FSRQ results could be interpreted by reconnection in pair-proton plasmas with $\Theta_e=3$,  $\sigma=1$, and $\kappa \sim 1-10$ (imagine the blue line with circles shifted to the right and upwards). The relevant range of multiplicities would be somewhere between $\sim 10-70$, for $\Theta_e=3$ and $\sigma=3$  (imagine the red line with circles shifted to the right and upwards). We find that reconnection in cold pair-proton plasmas ($\Theta_e \ll1$) with $\sigma \le 10$ typically results in slopes and mean lepton energies that are not compatible with the FSRQ results. 

BL Lac sources with $\langle\gamma_e\rangle \sim 10^2-10^3$ are compatible with our simulation results for reconnection in pair-proton plasmas with $\sigma \sim 3-10$, $\Theta_e\sim 10$ and $\kappa\sim 1-10$. The majority of BL Lac sources, however, requires mean Lorentz factors  $>10^3$. Reconnection in strongly magnetized plasmas ($\sigma > 10$) can lead to high values of the mean Lorentz factor, but at the same time produces hard power laws ($p<2$) above the peak Lorentz factor $\gamma_{inj}$ (see e.g., orange curves) that do not agree with the fitting results for  $\langle \gamma_e \rangle\gg 10^3$ (filled triangles). In this regime, however, we argue that $\gamma_{inj}$ could be interpreted as the maximum Lorentz factor of a hard power law with $p<2$, as found in our high-$\sigma$ models, with $p$ now corresponding to the index $s_1$ (see \eq{dist}). Because the determination of the maximum Lorentz factor from the simulation spectra is not trivial \citep[see e.g.][]{werner_18}, we refrain from drawing strong conclusions from the comparison of our results to the BL Lac sources in the sample of \cite{celotti_08}.

\begin{figure}
    \centering
    \includegraphics[width=0.47\textwidth]{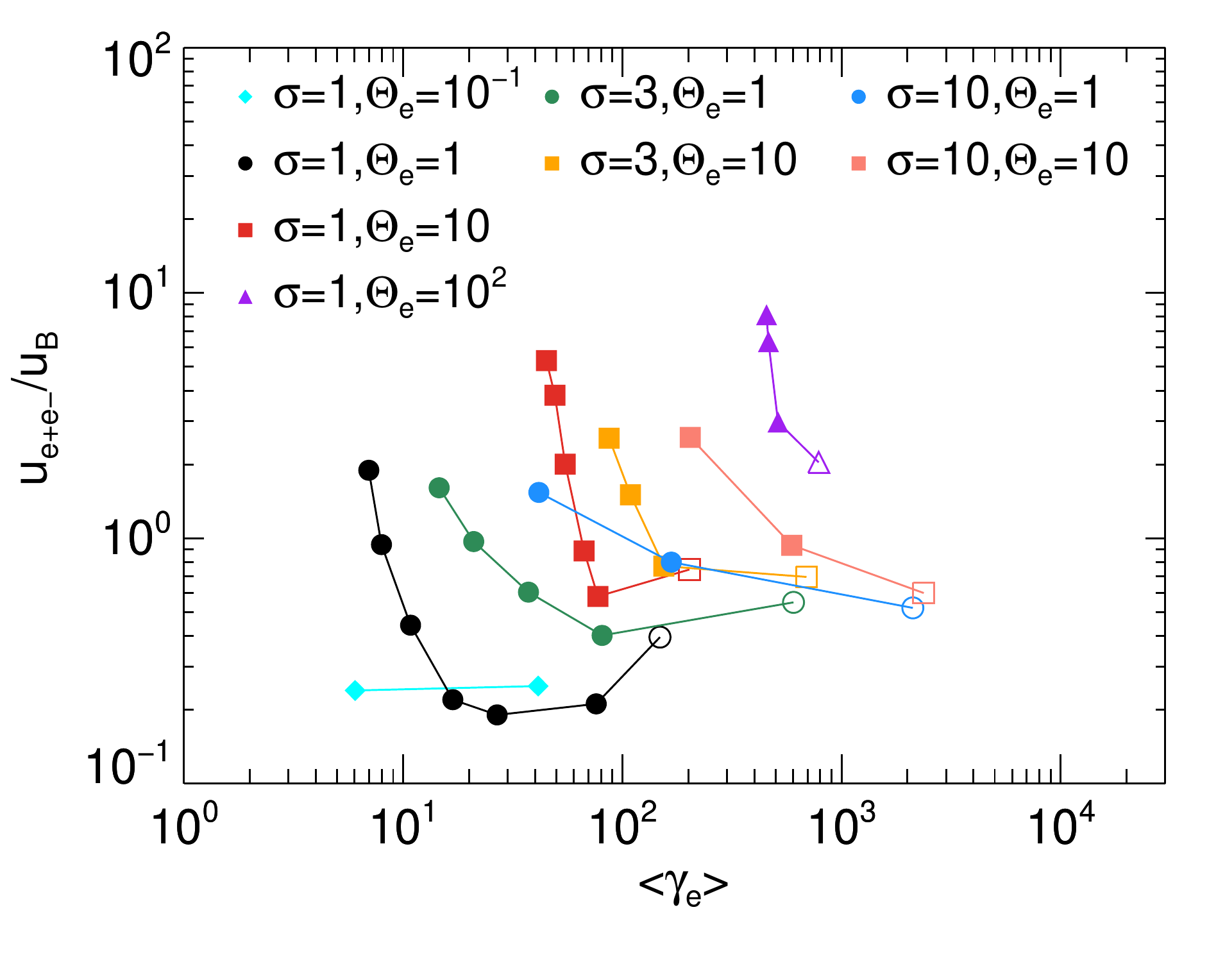}
    \caption{Ratio of post-reconnection lepton-to-magnetic energy densities plotted against the mean lepton Lorentz factor from our simulations (results from the box-size scaling runs are not included). Filled  and  open  symbols are  used  for  simulations  of reconnection in  pair-proton and electron-proton plasmas, respectively.}
    \label{fig:equipartition}
\end{figure}

\subsection{Equipartition conditions}
One of the reasons that makes the principle of energy equipartition between particles and magnetic fields attractive is that it leads to minimum power solutions for blazar jets \citep[e.g.,][]{dermer_14, petropoulou_16}. The energy density ratio of radiating particles and magnetic fields in the blazar emitting region is usually a free  parameter, which is determined by the fitting of photon spectra. Leptonic emission models typically find $0.03\lesssim u_{e^\pm}/u_B\lesssim 30$, although specific sources may require even higher values \citep[e.g.,][]{celotti_08, tavecchio_10, ghisellini_14}. Alternatively, one can impose the constraint of rough energy equipartition between pairs and magnetic fields while searching for the best fit model, as demonstrated successfully by \cite{cerutti_13b, dermer_14, dermer_15}.

The post-reconnection ratio $u_{e^\pm}/u_B$ obtained from our simulations is plotted in \fign{equipartition} as a function of the mean  lepton Lorentz factor $\langle \gamma_e\rangle$. We find that $0.2 \lesssim u_{e^\pm}/u_B \lesssim 10$, with higher values obtained for hotter upstream plasmas. Even larger ratios, as those inferred by modeling of TeV BL Lacs \citep[e.g.,][]{tavecchio_10}, would require a pool of ultra-relativistically hot particles entering the reconnection region. The presence of a guide field (i.e., of a magnetic field component that does not reconnect) would make the reconnection region more magnetically dominated, thus leading to $u_{e^\pm}/u_B < 0.2$. More specifically, for electron-proton reconnection it was demonstrated that the fraction of magnetic energy transferred to non-thermal electrons can decrease from $\sim50\%$ (in the absence of guide field) to $\sim 10\%$ for a guide field with strength comparable to that of the reconnecting field component \citep{sironi_15, werner_17}. Yet, dissipation efficiencies as low as a few percent are still compatible with the global energetic requirements for AGN emission \citep{ghisellini_14, sironi_15}. A systematic study of the effects of the guide-field in pair-proton reconnection will be the topic of a future study.

\section{Summary}\label{sec:summary}
For the first time, we have investigated magnetic reconnection in electron-positron-proton plasmas with a suite of large-scale 2D PIC simulations, covering a wide range of pair multiplicities ($\kappa=1-199$) for different values of the all-species plasma magnetization ($\sigma=1,3$ and 10) and plasma temperature ($\Theta_e=0.1,1,10$, and 100). In all cases we explored, protons in the upstream plasma have non relativistic temperatures and dominate the total mass. 

The inflow rate of plasma into the reconnection region (i.e., the reconnection rate) ranges between $\sim 0.05 v_A$ and $0.15 v_A$ for a wide range of values of the hot pair plasma magnetization $\sigeh$, with a weak trend towards higher rates for larger $\sigeh$ values. The motion of the plasma outflow in the reconnection region, which is governed by the proton inertia, is relativistic with a maximum four-velocity that approaches the expected asymptotic value of $\sqrt{\sigma}$. We found no significant dependence of the outflow four-velocity on the pair multiplicity or temperature. 

We showed that $\sim 1/3$ of the total energy remains in the post-reconnection magnetic field for $\sigeh \gtrsim 3$, with the remaining $2/3$ of the energy being shared between pairs and protons. Energy equipartition between protons and pairs is achieved for $\sigma\gg 1$ and $\sigeh \gtrsim 30$.  For $\sigeh \lesssim 3$, most of the energy in the reconnection region is carried by the pairs, with protons and magnetic fields contributing $\sim (1-10)\%$ to the total energy.

The reconnection process produces non-thermal particle energy distributions. We found that the mean Lorentz factor of the proton distribution (or, more accurately $\langle\gamma_i-1\rangle$) is almost independent of the pair multiplicity and plasma temperature, but it is approximately equal to $\sigma/3$.
The mean Lorentz factor of the pair distribution can be described by a simple analytical expression (see \eq{gammae}) for different values of $\sigeh$, $\sigma$, and $\Theta_e$. 

The electron and positron energy distributions in the reconnection region are similar and can be modeled as a power law with slope $p$ above a peak Lorentz factor, which, in most cases, is comparable with the mean Lorentz factor given by \eq{gammae}.
The energy distribution below the peak can be, in general, approximated by a flat power law (with index $<0$). We showed that $p$ is mainly controlled by $\sigeh$ (with harder power laws obtained for higher magnetizations) for a wide range of $\sigma, \Theta_e$, and $\kappa$ values. There is, however, a dependence of $p$ on pair multiplicity, with power laws getting steeper as $\kappa$ decreases from a few to unity.

We discussed the implications of our results in the context of AGN jets. We showed that reconnection in pair-proton plasmas naturally produces power-law pair distributions with slopes and average Lorentz factors similar to those  obtained by leptonic modeling of the broadband jet emission. In general, we find that the majority of the modeling results can be explained in the context of reconnection in pair plasmas with multiplicities $\kappa \sim 1-20$, magnetizations $\sigma \sim 1-10$, and temperatures $\Theta_e \sim 1-10$. 
\acknowledgments
\textit{Acknowledgments.} The authors thank the anonymous referee for their report that helped to improve the manuscript.
MP acknowledges support from the Lyman Jr.~Spitzer Postdoctoral Fellowship and the Fermi Guest Investigation grant 80NSSC18K1745. LS acknowledges support from DoE DE-SC0016542, NASA ATP NNX-17AG21G, NSF ACI-1657507, and NSF AST-1716567. AS is supported by NASA ATP 80NSSC18K1099 and the Simons Foundation (grant~267233). DG acknowledges support from the NASA ATP NNX17AG21G and the NSF AST-1816136 grants. The simulations were performed using computational resources at the TIGRESS high-performance computer center at Princeton University, the Texas Advanced Computing Center (TACC) at The University of Texas at Austin (TG-AST180038, TG-AST180001), and at the NASA High-End Computing (HEC) Program through the NASA Advanced Supercomputing (NAS) Division at Ames Research Center.
\software{TRISTAN-MP \citep{buneman_93, spitkovsky_05}, IDL version 8.6  (Exelis Visual Information Solutions, Boulder, Colorado).}

\appendix
\section{Parameter definitions}\label{sec:app1}
\begin{deluxetable}{ccc}
\tablecaption{Description, symbol, and definition of parameters used in this study.\label{tab:param}}
\tablehead{
\colhead{Parameter} & \colhead{Symbol} & \colhead{Definition} 
}
\startdata
 Pair multiplicity & $\kappa$ & $\nee/n_i$ \\
 Proton fraction & $q$ & $n_i/n_{e^-}=2/(\kappa+1)$ \\
 Lepton adiabatic index & $\hat{\gamma}_e$ & \citet{synge_57}\\
 Proton adiabatic index & $\hat{\gamma}_i$ & \citet{synge_57}\\
 Total plasma magnetization & $\sigma$ & \eq{stot} \\
 Hot pair plasma magnetization & $\sigeh$ & \eq{seh} \\
 Cold pair plasma magnetization & $\sigec$ & \eq{sec} \\
 Hot proton plasma magnetization & $\sigih$ & \eq{sih} \\
 Cold proton plasma magnetization & $\sigic$ & \eq{sic} \\ 
 Electron plasma $\beta$ & $\beta_e$ & \eq{betae} \\
 Plasma electron frequency & $\omega_{pe-}$ & \eq{ome} \\
 Plasma proton frequency & $\omega_{pi}$ & \eq{omp} \\
 Electron Larmor radius & $\rLe$ & \eq{rLe} \\
 Proton Larmor radius & $\rLi$ & \eq{rLi} \\
\enddata 
\end{deluxetable} 

\begin{figure*}
 \centering 
 \includegraphics[width=0.49\textwidth]{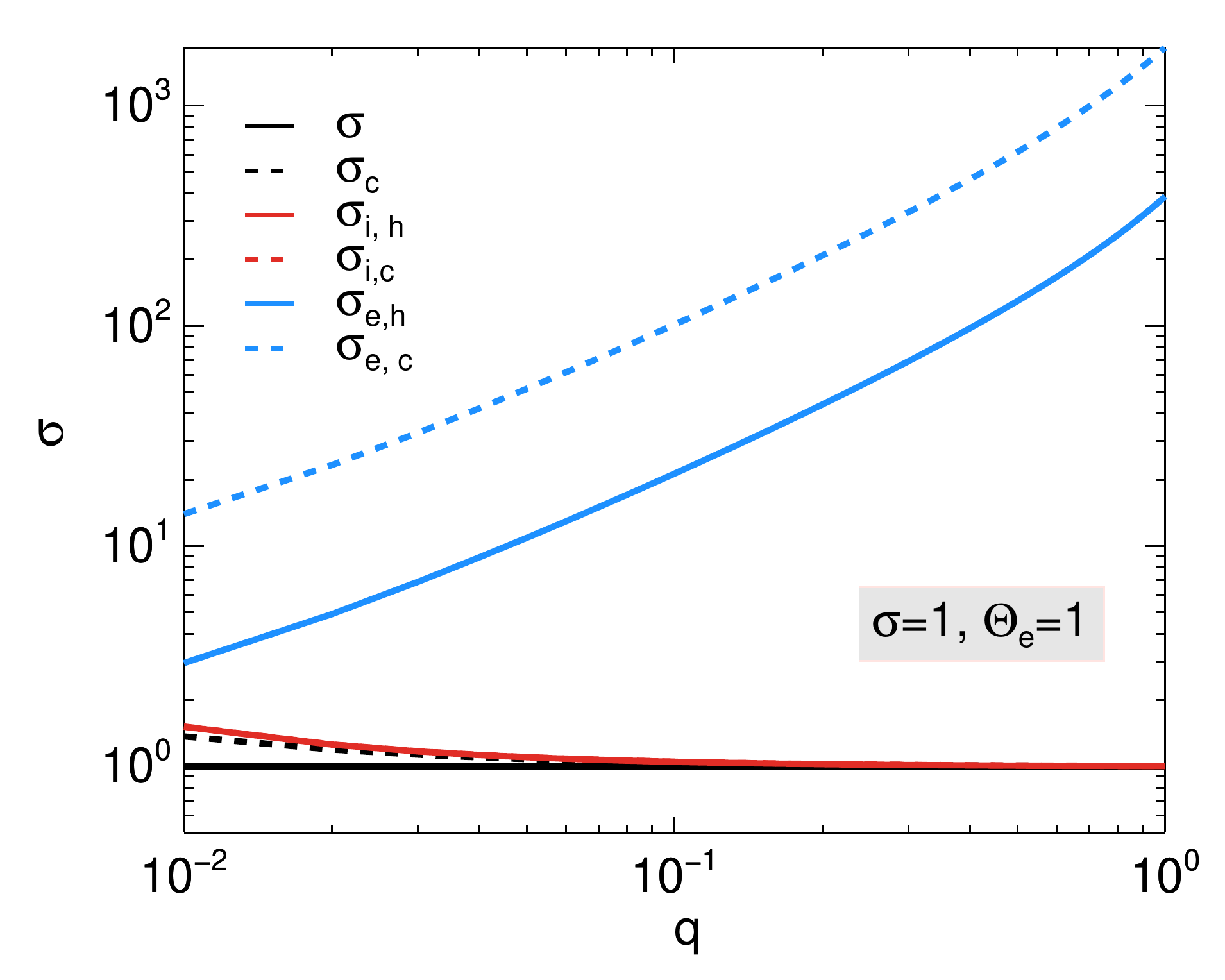}
 \includegraphics[width=0.49\textwidth]{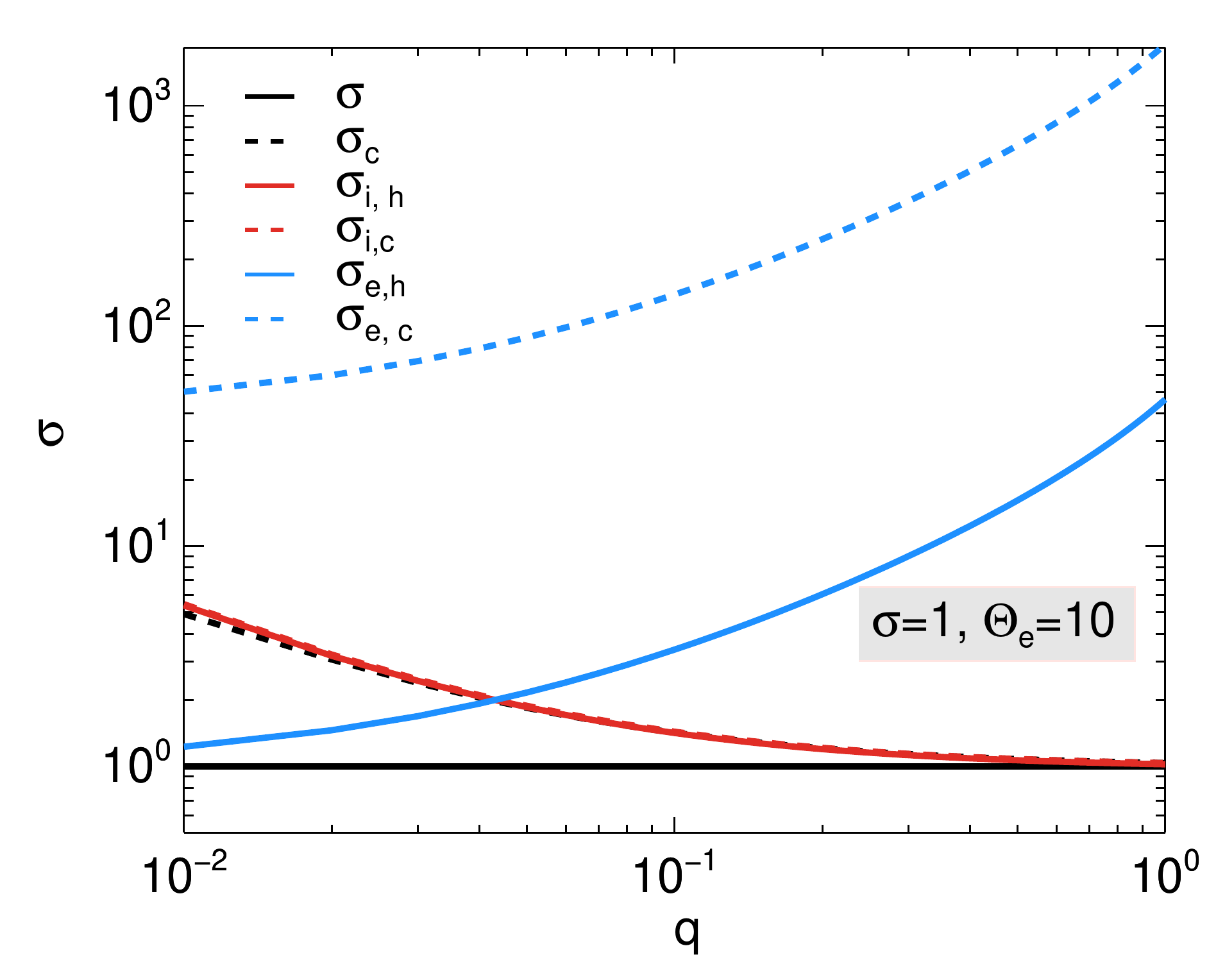}
 \caption{Various magnetizations -- defined in eqs.~(\ref{eq:stot})-(\ref{eq:sic}) -- plotted as a function of the proton fraction $q$ for $\sigma=1$ and two plasma temperatures: $\Theta_e=1$ (left panel) and $\Theta_e=10$ (right panel).} 
 \label{fig:sigma_ionfrac}
\end{figure*}

\begin{figure*}
 \centering 
 \includegraphics[width=0.49\textwidth]{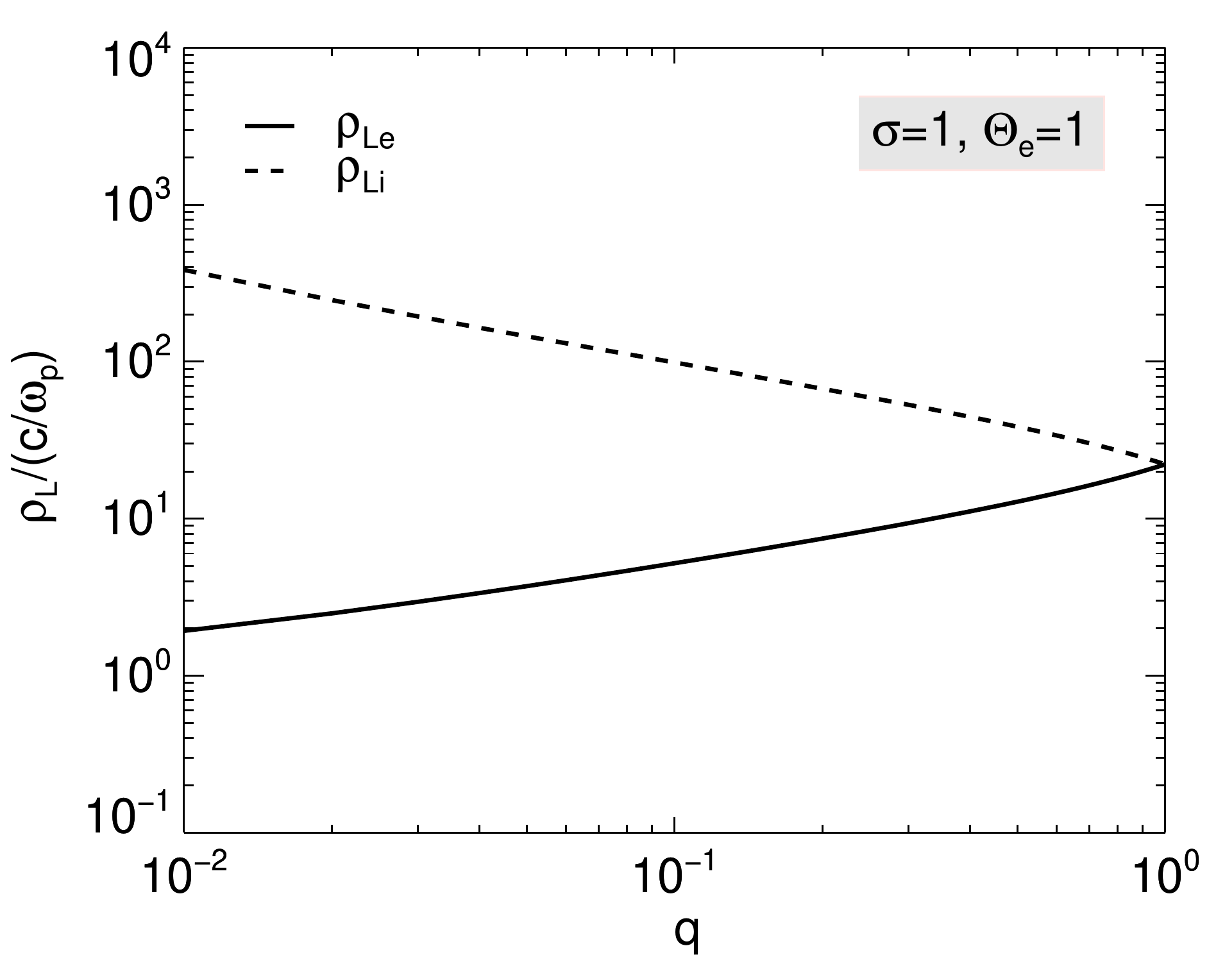}
 \includegraphics[width=0.49\textwidth]{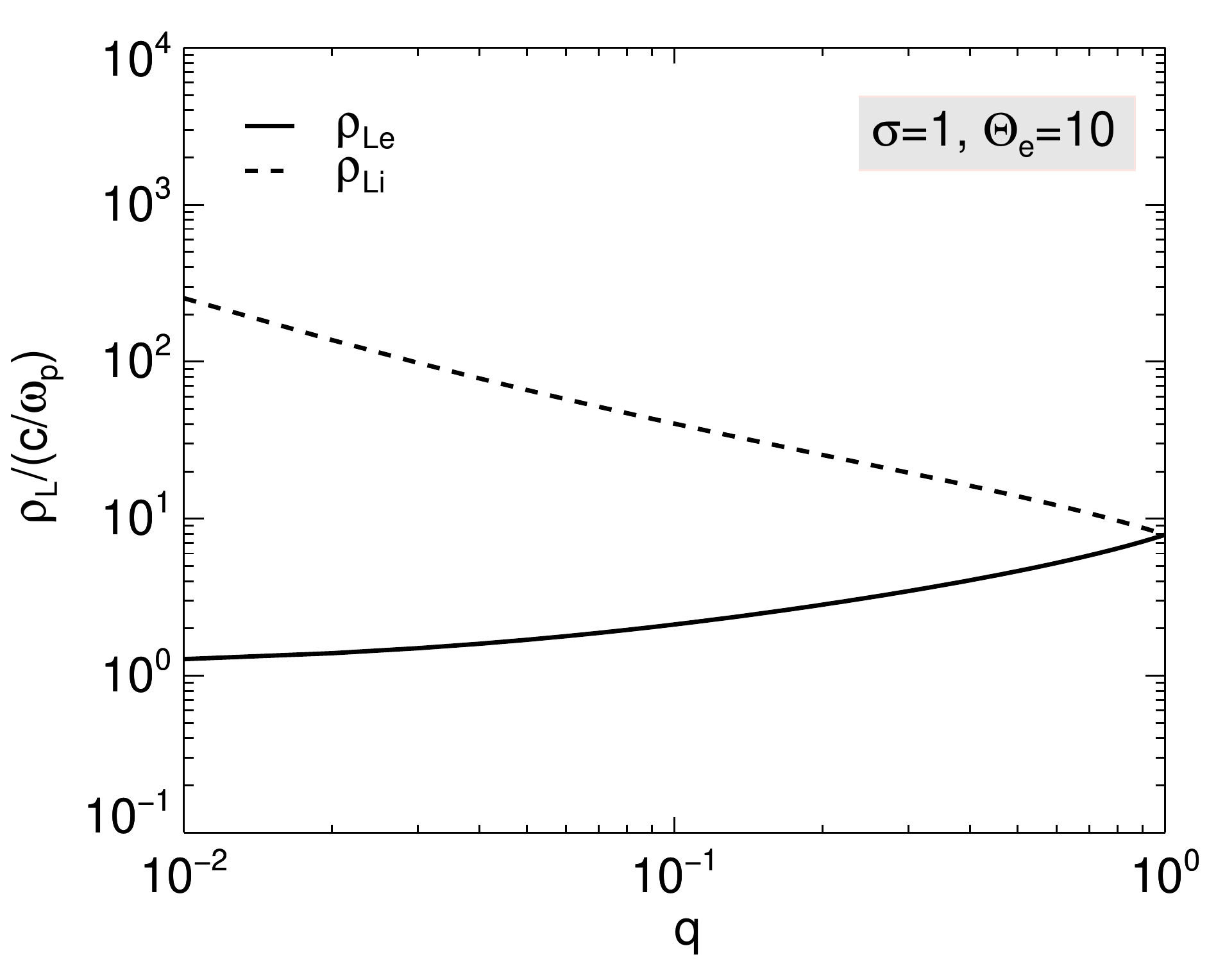}
 \caption{Electron and proton Larmor radii (see eqs.~(\ref{eq:rLe}) and ({\ref{eq:rLi})}, respectively) normalized to the all-species plasma skin depth plotted as a function of the proton fraction $q$   for $\sigma=1$ and two plasma temperatures: $\Theta_e=1$ (left panel) and $\Theta_e=10$ (right panel).} 
 \label{fig:rho_ionfrac}
\end{figure*}
We summarize the basic physical parameters that are relevant for this study (see Table~\ref{tab:param}) and provide their definitions below. The total (all-species) plasma magnetization is defined as:
\eqb
\sigma &=& \frac{B_0^2/4\pi }{n_i m_i c^2 + \frac{\hat{\gamma}_i}{\hat{\gamma}_i-1} n_i k T_i + \nee m_e c^2 + \frac{\hat{\gamma}_e}{\hat{\gamma}_e-1} \nee k T_e},
\label{eq:stot}
\eqe 
where $B_0$ is the upstream magnetic field strength and  $n_{i}, n_{e^\pm}$ are the number densities of protons and pairs, respectively, in the upstream region. Particles are initialized with temperatures $T_i=T_e$. The adiabatic indices for pairs and protons are computed iteratively using the equation of state by \citet{synge_57}. We find that $\hat{\gamma}_e\approx 4/3$
$\hat{\gamma}_i  \approx 5/3$, except for  $\Theta_e=0.1$ where $\hat{\gamma}_e\approx 1.5$. The cold plasma magnetization, which neglects the enthalpy terms is defined by:
\eqb
\sigma_c =\frac{B_0^2}{ 4\pi \left(n_i m_i c^2 + \nee m_e c^2 \right)}\cdot
\label{eq:stot_c}
\eqe 
A key parameter in the study of the post-reconnection particle energy distributions (see \sect{energy} and \sect{spectra}) is the hot pair plasma magnetization, which relates to the total $\sigma$ as:
\eqb
\sigeh = \sigma\frac{q \left(\frac{m_i}{m_e} + \frac{\hat{\gamma}_i \Theta_e}{\hat{\gamma}_i-1}\right) +(2-q)\left(1+\frac{\hat{\gamma}_e \Theta_e}{\hat{\gamma}_e-1} \right)}{(2-q) \left(1+\frac{\hat{\gamma}_e\Theta_e}{\hat{\gamma}_e-1}\right)}\cdot
\label{eq:seh-app}
\eqe 
The cold pair plasma magnetization is identical to $\sigeh$ only for non-relativistically hot plasmas ($\Theta_e \ll 1$) and is defined as:
\eqb
\sigec = \sigma \frac{q \left(\frac{m_i}{m_e} + \frac{\hat{\gamma}_i\Theta_e}{\hat{\gamma}_i-1}\right) +(2-q)\left(1+\frac{\hat{\gamma}_e\Theta_e}{\hat{\gamma}_e-1} \right)}{2-q}\cdot
\label{eq:sec}
\eqe 
Similar to the pair plasma, one can define the hot proton plasma magnetization:
\eqb
\sigih = \sigma \frac{q \left(\frac{m_i}{m_e} + \frac{\hat{\gamma}_i\Theta_e}{\hat{\gamma}_i-1}\right) +(2-q)\left(1+\frac{\hat{\gamma}_e\Theta_e}{\hat{\gamma}_e-1} \right)}{q \left(\frac{m_i}{m_e} + \frac{\hat{\gamma}_i\Theta_e}{\hat{\gamma}_i-1}\right)},
\label{eq:sih}
\eqe 
which is $\approx \sigma$ for all our cases. The cold proton plasma magnetization is written as:
\eqb
\sigic \!=\!\!\frac{\sigma m_e}{q m_i}\!\!\left[q \!\left(\frac{m_i}{m_e}   \!+\! \frac{\hat{\gamma}_i\Theta_e}{\hat{\gamma}_i-1}\right)\!\!+\!(2\!-\!q)\left(1\!+\!\frac{\hat{\gamma}_e\Theta_e}{\hat{\gamma}_e-1} \right)\!\right],
\label{eq:sic}
\eqe 
and it is the same as $\sigih$ as long as $\Theta_e \ll m_i/m_e$.

The ratio of the electron plasma pressure and the magnetic pressure (plasma $\beta_e$), which is a key parameter in studies of electron-proton reconnection, relates to $\sigeh$ as:
\eqb 
\beta_e \equiv \frac{8\pi n_{e^{-}}kT_e}{B_0^2} = \frac{2\Theta_e}{\sigeh (2-q) \left(1+ \frac{\hat{\gamma}_e}{\hat{\gamma}_e-1}\Theta_e \right)}\cdot
\label{eq:betae}
\eqe 
If all particle species are relativistically hot ($\Theta_e \gg  m_i/m_e$), then  $\sigeh \approx 2\sigma/(2-q)$ and $\beta_e$ reaches its maximum value $\approx 1/ 4\sigma$. 

Let $\omega_p$ denote the all-species plasma frequency:
\eqb 
\omega^2_p = \omega^2_{pe^-} + \omega^2_{pe^+} + \omega^2_{pi}.
\label{eq:omp_tot}
\eqe 
where the electron, positron, and proton plasma frequencies are given by:
\eqb 
\label{eq:ome}
\omega^2_{pe^-} & = & \frac{4\pi n_{e^-} e^2}{m_e \left(1+\frac{\Theta_e}{\hat{\gamma}_e-1} \right)},
\eqe 
\eqb
\omega^2_{pe^+} & = & \omega^2_{pe^-}(1-q), 
\label{eq:ompos}
\eqe 
and 
\eqb 
\omega^2_{pi} = \omega^2_{pe^-} \frac{m_e}{m_i}q\frac{1+\frac{\Theta_e}{\hat{\gamma}_e-1}}{1+\frac{\Theta_i}{\hat{\gamma}_i-1}}.
\label{eq:omp}
\eqe 
Finally, we define the Larmor radius of electrons and protons with Lorentz factors $\sigec$ and $\sigic$, respectively, assuming that all the magnetic energy is transferred to the particles:
\eqb 
\rLe \equiv \frac{\sigec m_e c^2}{e B_0} = \frac{c}{\omega_{pe^-}}\left(\frac{\sigec}{2-q}\right)^{1/2}\left(1+\frac{\Theta_e}{\hat{\gamma}_e-1}\right)^{-1/2}
\label{eq:rLe}
\eqe 
and 
\eqb 
\rLi \equiv \frac{\sigic m_i c^2}{e B_0} = \frac{c}{\omega_{pe^-}}\left(\frac{\sigic}{q}\frac{m_i}{m_e}\right)^{1/2}\left(1+\frac{\Theta_e}{\hat{\gamma}_e-1}\right)^{-1/2}.
\label{eq:rLi}
\eqe 
Figures \ref{fig:sigma_ionfrac} and \ref{fig:rho_ionfrac} show the various magnetizations and particle Larmor radii as a function of the proton fraction $q$, which is related to the pair multiplicity $\kappa$ as $q=2/(\kappa+1)$.
\section{Appearance of the reconnection layer}\label{sec:layer-boxsize}
\begin{figure*}
 \centering
 \includegraphics[width=0.49\textwidth, trim= 0 0 20 0]{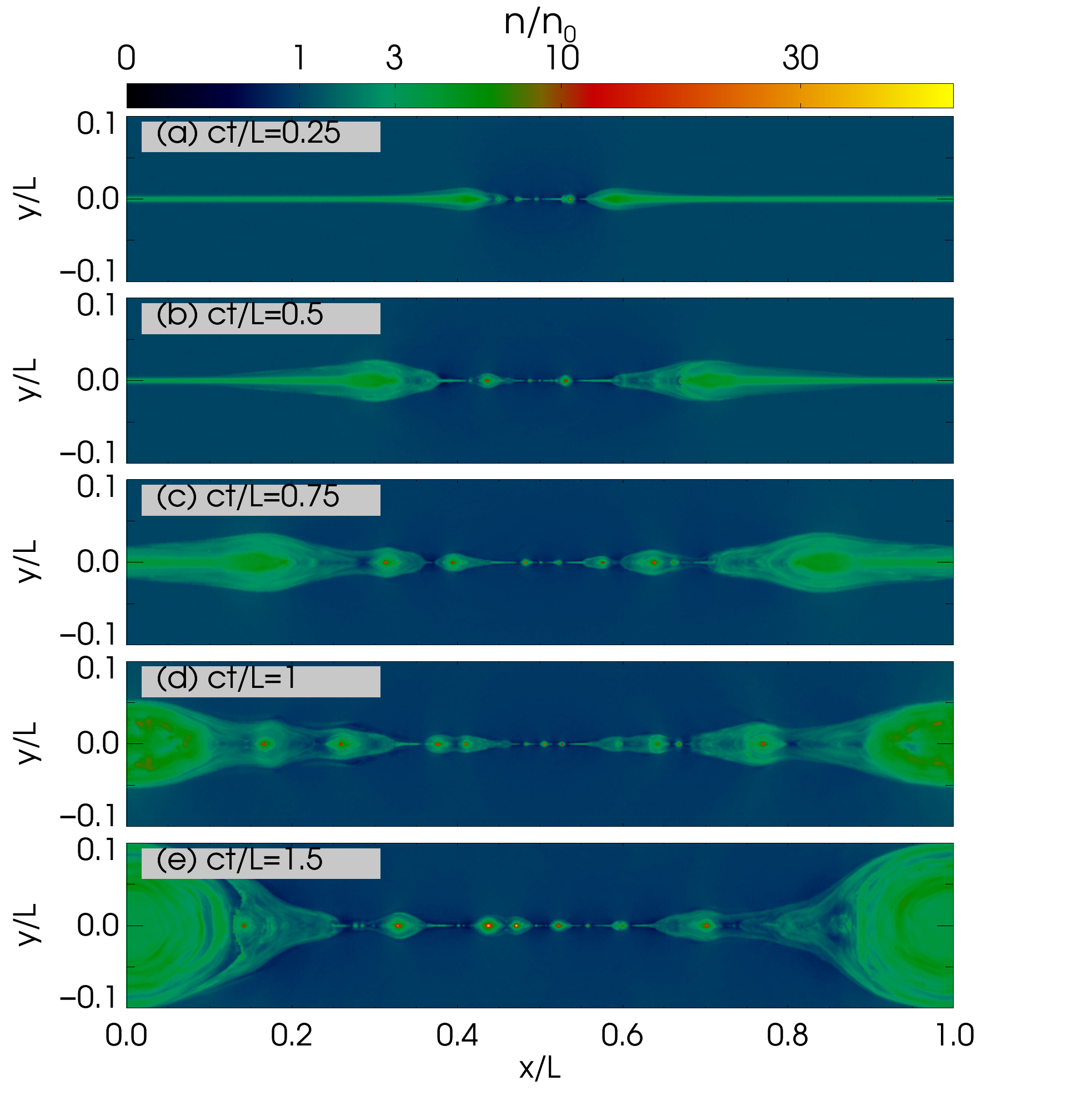}
  \includegraphics[width=0.49\textwidth, trim= 0 0 20 0]{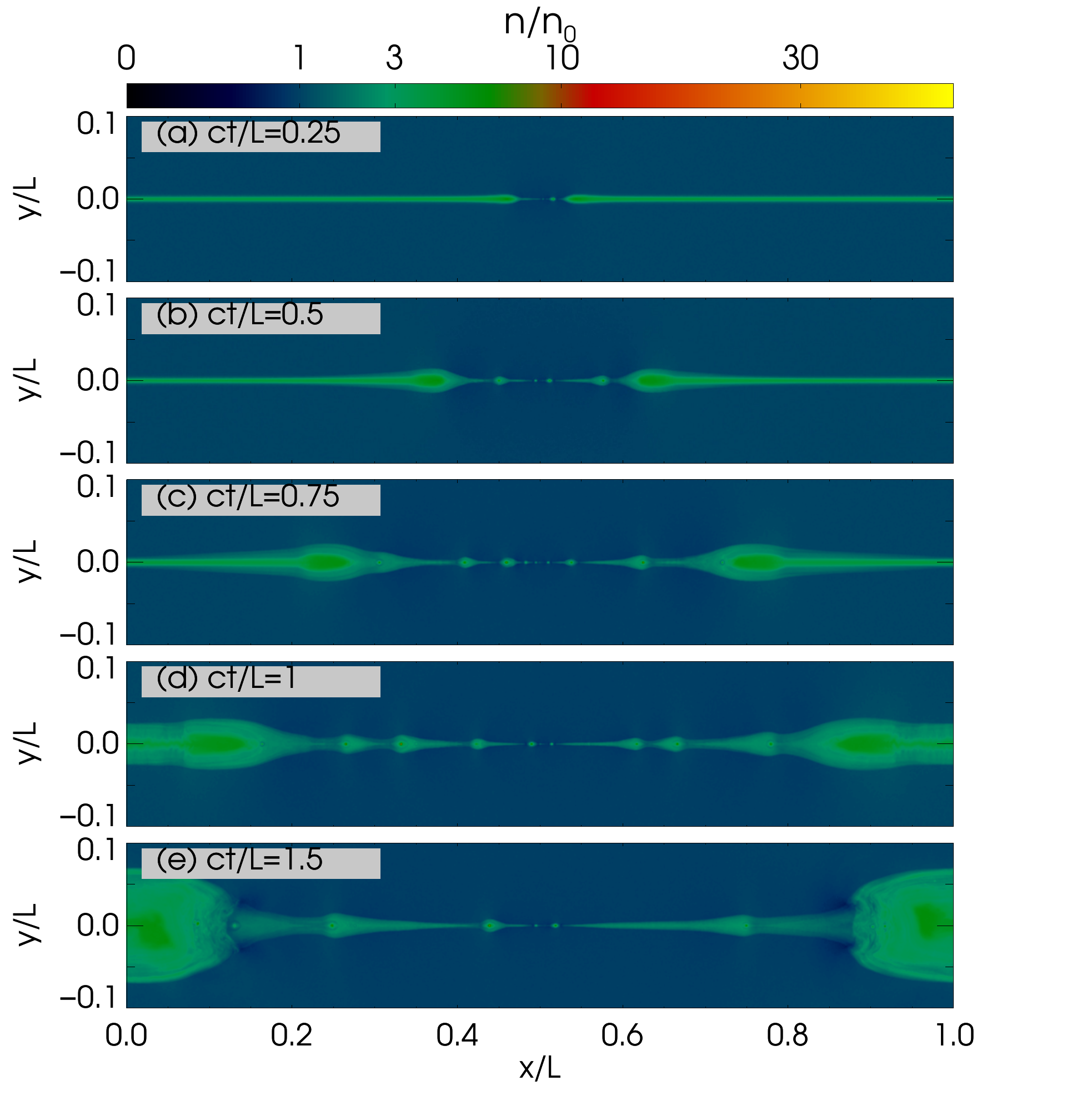}
 \caption{Snapshots of the 2D structure of the all-species particle number density $n$ (normalized to the number density $n_0$ far from the reconnection layer) from two simulations with different physical conditions, but similar box size in terms of $\rLi$ (see runs A2 and B1 in \tab{setup}): $\sigma=1, \Theta_e=1$, $\kappa=19$, $L/\rLi\simeq53$ (left) and  $\sigma=1, \Theta_e=10$, $\kappa=66$, $L/\rLi\simeq53$ (right).
 } 
 \label{fig:layer-1}
\end{figure*}

\begin{figure*}
 \centering
 \includegraphics[width=0.49\textwidth, trim= 0 0 20 0]{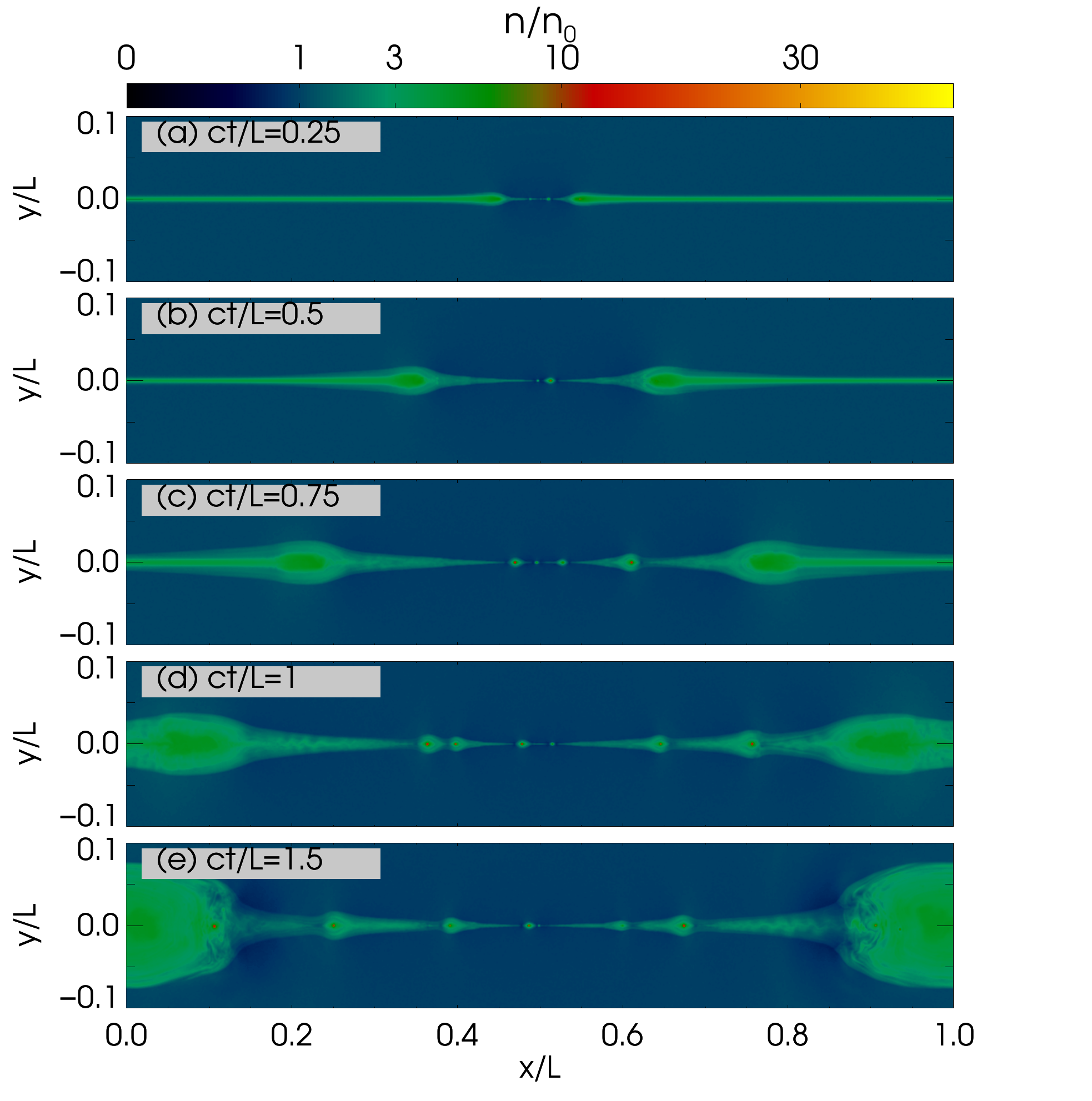}
  \includegraphics[width=0.49\textwidth, trim= 0 0 20 0]{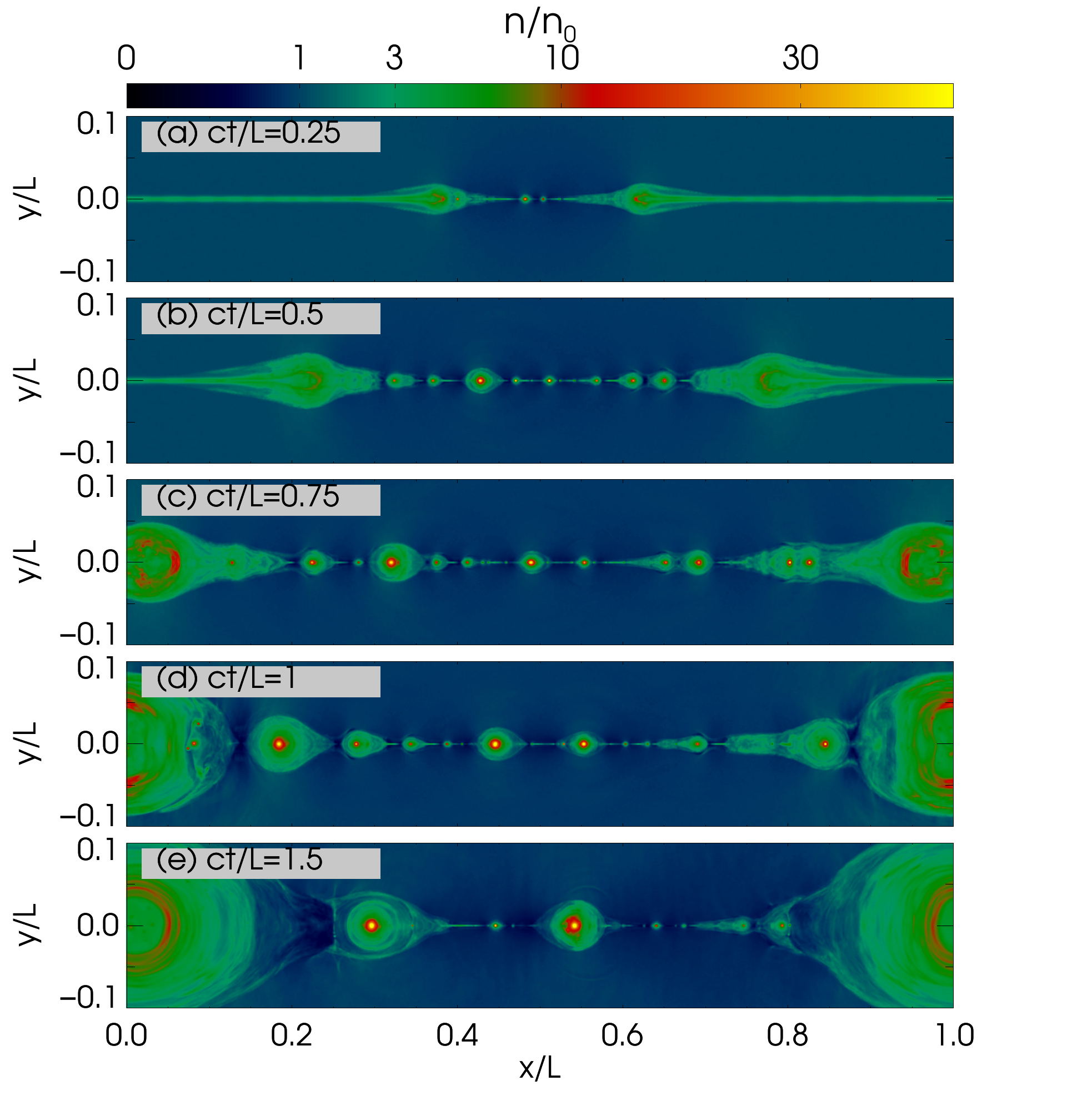}
 \caption{Same as in \fign{layer-1} but for $\sigma=1, \Theta_e=10$, $\kappa=19$, $L/\rLi\simeq130$ (left) and  $\sigma=3, \Theta_e=1$, $\kappa=6$, $L/\rLi\simeq122$ (right).
 } 
 \label{fig:layer-2}
\end{figure*}

\begin{figure*}
 \centering
 \includegraphics[width=0.49\textwidth, trim= 0 0 20 0]{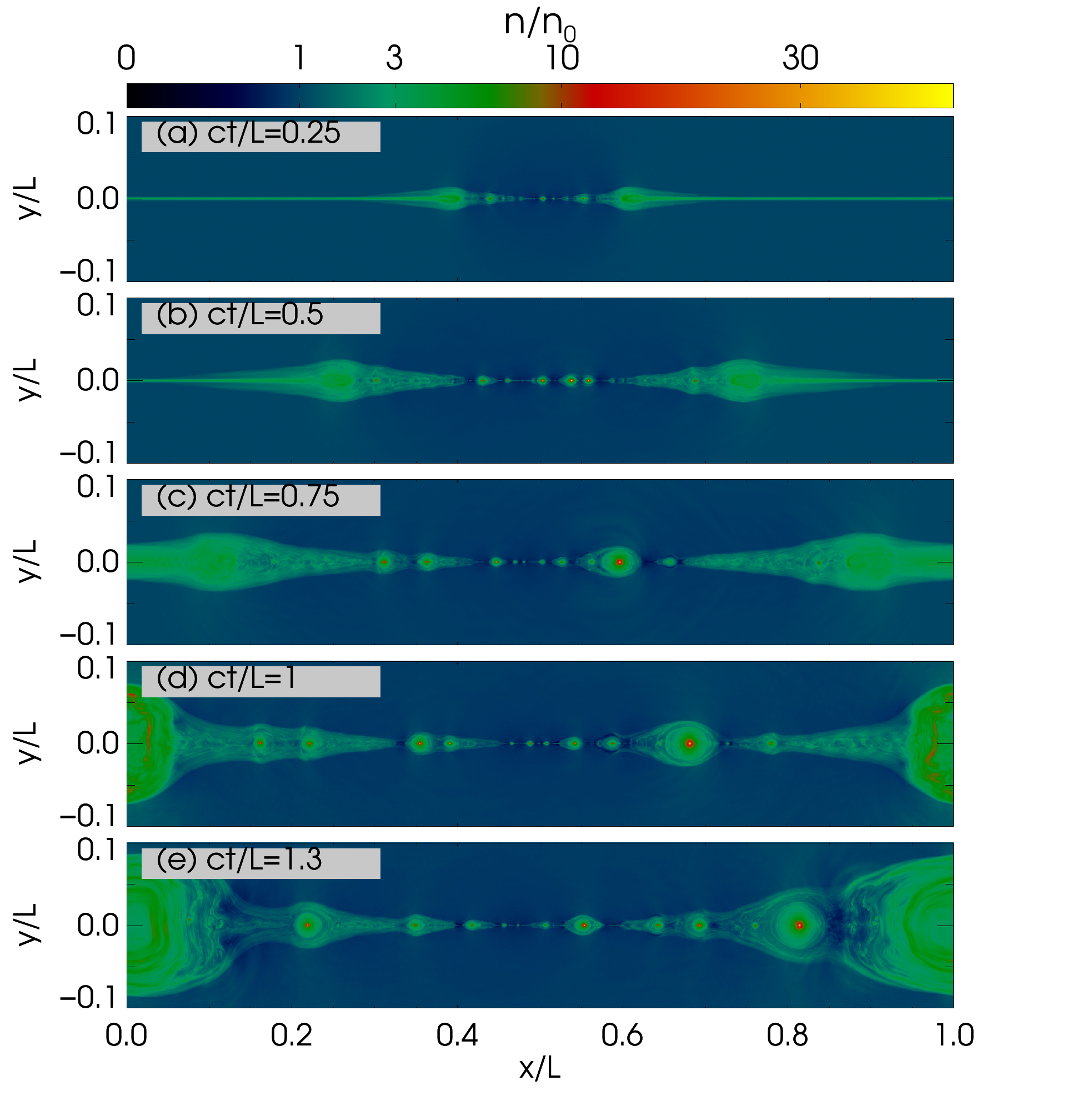}
  \includegraphics[width=0.49\textwidth, trim= 0 0 20 0]{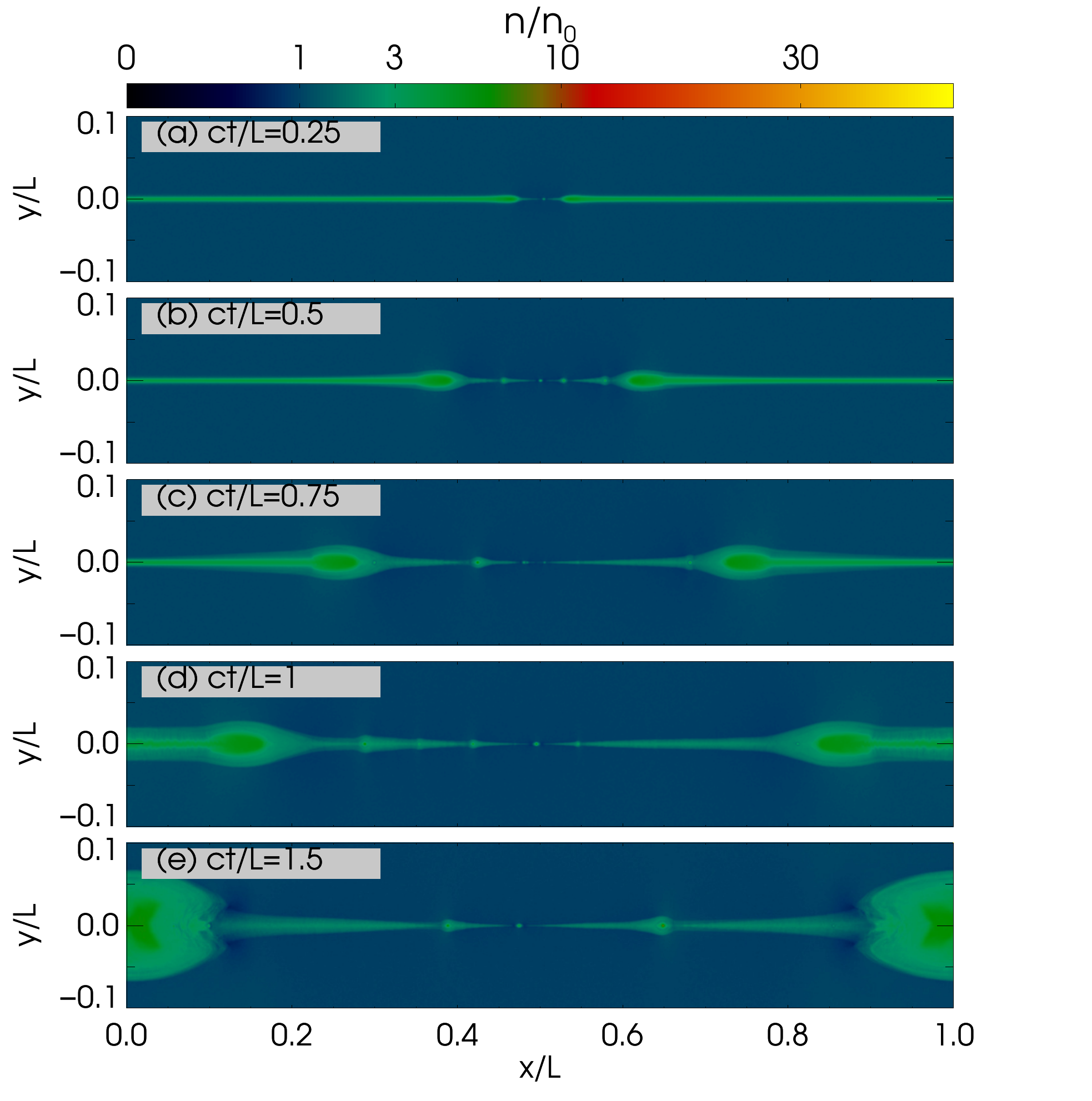}
 \caption{Same as in \fign{layer-1} but for $\sigma=1, \Theta_e=1$, $\kappa=6$, $L/\rLi\simeq211$ (left) and  $\sigma=1, \Theta_e=100$, $\kappa=19$, $L/\rLi\simeq206$ (right).
 } 
 \label{fig:layer-3}
\end{figure*}

In \sect{layer-ionfrac} we explored the effects of the pair multiplicity on the appearance of the layer. More specifically, we showed that the layer becomes more structured (i.e., more secondary plasmoids) as the pair multiplicity decreases, for all other parameters kept the same. One could argue that these differences are merely a result of the different box sizes in terms of the proton skin depth. A straightforward way of checking this possibility is to compare cases with different physical conditions, but similar box sizes in terms of $\rLi$. Snapshots of the density structure from three such pairs of simulations are presented in Figs.~\ref{fig:layer-1}-\ref{fig:layer-3}. 
These comparative plots clearly show that the appearance of the layer is significantly affected by the plasma conditions.

\begin{figure*}
 \centering 
 \includegraphics[width=0.49\textwidth, trim=0 0 0 0]{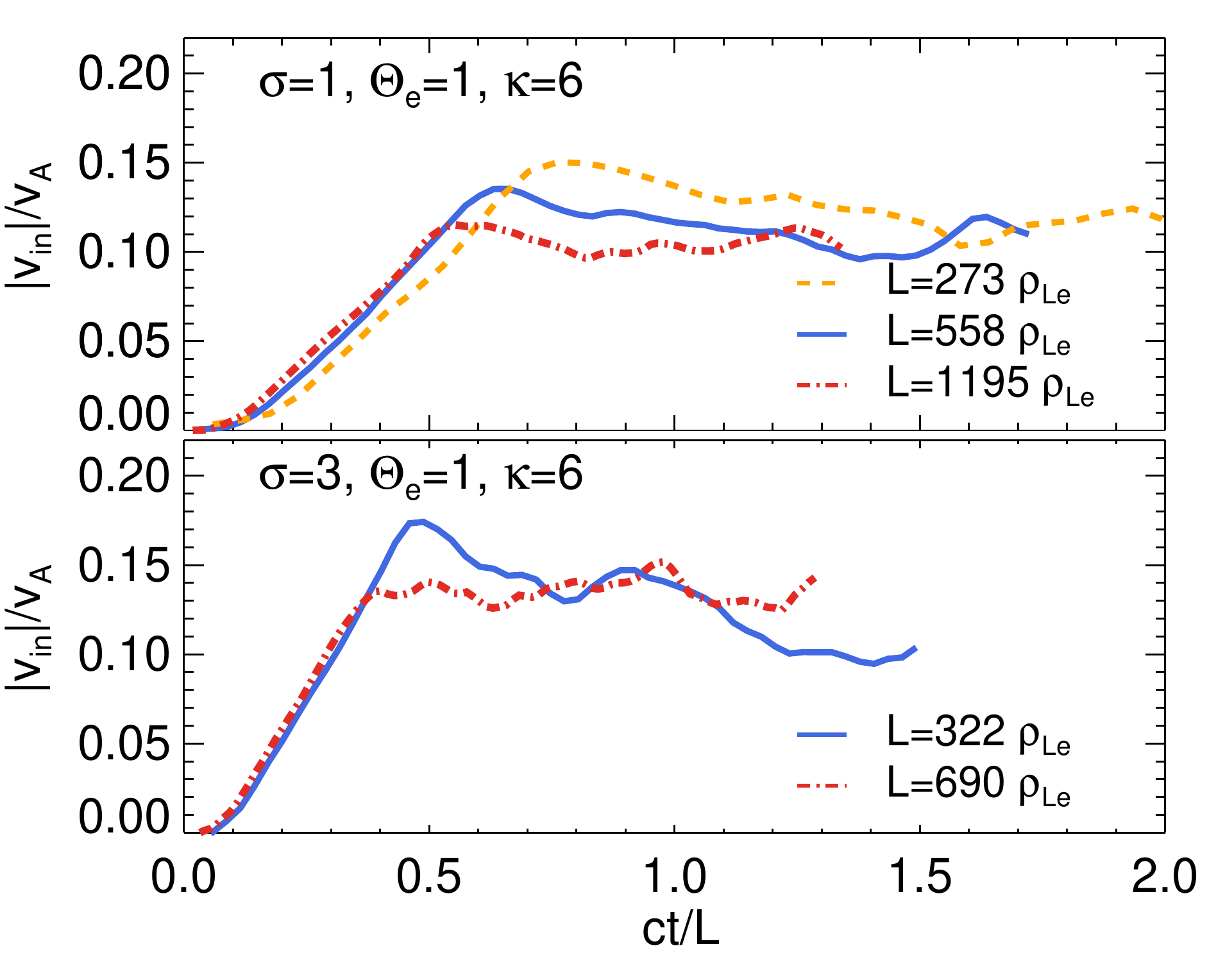}
 \includegraphics[width=0.49\textwidth, trim=0 0 0 0]{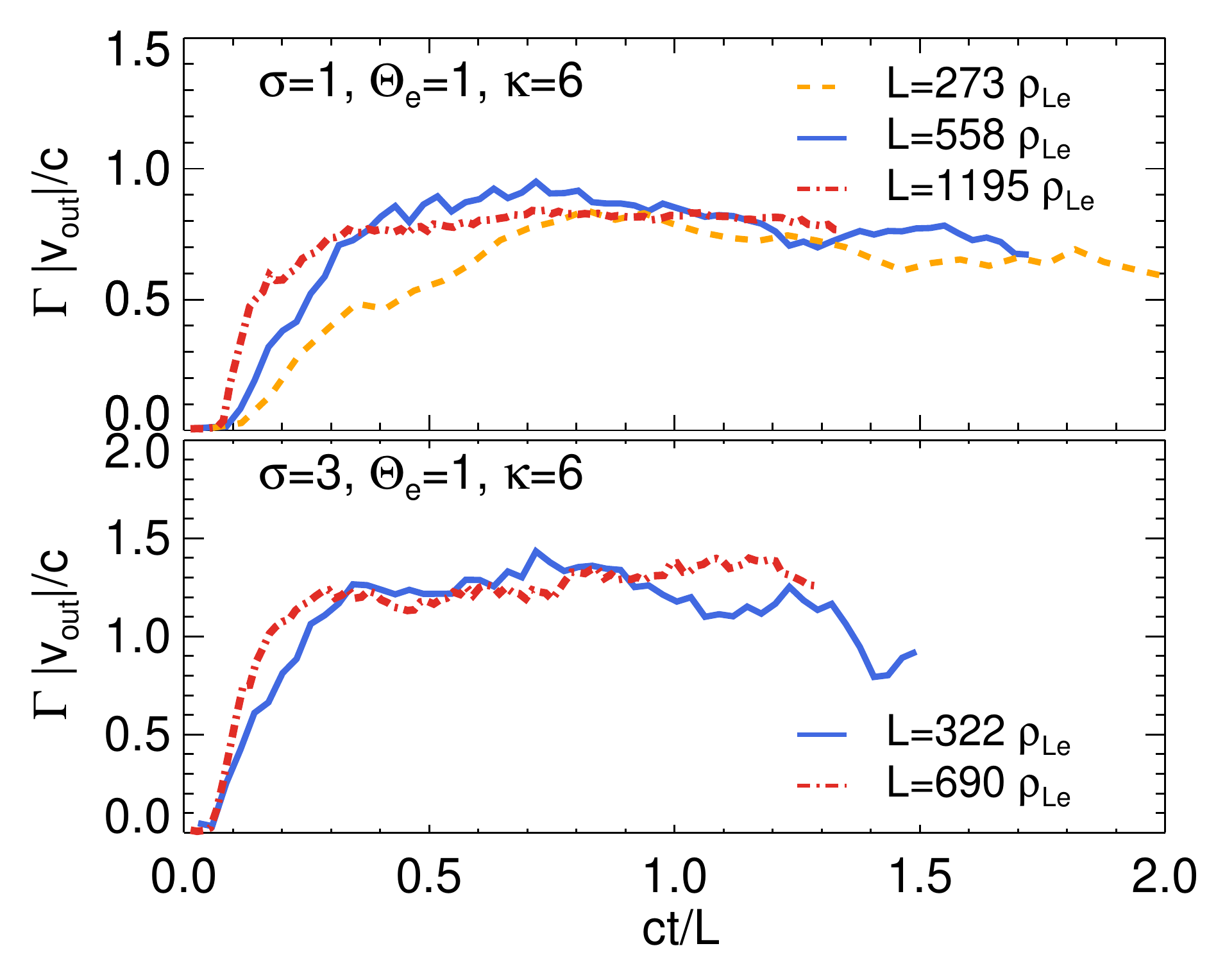}
 \caption{Temporal evolution of the inflow speed  (left panel) and the outflow four-velocity (right panel) from simulations of reconnection in plasmas with $\sigma=1, \Theta_e=1, \kappa=6$ (top panels) and $\sigma=3, \Theta_e=1, \kappa=6$ (bottom panels) for different box sizes marked on the plot (see runs A3-A5, C4-C5 in \tab{setup}).}
 \label{fig:flow-boxsize}
\end{figure*}

\begin{figure*}
 \centering 
 \includegraphics[width=0.32\textwidth, trim=0 0 0 0]{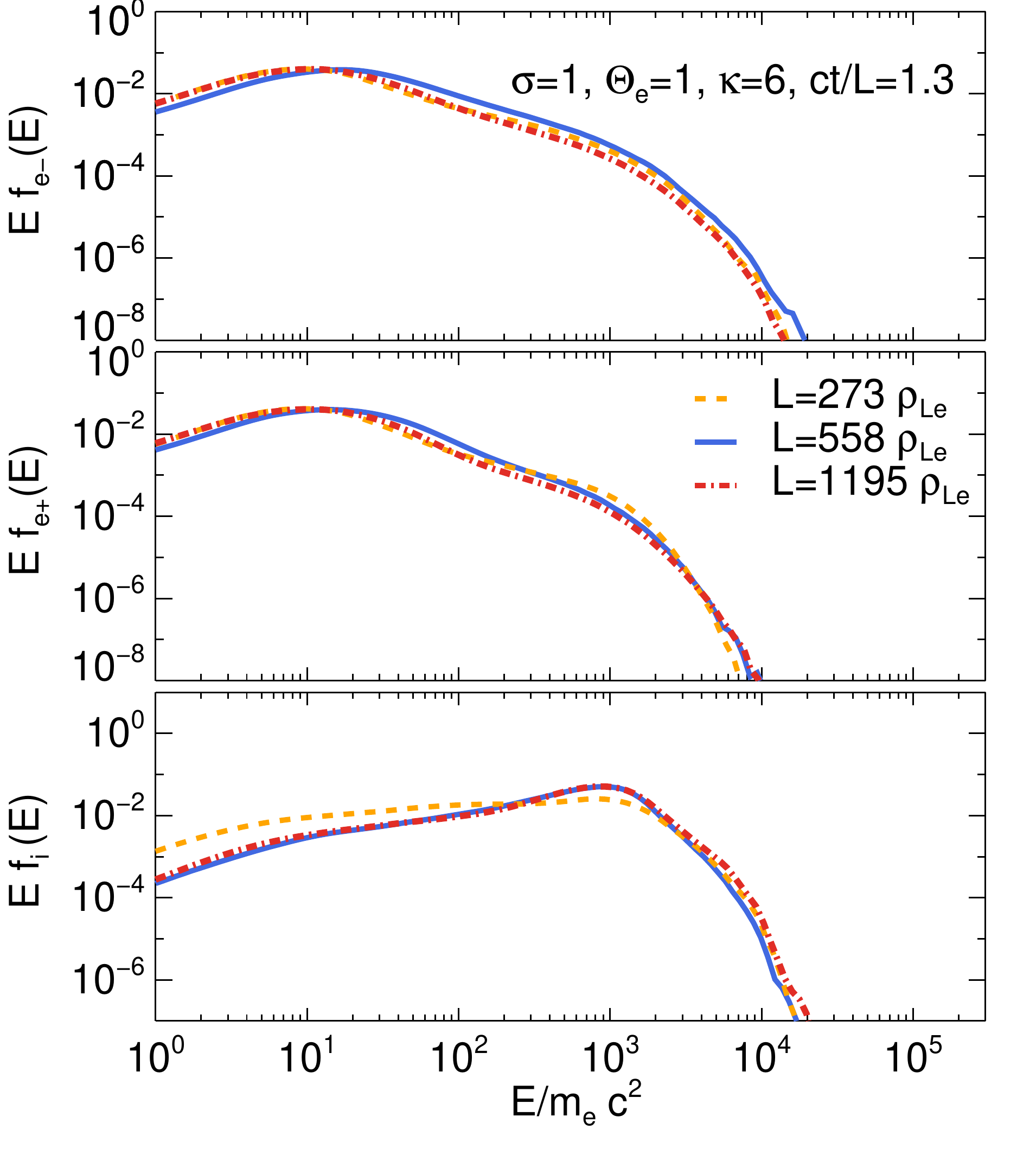}
 \includegraphics[width=0.32\textwidth, trim=0 0 0 0]{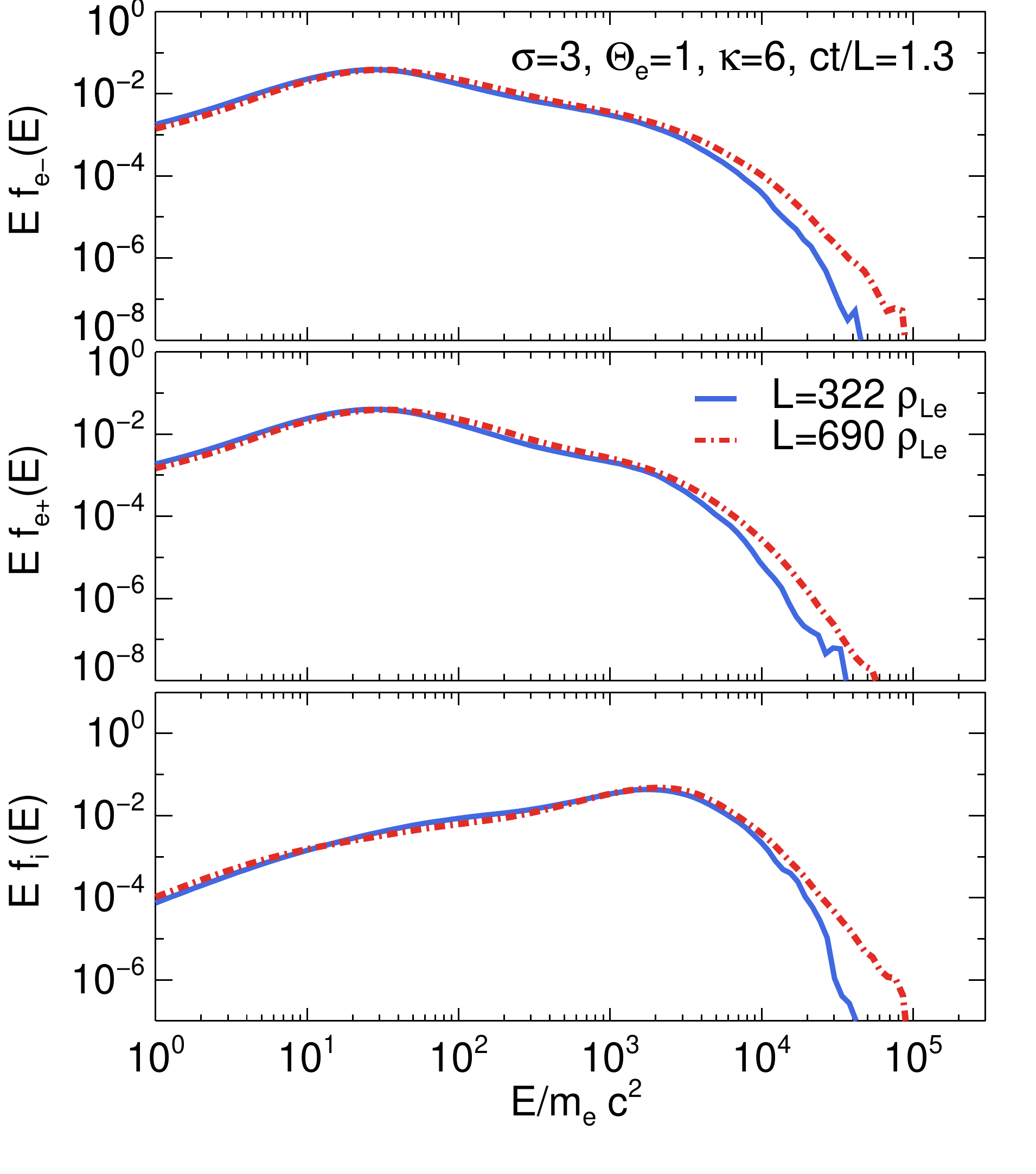}
  \includegraphics[width=0.31\textwidth, trim=0 0 0 0]{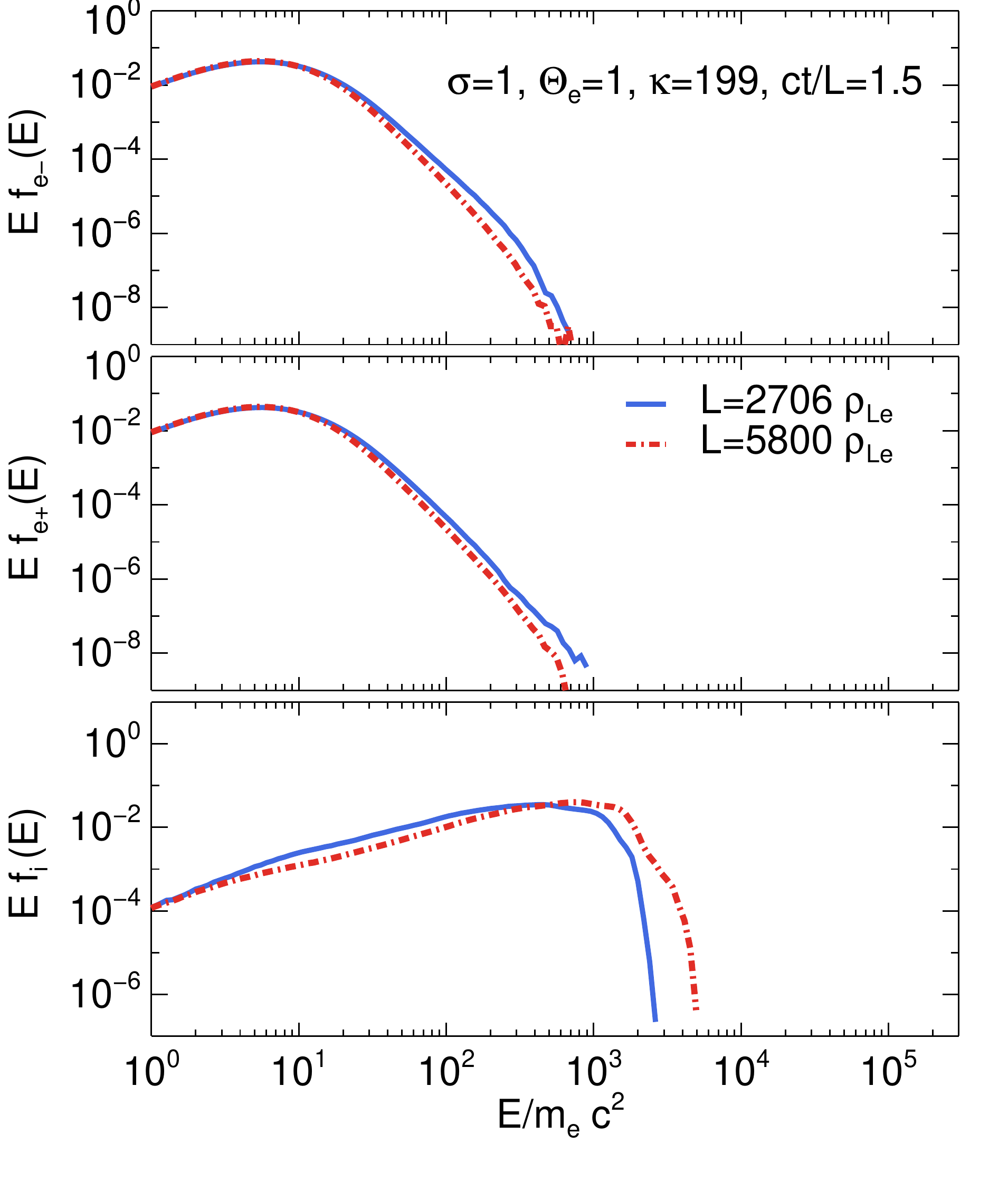}
 \caption{Post-reconnection electron, positron, and proton energy distributions computed from simulations of reconnection in plasmas with $\sigma=1, \Theta_e=1, \kappa=6$ (left panel), $\sigma=3, \Theta_e=1, \kappa=6$ (middle panel), and $\sigma=1, \Theta_e=1, \kappa=199$ (right panel) for different box sizes marked on the plot (see runs A3-A5, C4-C5, A0 and A9 in \tab{setup}). The spectra are computed at the same time (in units of $L/c$) and are normalized to the total number of protons within the reconnection region at that time.}
 \label{fig:spec-boxsize}
\end{figure*}

\section{Effects of box size}\label{sec:app2}
We discuss the effect of the box size on the inflow and outflow rates as well as on the post-reconnection particle energy distributions. 

We selected two simulations (see runs A3-A5, C4-C5 in \tab{setup}) and varied the box size in the $x$-direction, as indicated in \fign{flow-boxsize}. Although the peak inflow rate is systematically higher for smaller box sizes, the difference is less than $\sim 3-5\%$. The temporal evolution of the reconnection rate is similar for all box sizes  (top panel in \fign{flow-boxsize}), until  the formation of the boundary island inhibits the inflow of plasma in the reconnection region, as shown in the bottom panel (blue line). The asymptotic outflow four-velocity is independent of the box size, even for layer lengths of only a few hundred $\rLe$.

Snapshots of the post-reconnection particle energy distributions from simulations with different box sizes are shown in \fign{spec-boxsize}. The power-law segment of the pair energy spectra is similar for the different cases, suggesting a saturation of the power-law slope already for boxes as small as $L\sim 300 \rLe$ \citep[see also][]{ball_18}. Thus, we are confident that the power-law slopes we report in \sect{spectra-slope} (\fign{index}), which were obtained for the spectra plotted with blue lines in \fign{spec-boxsize}, are robust. The high-energy cutoff of the pair distribution, however, increases (almost linearly) with increasing box size, as shown more clearly in the right plot of \fign{spec-boxsize}. Even larger domains are needed for capturing the asymptotic temporal evolution of the cutoff energy. The proton distribution depends strongly on the box size, for both $\sigma$ values we considered. A well-developed power-law forms above the peak proton energy in the largest simulations, thus supporting the argument that reconnection results in extended non-thermal proton distributions (see also \sect{spectra-slope}). 

\begin{figure}
    \centering
    \includegraphics[width=0.45\textwidth]{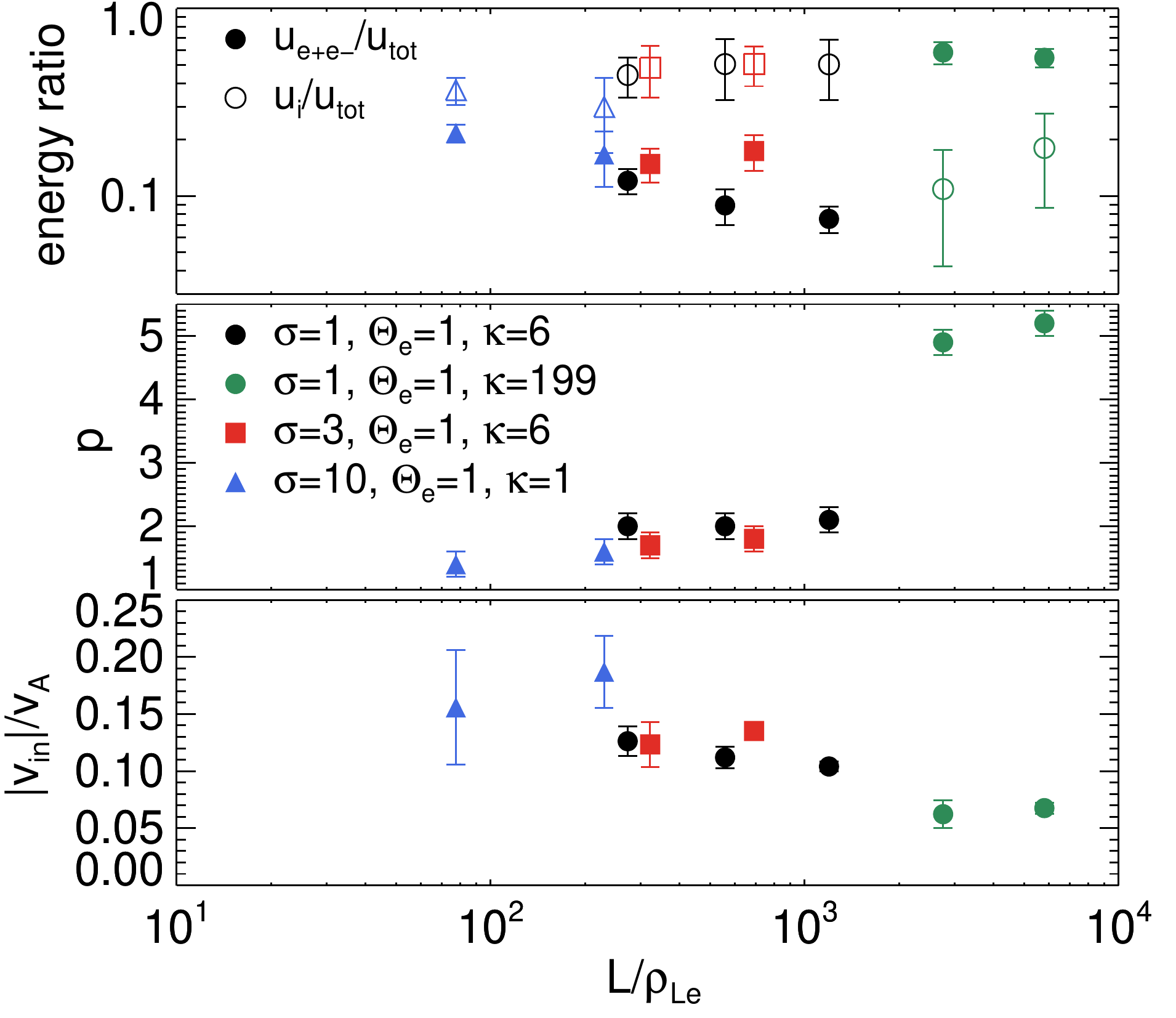}
    \caption{Summary plot showing the dependence of various quantities on the size of the simulation box. From top to bottom: energy ratios for pairs (filled symbols) and protons (open symbols), power-law slope of the lepton energy spectrum (as computed at the end of each simulation), and time-averaged reconnection rate. Error bars in the top and bottom panels indicate the standard deviation during the course of the simulation. A systematic error of $\pm 0.2$ is assigned in all power-law slopes (middle panel).}
    \label{fig:summary-boxsize}
\end{figure}

The effects of the box size on the quantities discussed above and in \sect{energy} are summarized in \fign{summary-boxsize}. The outflow four-velocities are not included in this plot, because they are almost the same for the box sizes we considered.

\section{Dependence of the mean lepton Lorentz factor on physical parameters}\label{sec:app3}
The mean energy of the relativistic pair distribution is of astrophysical importance, as it can be imprinted on the radiated non-thermal photon spectra (for details, see \sect{discussion}). We therefore attempted to quantify the dependence of mean lepton Lorentz factor on the physical parameters ($\sigma,\Theta_e$, and $\kappa$) using a proxy of $\langle\gamma_e-1\rangle$, as defined in \sect{energy}. We caution the reader that the latter does not necessarily refer to a pure power-law energy distribution. In fact, the definition of $\langle\gamma_e-1\rangle$ is agnostic to the shape of the lepton energy distribution.

In general, we find that $\langle\gamma_e-1\rangle$ can be described by:
\eqb 
\langle\gamma_e-1\rangle = a(\sigma, \Theta) \kappa^{-\chi(\sigma, \Theta)}+b(\sigma,\Theta),
\label{eq:gammae_fit}
\eqe 
where $a, b$, and $\chi$ are obtained from a $\chi^2$ fit to the data. The best-fit values and the associated $1\sigma$ statistical errors are summarized in \tab{fit}. We note that cases with $\sigma=1, \Theta_e=0.1$; $\sigma=10, \Theta_e=1$; and $\sigma=10, \Theta_e=10$ are excluded from the fit, since the number of $\kappa$ values is the same or less than the free parameters of \eq{gammae_fit}. Nevertheless, we still find that $\langle\gamma_e-1\rangle \propto \kappa^{-1}$.

\begin{table}
    \centering
    \caption{Parameter values (with their 1$\sigma$ statistical errors) obtained from a $\chi^2$ fit of \eq{gammae_fit} to the mean lepton Lorentz factor derived from our simulations for different $\sigma$ and $\Theta_e$ values. We exclude cases with less data points than the number of free model parameters.}
    \begin{tabular}{ccccc}
    \hline
    $\sigma$ & $\Theta_e$ & $a$ & $\chi$ & $b$ \\
    \hline 
    1   &   1 &  $114.9\pm18.1$ & $1.5\pm0.2$ & $6.5\pm0.7$\\
    1   &  10 &  $149.7\pm17.6$ & $1.3\pm0.2$ & $46.8\pm2.2$\\
    1   &  100 & $328.9\pm7.1$ & $1.6\pm0.1$ & $456.7\pm3.4$\\
    3   & 1  & $565.7\pm 73.6$ & $1.1\pm0.1$ & $13.5\pm2.1$\\
    3   & 10 & $603.7\pm 36.7$ & $1.2\pm 0.1$ & $85.7\pm 7.0$\\
    \hline
    \end{tabular} 
    \label{tab:fit}
\end{table}
\begin{figure*}
 \centering 
 \includegraphics[width=0.31\textwidth, trim=0 0 0 0]{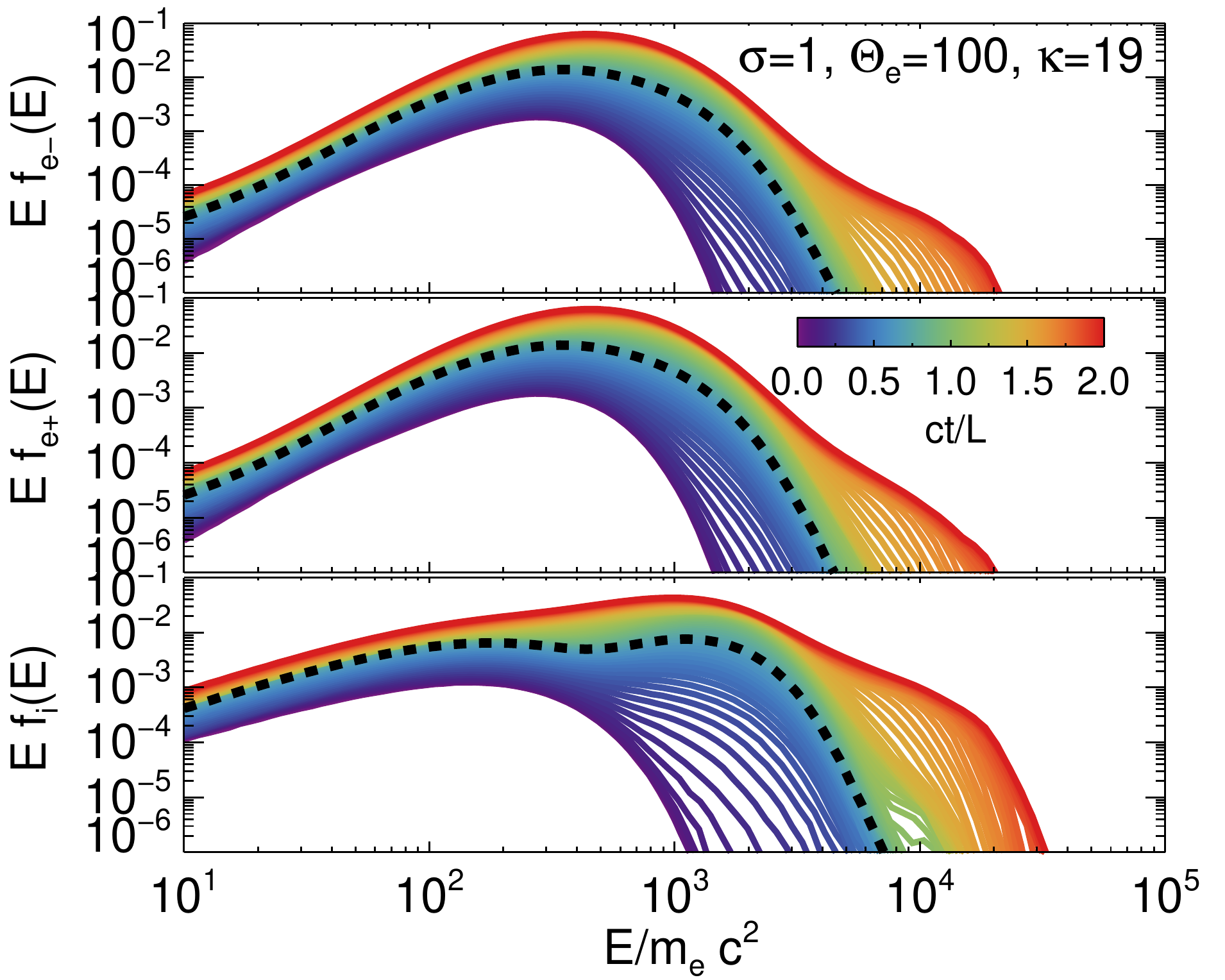}
  \includegraphics[width=0.32\textwidth, trim=0 20 0 0]{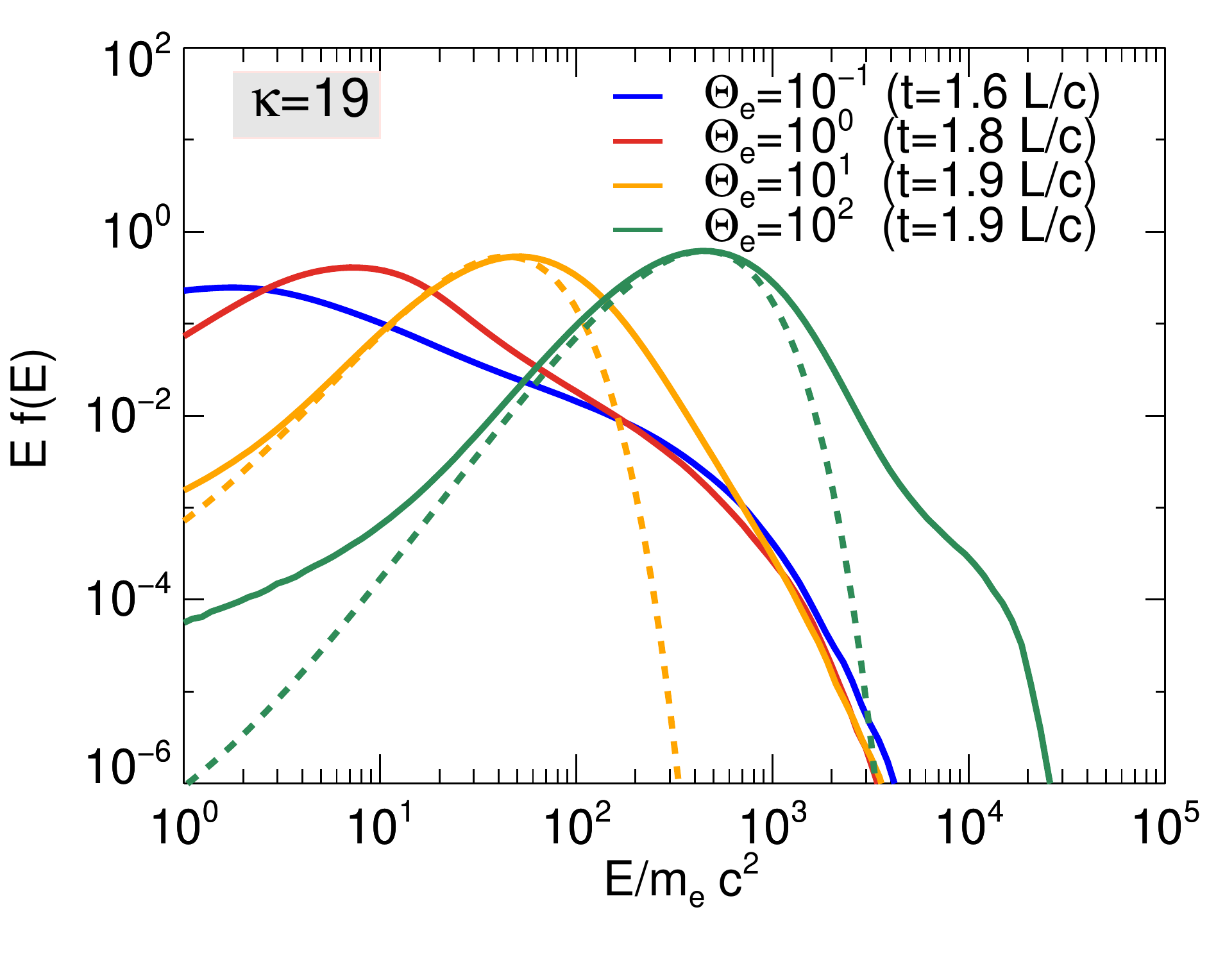} 
    \includegraphics[width=0.32\textwidth, trim=0 20 0 0]{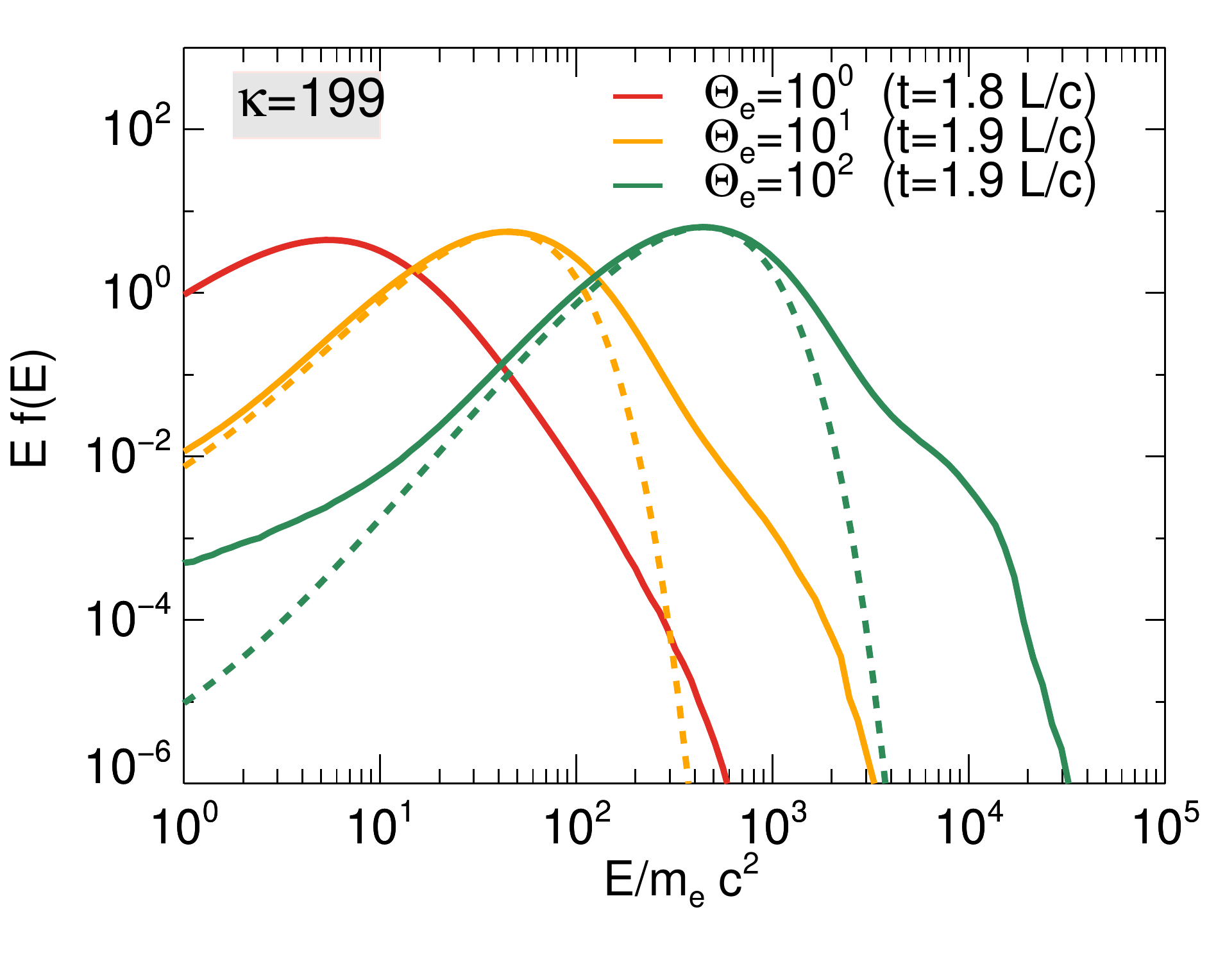}
 \caption{Left panel: Same as in \fign{spec-time} but for  $\Theta_e=100$, $\sigma=1$, and $\kappa=19$. Middle panel: Electron energy distributions from the post-reconnection region (solid lines) for simulations with $\sigma=1$, $\kappa=19$, and different temperatures marked on the plot. All spectra are computed at the end of each simulation (see inset legend) and are normalized to the total number of protons within the reconnection region at that time. A relativistic Maxwell-J{\"u}ttner distribution with temperature $1.5\Theta_e$ (normalized to match the maximum value of the respective electron energy distribution) is also plotted for comparison for $\Theta_e=10$ and 100 (dashed lines).  Right panel: Same as in the middle panel, but for $\kappa=199$.}
 \label{fig:spec-time-hot}
\end{figure*}  

\section{Effects of plasma temperature on pair energy spectra}\label{sec:app4}
The post-reconnection particle energy distributions obtained for the highest temperature simulations (H1-H3 in \tab{setup}) show a high-energy component that forms at late times, as illustrated in \fign{spec-time-hot} (left panel). This can be described by a power law with slope $p\sim 3.2-3.6$ (see \tab{index-gammae}), which is harder than the power laws obtained for lower temperatures but similar $\sigeh$ values (see \fign{index}). 

Snapshots of the pair energy distributions from simulations with the same magnetization and multiplicity, but different plasma temperatures, are shown in the middle and right panels of \fign{spec-time-hot}. Although for $\Theta_e=100$ there is a prominent high-energy component in the distributions that is independent of $\kappa$, we see a hint of this component at lower temperatures ($\Theta_e=10$) only at $\kappa=199$ (right panel). These results imply that the high-energy component of the spectrum is not just related to the  plasma temperature. The common denominator in all the cases that show the high-energy component is the high $\beta_e$ (i.e., $\beta_e >0.1$; see \tab{index-gammae}). 

Similar results have been reported by \cite{ball_18} for trans-relativistic reconnection in electron-proton plasmas with high $\beta_e$ approaching the maximum value $1/4\sigma$ (when both electrons and protons start as relativistically hot). The formation of the high-energy component was attributed to a Fermi-like acceleration of particles with initial energy $\sim k T_e$ bouncing between the reconnection outflow and the stationary boundary island (see Sect.~6.3 in \cite{ball_18}). The fact that it takes some time for the boundary island to grow, it is in agreement with the late-time formation of the high-energy component in the spectrum. 

\bibliographystyle{aasjournal} 
\bibliography{rec.bib}
\end{document}